# DARK MATTER, A NEW PROOF OF THE PREDICTIVE POWER OF GENERAL RELATIVITY

*Stéphane Le Corre (E-mail : le.corre.stephane@hotmail.fr)*
*No affiliation*

Without observational or theoretical modifications, Newtonian and general relativity seem to be unable to explain gravitational behavior of large structure of the universe. The assumption of dark matter solves this problem without modifying theories. But it implies that most of the matter in the universe must be unobserved matter. Another solution is to modify gravitation laws. In this article, we study a third way that doesn't modify gravitation neither matter's distribution, by using a new physical assumption on the clusters. Compare with Newtonian gravitation, general relativity (in its linearized approximation) leads to add a new component without changing the gravity field. As already known, this component for galaxies is too small to explain dark matter. But we will see that the galaxies' clusters can generate a significant component and embed large structure of universe. We show that the magnitude of this embedding component is small enough to be in agreement with current experimental results, undetectable at our scale, but detectable at the scale of the galaxies and explain dark matter, in particular the rotation speed of galaxies, the rotation speed of dwarf satellite galaxies, the expected quantity of dark matter inside galaxies and the expected experimental values of parameters $\Omega_{\mathrm{dm}}$ of dark matter measured in CMB. This solution implies testable consequences that differentiate it from other theories: decreasing dark matter with the distance to the cluster's center, large quantity of dark matter for galaxies close to the cluster's center, isolation of galaxies without dark matter, movement of dwarf satellite galaxies in planes close to the supergalactic plane, close orientations of spin's vectors of two close clusters, orientation of nearly all the spin's vector of galaxies of a same cluster in a same half-space, existence of very rare galaxies with two portions of their disk that rotate in opposite directions…

## 1. Overview

### 1.1. Current solutions

Why is there a dark matter assumption? Starting with the observed matter, the gravitation theories (Newtonian and general relativity) seem to fail to explain several observations. The discrepancies between the theory and the observations appear only for large astrophysical structure and with orders of magnitude that let no doubt that something is missing in our knowledge of gravitation interaction. There are mainly three situations that make necessary this assumption. At the scale of the galaxies, the rotation's speeds of the ends of the galaxies can be explained if we suppose the presence of more matter than the observed one. This invisible matter should represent more than 90% (RUBIN et al., 1980) of the total matter (sum of the observed and invisible matters). At the scale of the clusters of galaxies, the gravitational lensing observations can be explained with the presence of much more invisible matter (ZWICKY, 1937; TAYLOR et al., 1998; WU et al., 1998), at least 10 times the observed matter. At the scale of the Universe, cosmological equations can very well explain the dynamics of our Universe (for example the inhomogeneities of the microwave background) with the presence of more matter than the observed one. This invisible matter should, in this third situation, represent about 5 times the observed matter (PLANCK Collaboration, 2014).
How can we explain the origin of this dark matter? In fact, a more objective question should be how we can explain these discrepancies between theories and observations. There are two ways to solve this problem, supposing that what we observe is incomplete or supposing that what we idealize is incomplete. The first one is to suppose the existence of an invisible matter (just like we have done previously to quantify the discrepancies). It is the famous dark matter assumption that is the most widely accepted assumption. This assumption has the advantage of keeping unchanged the gravitation theories (NEGI, 2004). The inconvenient is that until now, no dark matter has been observed, WIMPS (ANGLOHER et al., 2014; DAVIS, 2014), neutralinos (AMS Collaboration, 2014), axions (HARRIS & CHADWICK, 2014). More than this the dynamics with the dark matter assumption leads to several discrepancies with observations, on the number of dwarf galaxies (MATEO, 1998; MOORE et al., 1999; KLYPIN et al., 1999) and on the distribution of dark matter (DE BLOK, 2009; HUI, 2001). The second one is to modify the gravitation theories. The advantage of this approach is to keep unchanged the quantity of observed matter. The inconvenient is that it modifies our current theories that are very well verified at our scale (planet, star, solar system…). One can gather these modified theories in two categories, one concerning Newtonian idealization and the other concerning general relativity. Briefly, in the Newtonian frame, one found MOND (MILGROM, 1983) (MOdified Newtonian Dynamics) that essentially modified the inertial law and another approach that modifies gravitational law (DISNEY, 1984; WRIGHT & DISNEY, 1990). In the general relativity frame, one found some studies that don't modify Einstein's equations but that looking for specific metrics (LETELIER, 2006; COOPERSTOCK & TIEU, 2005; CARRICK & COOPERSTOCK, 2010). And one found some other studies that modified general relativity. Mainly one has a scalar-tensor-vector gravity (MOFFAT, 2006) (MOG), a 5D general relativity taking into account Hubble expansion (CARMELI, 1998; HARTNETT, 2006), a quantum general relativity (RODRIGUES et al., 2011), two generalizations of MOND, TeVeS (BEKENSTEIN, 2004; MCKEE, 2008) and BSTV (SANDERS, 2005), a phenomenological covariant approach (EXIRIFARD, 2010)…

## 1.2. Solution studied in this paper

In our study, we will try a third way of explanation of these discrepancies. This explanation doesn't modify the quantities of observed matter and doesn't modify general relativity. To demonstrate the capacity of the general relativity to explain dark matter, we are going to use, in our paper, the native metric of linearized general relativity (also called gravitoelectromagnetism) in agreement with the expected domain of validity of our study ($\varphi \ll c^2$ and $v \ll c$). It doesn't modify the gravity field of Newtonian approximation but defines a better approximation by adding a component (called gravitic field, similar to magnetic field of Maxwell idealization) correcting some imperfections of the Newtonian idealization (in particular the infinite propagation speed of gravitation). In a first paragraph, we will recall the linearization of general relativity and give some orders of magnitude of this component, the gravitic field. In a second paragraph, we will focus first on the problem of the rotation of the extremities of galaxies on sixteen galaxies. We will retrieve a published result (LETELIER, 2006), showing that own gravitic field of a galaxy cannot explain dark matter of galaxies. But one will see that clusters' gravitic field could replace dark matter assumption. Before talking about its origin, one will see the orders of magnitude of this component, expected to obtain the "flat" rotation speed of galaxies. Its magnitude will be small enough to be in agreement with the fact that it is undetectable and that the galaxies are randomly oriented. In a third paragraph, we will talk about the possible origins of this component. One will see that the value of the gravitic field explaining dark matter for galaxies could likely come from the cluster of galaxies (in agreement with a recent observation showing that dark matter quantity decreases with the distance to the center of galaxies' cluster). In the next paragraphs, one will use these values of gravitic field on several situations. It will permit to retrieve many observations (some of them not explained). These values (if interpreted in the hypothesis of dark matter) will give the expected quantity of dark matter inside the galaxies. These same values will explain the speed of dwarf satellite galaxies. Its theoretical expression will explain their planar movement (recently observed and unexplained). We will also show that we can obtain a good order of magnitude for the quantity of dark matter $\Omega_{dm}$ (obtained from the inhomogeneities of the microwave background, CMB). One will still make some predictions and in particular an original and necessary consequence on the dwarf galaxies that can differentiate our solution compared to the dark matter assumption. We will end with a comparison between our solution and the dark matter assumption. Some recent articles (CLOWE et al., 2006; HARVEY et al., 2015) show that if dark matter is actually matter, it cannot be ordinary matter. We will see at the end that these studies do not invalidate the solution proposed in this paper. Instead they could even allow testing our solution.

The goal of this study is to open a new way of investigation to explain dark matter. This work has not the pretention to produce a definitive proof (in particular it will need to be extended to the general frame of general relativity and not only in its linearization approximation). But it nevertheless provides a body of evidence that confirm the relevance of the proposed solution and leads to several significant predictions that differentiate it from the other explanations.

To end this introduction, one can insist on the fact that this solution is naturally compliant with general relativity because it is founded on a component coming from general relativity (gravitic field), traditionally neglected, on which several papers have been published and some experimental tests have been realized. The only assumption that will be made in this paper is that there are some large astrophysical structures that can generate a significant value of gravitic field. And we will see that the galaxies' cluster can generate it.

## 2. Gravitation in linearized general relativity

Our study will focus on the equations of general relativity in weak field. These equations are obtained from the linearization of general relativity (also called gravitoelectromagnetism). They are very close to the modeling of electromagnetism. Let's recall the equations of linearized general relativity.

### 2.1. Theory

From general relativity, one deduces the linearized general relativity in the approximation of a quasi-flat Minkowski space ($g^{\mu\nu} = \eta^{\mu\nu} + h^{\mu\nu}$ ; $|h^{\mu\nu}| \ll 1$). With following Lorentz gauge, it gives the following field equations (HOBSON et al., 2009) (with $\square = \frac{1}{c^2}\frac{\partial^2}{\partial t^2} - \Delta$ ):

$$\partial_\mu \bar{h}^{\mu\nu} = 0 \quad ; \quad \square \bar{h}^{\mu\nu} = -2\frac{8\pi G}{c^4}T^{\mu\nu} \qquad (I)$$

With:

$$\bar{h}^{\mu\nu} = h^{\mu\nu} - \frac{1}{2}\eta^{\mu\nu}h \ ; \ h \equiv h^\sigma_\sigma \ ; \ h^\mu_\nu = \eta^{\mu\sigma}h_{\sigma\nu} \ ; \ \bar{h} = -h \qquad (II)$$

The general solution of these equations is:

$$\bar{h}^{\mu\nu}(ct, \vec{x}) = -\frac{4G}{c^4}\int \frac{T^{\mu\nu}(ct - |\vec{x}-\vec{y}|, \vec{y})}{|\vec{x}-\vec{y}|}d^3\vec{y}$$

In the approximation of a source with low speed, one has:

$$T^{00} = \rho c^2 \ ; \ T^{0i} = c\rho u^i \ ; \ T^{ij} = \rho u^i u^j$$

And for a stationary solution, one has:

$$\bar{h}^{\mu\nu}(\vec{x}) = -\frac{4G}{c^4}\int \frac{T^{\mu\nu}(\vec{y})}{|\vec{x}-\vec{y}|}d^3\vec{y}$$

At this step, by proximity with electromagnetism, one traditionally defines a scalar potential $\varphi$ and a vector potential $H^i$. There are in the literature several definitions (MASHHOON, 2008) for the vector potential $H^i$. In our study, we are going to define:

$$\bar{h}^{00} = \frac{4\varphi}{c^2} \ ; \ \bar{h}^{0i} = \frac{4H^i}{c} \ ; \ \bar{h}^{ij} = 0$$

With gravitational scalar potential $\varphi$ and gravitational vector potential $H^i$:

$$\varphi(\vec{x}) \equiv -G\int \frac{\rho(\vec{y})}{|\vec{x}-\vec{y}|}d^3\vec{y}$$

$$H^i(\vec{x}) \equiv -\frac{G}{c^2}\int \frac{\rho(\vec{y})u^i(\vec{y})}{|\vec{x}-\vec{y}|}d^3\vec{y} = -K^{-1}\int \frac{\rho(\vec{y})u^i(\vec{y})}{|\vec{x}-\vec{y}|}d^3\vec{y}$$

With $K$ a new constant defined by:

$$GK = c^2$$

This definition gives $K^{-1} \sim 7.4 \times 10^{-28}$ very small compare to $G$.

The field equations $(I)$ can be then written (Poisson equations):

$$\Delta\varphi = 4\pi G\rho \ ; \ \Delta H^i = \frac{4\pi G}{c^2}\rho u^i = 4\pi K^{-1}\rho u^i \quad (III)$$

With the following definitions of $\vec{g}$ (gravity field) and $\vec{k}$ (gravitic field), those relations can be obtained from following equations:

$$\vec{g} = -\overrightarrow{grad}\varphi \ ; \ \vec{k} = \overrightarrow{rot}\,\vec{H}$$
$$\overrightarrow{rot}\vec{g} = 0 \ ; \ div\,\vec{k} = 0 \ ;$$
$$div\,\vec{g} = -4\pi G\rho \ ; \ \overrightarrow{rot}\vec{k} = -4\pi K^{-1}\vec{j_p}$$

With relations $(II)$, one has:

$$h^{00} = h^{11} = h^{22} = h^{33} = \frac{2\varphi}{c^2} \ ; \ h^{0i} = \frac{4H^i}{c} \ ; \ h^{ij} = 0 \quad (IV)$$

The equations of geodesics in the linear approximation give:

$$\frac{d^2x^i}{dt^2} \sim -\frac{1}{2}c^2\delta^{ij}\partial_j h_{00} - c\delta^{ik}(\partial_k h_{0j} - \partial_j h_{0k})v^j$$

It then leads to the movement equations:

$$\frac{d^2\vec{x}}{dt^2} \sim -\overrightarrow{grad}\varphi + 4\vec{v} \wedge (\overrightarrow{rot}\vec{H}) = \vec{g} + 4\vec{v} \wedge \vec{k}$$

One will need another relation for our next demonstration. In agreement with previous Poisson equations $(III)$, we deduce that gravitic field evolves with $r^{-2}$ ($k \propto r^{-2}$). More precisely, just like in electromagnetism, one can deduce from Poisson equation that $\vec{k} \sim \left(-\frac{1}{K}\right) m_p \left(\frac{1}{r^2}\right) \vec{v_s} \wedge \vec{u}$ but it is its dependence in $r^{-2}$ that is pertinent for our study.

From relation $(IV)$, one deduces the metric in a quasi flat space:

$$ds^2 = \left(1 + \frac{2\varphi}{c^2}\right)c^2dt^2 + \frac{8H_i}{c}cdtdx^i - \left(1 - \frac{2\varphi}{c^2}\right)\sum (dx^i)^2$$

In a quasi-Minkowski space, one has:

$$H_i dx^i = -\delta_{ij}H^j dx^i = -\vec{H}.\overrightarrow{dx}$$

We retrieve the known expression (HOBSON et al., 2009) with our definition of $H_i$:

$$ds^2 = \left(1 + \frac{2\varphi}{c^2}\right)c^2dt^2 - \frac{8\vec{H}.\overrightarrow{dx}}{c}cdt - \left(1 - \frac{2\varphi}{c^2}\right)\sum (dx^i)^2 \quad (V)$$

Remark: Of course, one retrieves all these relations starting with the parametrized post-Newtonian formalism. From (CLIFFORD M. WILL, 2014) one has:

$$g_{0i} = -\frac{1}{2}(4\gamma + 4 + \alpha_1)V_i \ ; \ V_i(\vec{x}) = \frac{G}{c^2}\int \frac{\rho(\vec{y})u_i(\vec{y})}{|\vec{x} - \vec{y}|}d^3\vec{y}$$

The gravitomagnetic field and its acceleration contribution are:

$$\overrightarrow{B_g} = \vec{\nabla} \wedge (g_{0i}\overrightarrow{e^i}) \ ; \ \overrightarrow{a_g} = \vec{v} \wedge \overrightarrow{B_g}$$

And in the case of general relativity (that is our case):

$$\gamma = 1 \ ; \ \alpha_1 = 0$$

It then gives:

$$g_{0i} = -4V_i \ ; \ \overrightarrow{B_g} = \vec{\nabla} \wedge (-4V_i\overrightarrow{e^i})$$

And with our definition:

$$H_i = -\delta_{ij}H^j = \frac{G}{c^2}\int \frac{\rho(\vec{y})\delta_{ij}u^j(\vec{y})}{|\vec{x} - \vec{y}|}d^3\vec{y} = V_i(\vec{x})$$

One then has:

$$g_{0i} = -4H_i \ ; \ \overrightarrow{B_g} = \vec{\nabla} \wedge (-4H_i\overrightarrow{e^i}) = \vec{\nabla} \wedge (4\delta_{ij}H^j\overrightarrow{e^i})$$
$$= 4\vec{\nabla} \wedge \vec{H}$$
$$\overrightarrow{B_g} = 4\overrightarrow{rot}\,\vec{H}$$

With the following definition of gravitic field:

$$\vec{k} = \frac{\overrightarrow{B_g}}{4}$$

One then retrieves our previous relations:

$$\vec{k} = \overrightarrow{rot}\,\vec{H} \ ; \ \overrightarrow{a_g} = \vec{v} \wedge \overrightarrow{B_g} = 4\vec{v} \wedge \vec{k}$$

A last remark: The interest of our notation is that the field equations are strictly equivalent to Maxwell idealization. Only the movement equations are different with the factor "4". But of course, all the results of our study could be obtained in the traditional notation of gravitomagnetism with the relation $\vec{k} = \frac{\overrightarrow{B_g}}{4}$.

To summarize Newtonian gravitation is a traditional approximation of general relativity. But linearized general relativity shows that there is a better approximation, equivalent to Maxwell idealization in term of field equation, by adding a gravitic field very small compare to gravity field at our scale. And, as we are going to see it, this approximation can also be approximated by Newtonian gravitation for many situations where gravitic field can be neglected. In other words, linearized general relativity explains how, in weak field or quasi flat space, general relativity improves Newtonian gravitation by adding a component (that will become significant at the scales of clusters of galaxies as we will see it).

In this approximation (linearization), the non linear terms are naturally neglected (gravitational mass is invariant and gravitation doesn't act on itself). This approximation is valid only for low speed of source and weak field (domain of validity of our study).

All these relations come from general relativity and it is in this theoretical frame that we will propose an explanation for dark matter.

## 2.2. Orders of magnitude

The theory used in this study is naturally in agreement with general relativity because it is the approximation of linearized general relativity. But it is interesting to have orders of magnitude for this new gravitic field.

### 2.2.1. Linearized general relativity and classical mechanics

In the classical approximation ($\|\vec{v}\| \ll c$), the linearized general relativity gives the following movement equations ($m_i$ the inertial mass and $m_p$ the gravitational mass):

$$m_i\frac{\overrightarrow{dv}}{dt} = m_p[\vec{g} + 4\vec{v} \wedge \vec{k}]$$

A simple calculation can give an order of magnitude to this new component of the force due to the gravitic field. On one hand, gravity field gives $\|\vec{g}\| \propto G\frac{m_p}{r^2} \sim 6.67 \times 10^{-11}\frac{m_p}{r^2}$, on the other hand, for a speed $\|\overrightarrow{v_s}\| \sim 1m.s^{-1} \ll c$ (speed of the source that generates the field) one has $\|\vec{k}\| \propto \|\overrightarrow{v_s}\|.K^{-1}\frac{m_p}{r^2} \sim 7.4 \times 10^{-28}\frac{m_p}{r^2} \sim 10^{-17}G\frac{m_p}{r^2}$. That is to say that for a test particle speed $\|\vec{v}\| \sim 1m.s^{-1} \ll c$, the gravitic force $\vec{F_k}$ compared to the gravity force $\vec{F_g}$ is about $\|\vec{F_k}\| \sim 10^{-17}\|\vec{F_g}\|$. This new term is extremely small and undetectable with the current precision on Earth.

### 2.2.2. Linearized general relativity and special relativity

Linearized general relativity is valid in the approximation of low speed for source only. But, there isn't this limitation for test particle. One can then consider a test particle of high speed. In the special relativity approximation ($\|\vec{v}\| \sim c$), with $\vec{v}$ the speed of the test particle, the linearized general relativity gives the following movement equations:

$$\frac{\overrightarrow{dp}}{dt} = m_p[\vec{g} + 4\vec{v} \wedge \vec{k}]$$

It is the same equation seen previously, with here the relativistic momentum. We have seen that $\|\vec{k}\| \sim 10^{-17} \|\vec{g}\|$ in classical approach. If $v \sim c$ (speed of the test particle), one always has $\|\vec{v} \wedge \vec{k}\| \sim 10^{-9} \|\vec{g}\|$ which is always very weak compared to the force of gravity ($\|\vec{F_k}\| = 10^{-9} \|\vec{F_g}\|$). Once again, even in the domain of special relativity (high speed of test particle), the gravitic term is undetectable.

To explain dark matter, we are going to see that we need to have $\|\vec{k}\| \sim 10^{-16}$ inside the galaxies. It means that, for our galaxy, our solar system must be embedded in such a gravitic field. For a particle test of speed $v \sim c$, it gives a gravitic acceleration $\|\vec{v} \wedge \vec{k}\| \sim 10^{-8} m.s^{-2}$. Compared to gravity acceleration ($\|\vec{g}\| \sim 300\, m.s^{-2}$) near Sun, it represents an undetectable correction of about $10^{-11}\, \|\vec{g}\|$ (on the deviation of light for example).

## 3. Gravitic field: an explanation of dark matter

One of clues which push to postulate the existence of dark matter is the speed of the ends of the galaxies, higher than what it should be. We will see that the gravitic field can explain these speeds without the dark matter assumption. For that, we will first consider an example (data from "Observatoire de Paris") to demonstrate all our principles. Next, we will apply the traditional computation (KENT, 1987) on sixteen measured curves: NGC 3198, NGC 4736, NGC 300, NGC 2403, NGC 2903, NGC 3031, NGC 5033, NGC 2841, NGC 4258, NGC 4236, NGC 5055, NGC 247, NGC 2259, NGC 7331, NGC 3109 and NGC 224.

### 3.1. Dark matter mystery

Some examples of curves of rotation speeds for some galaxies are given in Fig. 1.

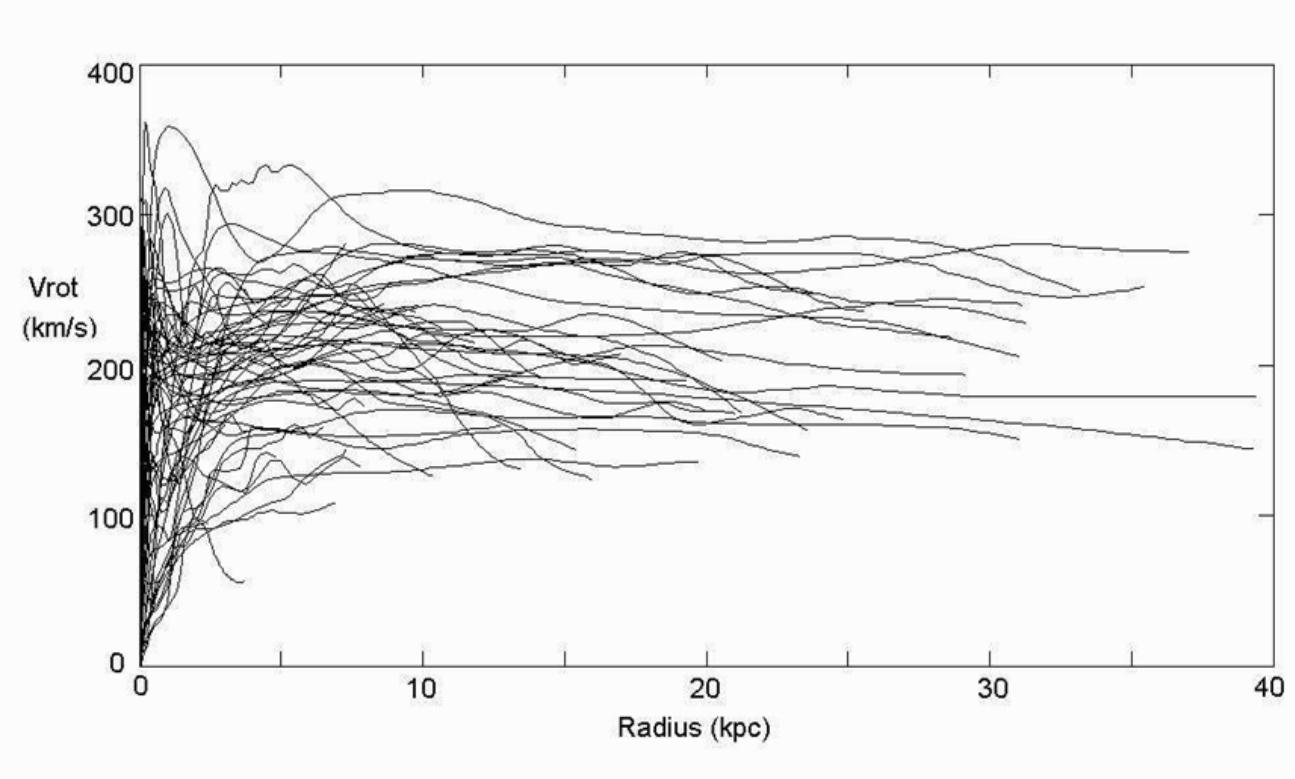

Fig. 1: Superposition of several rotation curves (Sofue et al., 1999)

One can roughly distinguish three zones:

- Zone (I) [0 ; 5 kpc]: close to the center of the galaxy, fast growth rotation speed
- Zone (II) [5 ; 10 kpc]: zone of transition which folds the curve, putting an end to the speed growth
- Zone (III) [10 kpc ; ..[: towards the outside of the galaxy, a "flat" curve with a relative constancy speed (contrary to the decreasing theoretical curve)

Gravitation, for which speed should decrease, failed to explain such curves in zone (III) without dark matter assumption. And these examples are not an exception; it is a general behavior.

### 3.2. Basis of our computation

The dark matter assumption is a way to increase the effect of the gravity force. To be able to have a gravitic force ($\overrightarrow{F_k} = m_p 4\vec{v} \wedge \vec{k}$) that can replace the dark matter, the gravitic field $\vec{k}$ should have a consistent orientation with the speed of the galactic matter to generate a centripetal force (just like the gravity force). This can be performed for example with the situation of the following figure:

Galaxy (schematically)

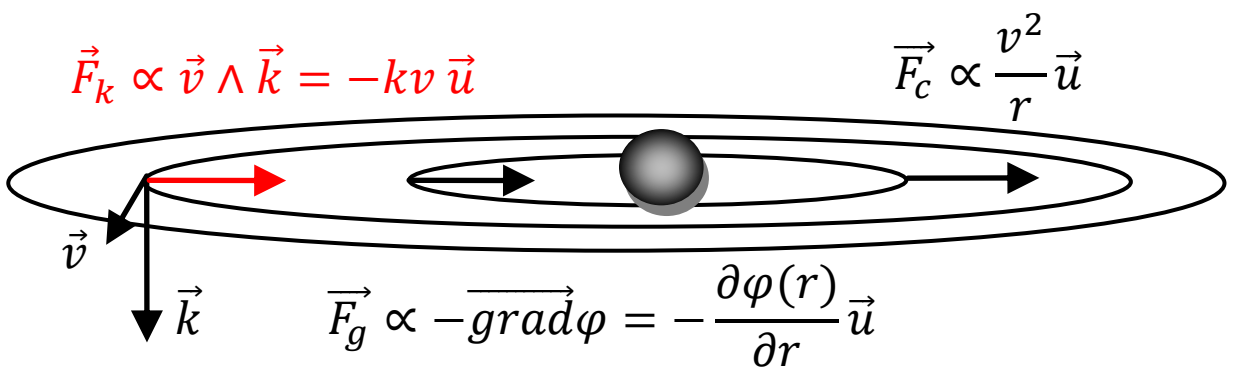

Fig. 2: Simplified representation of the equilibrium of forces in a galaxy

The simplified Fig. 2 can help us to visualize how the two components of the linearized general relativity intervene in the equilibrium of forces. The gravitic field, perpendicular to galaxy rotation plane, with the velocity of the matter generates a centripetal force increasing the Newtonian gravitation.

A first non trivial point should be explained: "the orientation of $\vec{k}$ generates a centripetal force" (and not a centrifugal one). At this step, one can consider it as an assumption of our calculation. But we will see that with our solution, $\vec{k}$ can't come from galaxies. Then, another origin will be studied. And we will see that, in this context, the centripetal orientation can be justified (§5.3), and furthermore it will lead to very important "sine qua none" predictions that will differentiate our solution from the traditional dark matter assumption and from MOND theories. So this assumption will be only a temporary hypothesis. It will be a very constraining non trivial condition, but some observations seem to validate it.

A second non trivial point to explain is that the gravitic field $\vec{k}$ is perpendicular to the speed of the matter. It can be justified independently of its origin. Our solution, for the galaxies, represents an equilibrium state of the physical equations. This means that the planes of movement must be roughly perpendicular to $\vec{k}$. It can be explained by a filtering of the speed over the time. If, initially, the speed of a body is not perpendicular to the gravitic field $\vec{k}$, the trajectory of this body won't be a closed trajectory (as a circle or an ellipse) but a helicoidally trajectory, meaning that this body will escape. Finally, at the equilibrium state (state of our studied galaxies), the gravitic field will have imposed the movement of the matter in roughly perpendicular planes to $\vec{k}(r)$. One can add that $\vec{k}(r)$ can change its direction with $r$ without modifying this expectation. This allows posing that in our calculations, we are in the very general configuration $\vec{v}(r) \perp \vec{k}(r)$ whatever the position $r$.

The traditional computation of rotation speeds of galaxies consists in obtaining the force equilibrium from the three following components: the disk, the bugle and the halo of dark matter. More precisely, one has (KENT, 1986):

$$\frac{v^2(r)}{r} = \left(\frac{\partial\varphi(r)}{\partial r}\right) \; with \; \varphi = \varphi_{disk} + \varphi_{bulge} + \varphi_{halo}$$

Then total speed squared can be written as the sum of squares of each of the three speed components:

$$v^2(r) = r\left(\frac{\partial\varphi_{disk}(r)}{\partial r}\right) + r\left(\frac{\partial\varphi_{bulge}(r)}{\partial r}\right) + r\left(\frac{\partial\varphi_{halo}(r)}{\partial r}\right)$$
$$= v_{disk}^2(r) + v_{bulge}^2(r) + v_{halo}^2(r)$$

From this traditional decomposition, we obtain the traditional graph of the different contributions to the rotation curve (just like in Fig. 3).

Disk and bulge components are obtained from gravity field. They are not modified in our solution. So our goal is now to obtain only the traditional dark matter halo component from the linearized general relativity. According to this idealization, the force due to the gravitic field $\vec{k}$ takes the following form $\|\overrightarrow{F_k}\| = m_p 4\|\vec{v} \wedge \vec{k}\|$ and it corresponds to previous term $m_p \frac{\partial\varphi_{halo}(r)}{\partial r} = \|\overrightarrow{F_k}\|$. I recall that we just have seen that the more general situation at the equilibrium state is to have the approximation $\vec{v} \perp \vec{k}$. In a second step, we will see some very important consequences due to the vector aspect $(\vec{v} \wedge \vec{k})$, it will lead to several predictions. This idealization gives the following equation:

$$\frac{v^2(r)}{r} = \left(\frac{\partial\varphi_{disk}(r)}{\partial r}\right) + \left(\frac{\partial\varphi_{bulge}(r)}{\partial r}\right) + 4k(r)v(r)$$
$$= \frac{v_{disk}^2(r)}{r} + \frac{v_{bulge}^2(r)}{r} + 4k(r)v(r)$$

Our idealization means that:

$$v_{halo}^2(r) = v^2(r) - v_{disk}^2(r) - v_{bulge}^2(r) = 4rk(r)v(r)$$

The equation of dark matter (gravitic field) is then:

$$v_{halo}(r) = 2\big(rk(r)v(r)\big)^{1/2} \qquad (VI)$$

This equation gives us the curve of rotation speeds of the galaxies as we wanted. The problem is that we don't know this gravitic field $\vec{k}$, but we know the curves of speeds that one wishes to have. We will thus reverse the problem and will look at if it is possible to obtain a gravitic field which gives the desired speeds curves.
From the preceding relation $(VI)$, let us write $k$ according to $v$, one has:

$$\frac{v_{halo}^2(r)}{4rv(r)} = k(r) \quad (VII)$$

## 3.3. Computation step I: a mathematical solution

To carry out our computation, we are going to take in account the following measured and theoretical curves due to the "Observatoire de Paris / U.F.E." (Fig. 3).

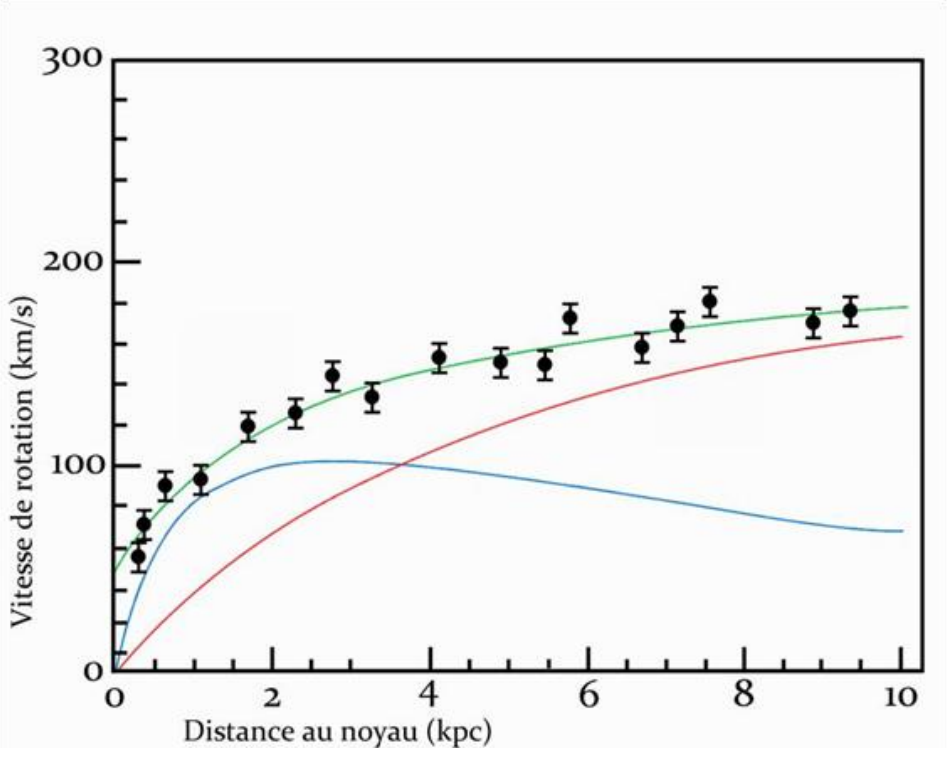

Fig. 3: Rotation curve of a typical spiral galaxy, showing measurement points with their error bar, the curve (green) adjusting the best data (black), the speed of the disk (in blue) and that of a halo of invisible matter needed to account for the observed points (in red). [Crédit "Astrophysique sur Mesure"]

On Fig. 4, one has the gravitic field computed with formula $(VII)$. The numerical approximation used for $v_{halo}(r)$ and $v(r)$ curves are given at the end of the paper in Tab.2:

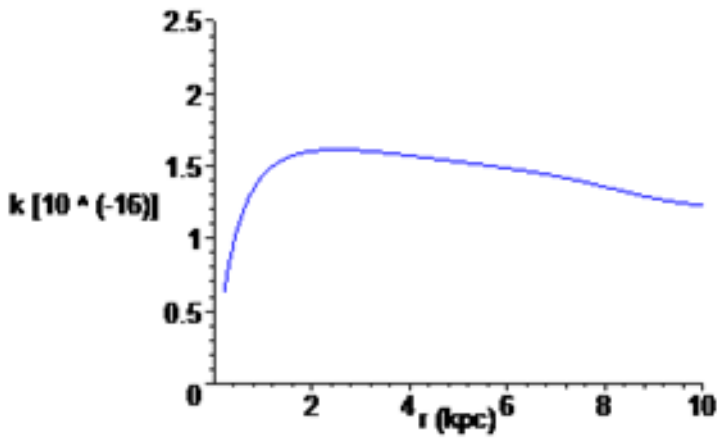

Fig. 4: Gravitic field which gives the expected rotational speed curve of Fig. 3.

This computed gravitic field is the one necessary to obtain the measured rotation speed of this galaxy without dark matter assumption. This curve plays the same role than the distribution of dark matter in the eponym assumption, it explains $v_{halo}$. But of course one must now study this solution in a more physical way. We will see that if we only consider the internal gravitic field of the galaxy it doesn't physically work but a solution can be imagined. This solution will be able to justify physically this curve, to be in agreement with observations and even to explain some unexplained observations.

## 3.4. Computation step II: a physical solution

At this step, we only mathematically solved the problem of the dark matter by establishing the form that the gravitic field $\vec{k}$ should take. The only physical assumption that we have made is that the force due to the gravitic field is written in the form $\overrightarrow{F_k} \propto \vec{v} \wedge \vec{k}$. And under this only constraint, we just come to show that it is possible to obtain a speeds curve like that obtained in experiments. To validate this solution, it is necessary to check the physical relevance of this profile of gravitic field and in particular to connect it to our gravitic definition of the field. Furthermore, at this step, with our approximation, this computed gravitic field doesn't represent only the gravitic field of the galaxy but also the others internal effects, in particular near the center (as frictions for example). But our goal will be to obtain a good physical approximation far from the galaxy center where the dark matter becomes unavoidable.
To physically study this solution far from galaxy center, the galaxy can be approximated as an object with all its mass at the origin point and some neglected masses around this punctual center. Such an idealization is an approximation in agreement with the profile of the mass distribution of a galaxy. Far from galaxy center, it's also the domain of validity of linearized general relativity. Far from galaxy center, one then should retrieve some characteristics of our "punctual" definition ($\vec{k} \approx \left(-\frac{1}{K}\right) m_p \left(\frac{1}{r^2}\right) \vec{v} \wedge \vec{u}$ in agreement with Poisson equations) in particular the three following characteristics: decreasing curve, curve tending to zero and $\vec{k} \propto \left(\frac{1}{r^2}\right)$.

*Positive point:* To be acceptable physically, it is necessary at least that this field is decreasing far from its source. For a galaxy, the large majority of its mass is in the neighborhoods of the center (zone (I)). Thus one must obtain a gravitic field which decreases far from the central area (zone (III)). The field obtained is globally in conformity with this expectation.

*Negative points:* Compared to the characteristic of the "punctual" definition ($\vec{k} \approx \left(-\frac{1}{K}\right) m_p \left(\frac{1}{r^2}\right) \vec{v} \wedge \vec{u}$), the previous gravitic field reveals two problems. First, our curve doesn't decrease near the galaxy's center. Secondly, our curve doesn't decrease to zero.

Let's see the first problem: The gravitic field graph does not always decrease. The curve starts to decrease only around 2 $kpc$. This problem can certainly be solved because our approximation take in account only gravitation effect. More precisely, gravitic field curve is computed starting with real value of rotation speed. But this measured rotation speed is due to gravitation but also to others phenomena (frictions for example). And our computation procedure takes in account only gravitation. So, previous computed "$k$" curve (obtained from measured speeds) doesn't represent only gravitation, in particular in zone (I) and (II). In these zones, the dynamic is dominated by others phenomena due to density of matter, in particular frictions and collisions, that could explain the graph. More than this, in these zones, the condition of validity of the approximation of general relativity in a quasi flat space is certainly not verified. So in this zone, our approximation underestimates the real internal gravitic field ("computed value = real value - dissipation due to the friction's effects").
Let's see now the second problem which will lead us to a possible explanation of the dark matter. The end of previous computed gravitic field curve should tend to our punctual definition of gravitic field. That is to say that gravitic field should decrease to zero. It clearly does not. More precisely, it should decrease to zero with a curve in $\frac{1}{r^2}$. One can try to approximate previous computed curve with a curve in $\frac{1}{r^2}$ (Fig. 5). But unfortunately, the approximation is so bad that it cannot be an idealization of the real situation as one can see it on the following graph.

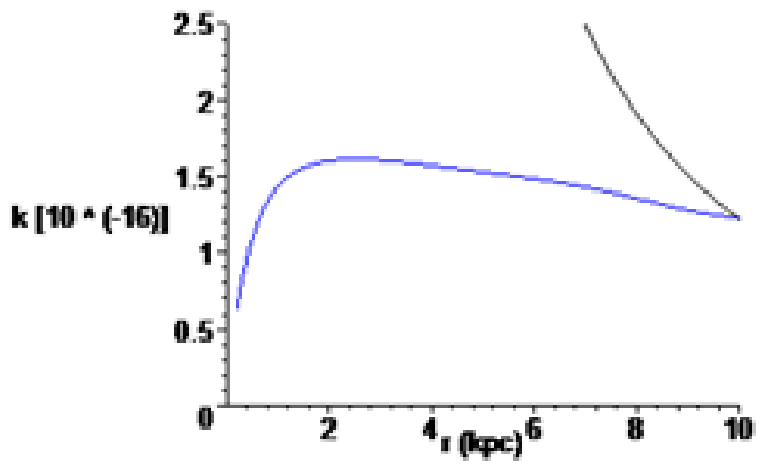

Fig. 5: Approximation of gravitic field with a curve in $r^{-2}$. It cannot represent reality.

This result means that the "internal" gravitic component ("internal" because due to the own galaxy) cannot explain the disk rotation speed at the end of the galaxies. One can note that this result is in agreement with the studies in (LETELIER, 2006), (CLIFFORD M. WILL, 2014) or (BRUNI et al., 2013) that also take into account non linear terms. It also means that, far from the center of the galaxy, even the non linear term cannot explain the dark matter.
But this gravitic field is generated by all moving masses. The clusters of galaxies, the clusters of clusters (superclusters) and so on have a gravitic field. The previous calculation only invalidates the galactic gravitic field as an explanation of dark matter. And as a galaxy is embedded in such astrophysical structures (cluster, supercluster…), one has to study if the gravitic field of these large structures could explain the rotation speeds of the ends of the galaxies. Instead of studying each structure one by one (that would be difficult because we don't know the value of their own gravitic fields) one simplifies the study by looking at, in a more general way, what value of gravitic field one should have to explain the rotation speed of the ends of galaxies. By this way, one will have the order of magnitude of the required gravitic field. Then, one will be able to show which large astrophysical structures could give such a value or not. We will see that the clusters (and their neighbors) are very good candidates. And also, from these values of gravitic field, one will be able to obtain lot of very interesting observational results and to make predictions to test our solution.

So, because general relativity implies that all large structures generate a gravitic field and because observations imply that galaxies are embedded in these large structures, we make the unique assumption of our study (and once again we will see that this assumption will be more likely a necessary condition of general relativity):

Assumption (I):

- Galaxies are embedded in a non negligible external gravitic field
- This external gravitic field, as a first approximation, is locally constant (at the scale of a galaxy).

Remarks: The assumption is only on the fact that the gravitic field is large enough (non negligible) to explain rotation curve and not on the existence of these gravitic fields that are imposed by general relativity (Lense-Thirring effect). And one can also note (to understand our challenge) that the constraints imposed by the observations imply that, in the same time, it must be small enough not to be directly detectable in our solar system and not to impose the orientation of the galaxies but large enough to explain dark matter. All these requirements will be verified. Furthermore, when we are going to look at the origin of this embedding gravitic field, this assumption will become more an unavoidable condition imposed by general relativity at the scale of clusters than a hypothesis. The second point is mainly to simplify the computation and we will see that it works very well.
But of course, this external gravitic field should be consistently derived from the rotation of matter within general relativity (it will be seen in §4.1). But we will also see (§6.2) that this approximation of a uniform gravitic field is compliant with linearized general relativity (just like it is in Maxwell idealization for electromagnetism).

Mathematically, this assumption means that $\vec{k} = \overrightarrow{k_{punctual}} + \overrightarrow{k_0}$ with the "internal" gravitic component of the galaxy $\|\overrightarrow{k_{punctual}}\| = \frac{K_1}{r^2}$ ("punctual" definition of the linearized general relativity) and an "external" gravitic component $\overrightarrow{k_0}$, the approximately constant gravitic field of the close environment (assumption (I)). If we take $\|\overrightarrow{k_0}\| = 10^{-15.97}$ and $K_1 = 10^{24.19}$ (values obtained to adjust this curve on the previous computed one), one has the curves on Fig. 6:

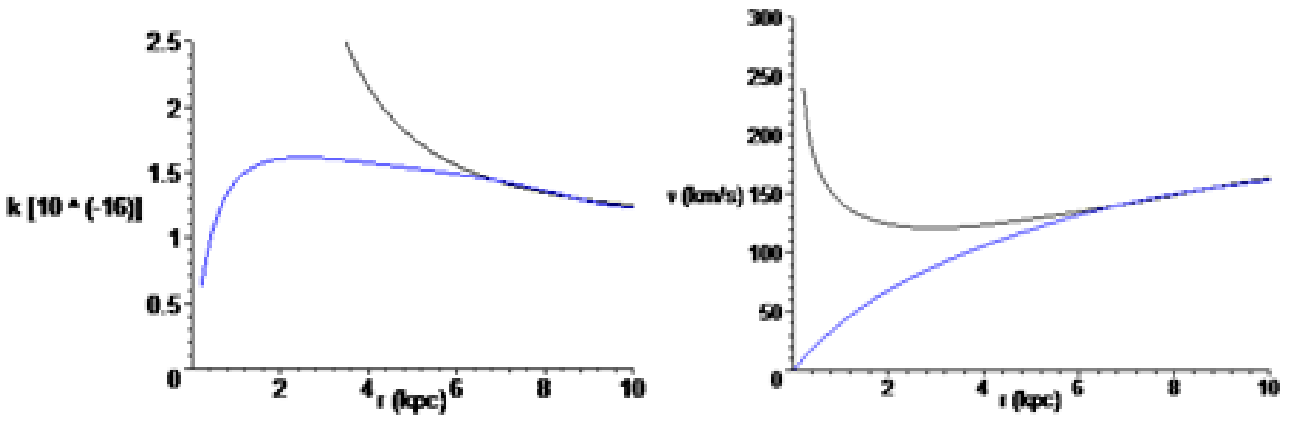

Fig. 6: Gravitic field approximation ($k = \frac{10^{24.19}}{r^2} + 10^{-15.97}$) and the rotation speed computed with it.

Left curves represent gravitic field and right curves the rotation speed obtained from formula $(VI)$. In the last part of the graph (domain of validity of linearized general relativity), the curves are indistinguishable. From about 6 $kpc$ to the end, the evolution in $\frac{1}{r^2}$ is excellent.

*Important remark*: To be clearer, our expression of "dark matter" is $\vec{v}(r) \wedge \vec{k}(r) = \left(\vec{v}(r) \wedge \frac{\overrightarrow{K_1}}{r^2} + \vec{v}(r) \wedge \overrightarrow{k_0}\right)$. To idealize our curve with $k(r) = \left(\frac{K_1}{r^2} + k_0\right)$ as previously, it doesn't imply that the internal gravitic field of the galaxies $\frac{\overrightarrow{K_1}}{r^2}$ and the external gravitic field $\overrightarrow{k_0}$ are collinear. Indeed, as one can see on the values of these two components, except in a transition zone (around a value noted $r_0$, as we are going to see it below, leading to very important predictions), the two vectors can be seen as spatially separated, $\overrightarrow{k}(r) \sim \frac{\overrightarrow{K_1}}{r^2}$ for $r < r_0$ and $\vec{k}(r) \sim \overrightarrow{k_0}$ for $r > r_0$. Concretely, it gives for $r < r_0$, $\|\vec{v}(r) \wedge \vec{k}(r)\| \sim \frac{K_1}{r^2} v.sin90° = \frac{K_1}{r^2} v(r)$ because in this area $\vec{v}(r) \perp \overrightarrow{K_1}$ at the equilibrium. I recall that, at the equilibrium (over the time), $\vec{k}$ has imposed the movement of the matter in roughly perpendicular plane to $\vec{k}(r)$. And for $r > r_0$, $\|\vec{v}(r) \wedge \vec{k}(r)\| \sim k_0 v.sin90° = k_0 v(r)$ because in this area $\vec{v}(r) \perp \overrightarrow{k_0}$ at the equilibrium. So the vectors don't need to be collinear to write $\|\vec{v}(r) \wedge \vec{k}(r)\| = \left(\frac{K_1}{r^2} + k_0\right) v(r)$. We will be in this situation for all the next studied galaxies (with two components very different in magnitude that allows seeing them as spatially separated). Our approximation is then very general, except in a transition zone that is not studied in our paper.
One can also add that the order of magnitude of $k_0$ and $\frac{K_1}{r^2}$ far from the center of the galaxy allows justifying the use of linearized general relativity.

To summarize our dark matter explanation, in the frame of linearized general relativity, the speeds of rotation at the ends of the galaxy can be obtained with:

- An internal gravitic field that evolves like $r^{-2}$ far from the center of galaxy (in agreement with the "punctual" definition of linearized general relativity)
- A constant external gravitic field embedding the galaxy.

One can write these contributions:

$$k(r) = \left(\frac{K_1}{r^2} + k_0\right) \qquad (VIII)$$

Remark: In theoretical terms, the use of the approximation $\frac{K_1}{r^2}$ allows extracting the pure internal gravitic field by neglecting the other non gravitic effects from the central area of the galaxy. But as it is obtained by fitting experimental curves that underestimate gravitational effects (because of the frictions for example, as said before), this term must always be underestimated. We will quantify this discrepancy. I recall that our main goal is to obtain an approximation in the ends of the galaxy (that is to say $k_0$), area where the other non gravitic effects vanish. Our approximation should be sufficient for this goal.
One can note that this contribution to be compliant with general relativity implies two constraints. First, the term $\frac{K_1}{r^2}$ that is the own (internal) gravitic field of the galaxy should be retrieved from simulations ever done in several papers (for example BRUNI et al., 2013). It will validate our approximation. Secondly, for each gravitic field, there should be a gravity field. For example, in our computation, associated to $\frac{K_1}{r^2}$, we take into account the gravity field for the galaxy (the term $\frac{\partial \varphi_{disk}(r)}{\partial r}$). But for our external gravitic field $k_0$, there should also be a gravity field. So to be compliant with our assumption, this gravity field should be negligible compare to $k_0$ (because we don't take it into account). These constraints will be verified in this study.
Remark 3: At the end of our study, we will generalize the expression $(VIII)$ for a source of gravitic field that won't be punctual but a "gravitic" dipole (just like in electromagnetism, with a magnetic dipole).

## 3.5. Application on several galaxies

We are now going to test our solution on different galaxies studied in (KENT, 1987). Because the linearized general relativity doesn't modify the components $v_{disk}(r)$ and $v_{bulge}(r)$, one can focus our study on the relation $v_{halo}^2(r) = 4k(r)v(r)r$.
For that we are going to determine $k_{exp}(r)$ curve, from experimental data with the relation, formula $(VII)$:

$$\frac{v_{halo}^2(r)}{4rv(r)} = k_{exp}(r)$$

I recall that we know the curve $v(r)$ which is given by experimental data and we know $v_{halo}(r)$, the resultant component obtained from relation $v_{halo}^2(r) = v^2(r) - v_{disk}^2(r) - v_{bulge}^2(r)$ where $v_{disk}^2(r)$ and $v_{bulge}^2(r)$ are also deduced from experimental data. The numerical approximation used for $v_{halo}(r)$ and $v(r)$ curves are given at the end of the paper in Tab. 3.

Then we are going to approach this $k_{exp}(r)$ curve (blue curves in Fig. 7) with our expected expression $k(r) = \left(\frac{K_1}{r^2} + k_0\right)$ which explains dark matter (black curves of right graphs in Fig. 7).

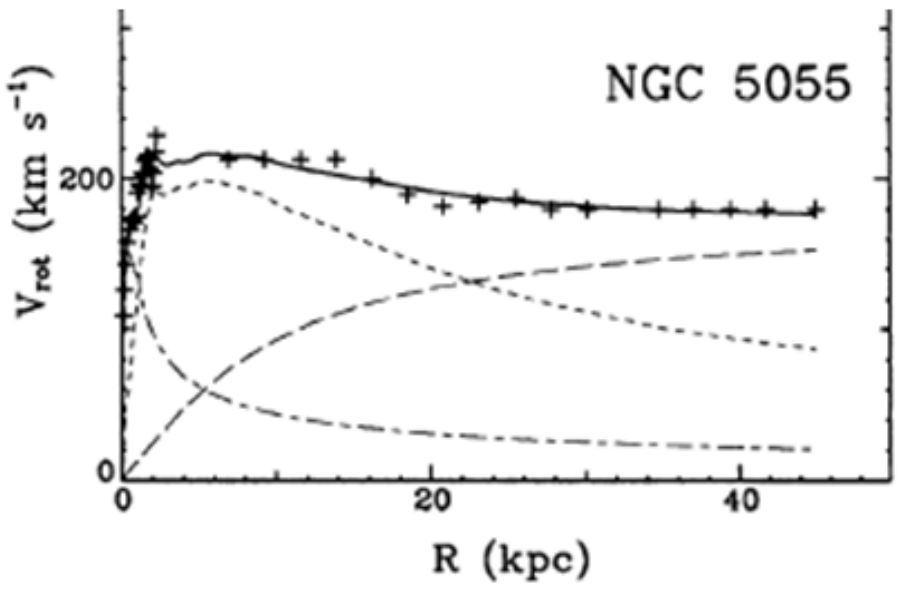

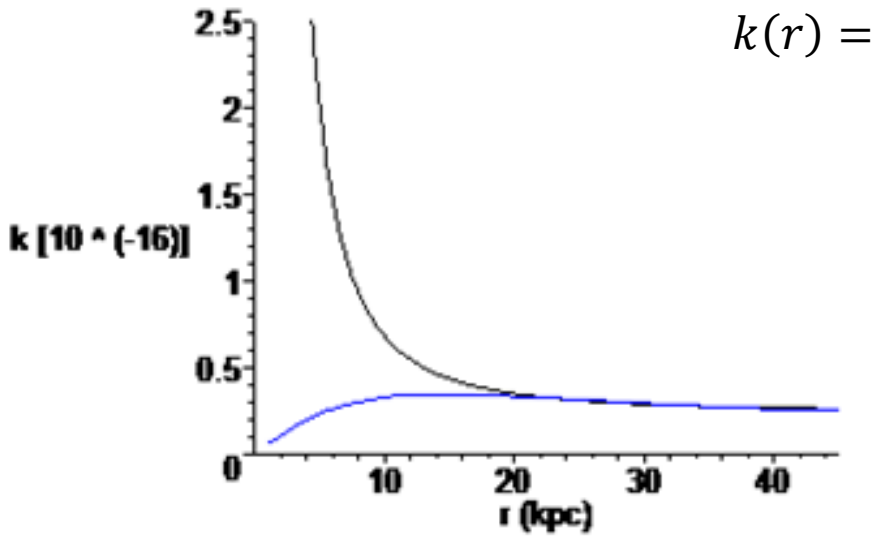

$$k(r) = \frac{10^{24.6}}{r^2} + 10^{-16.62}$$

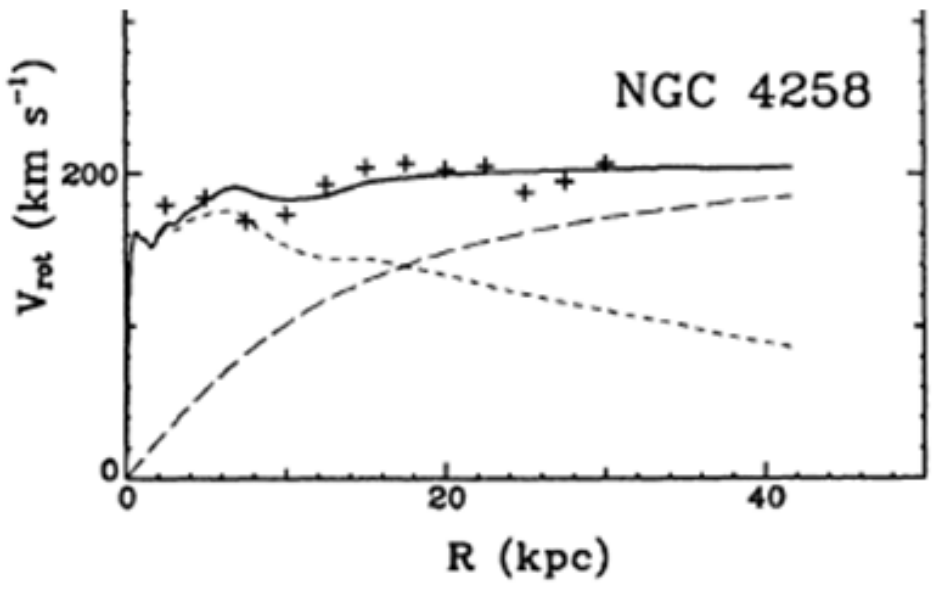

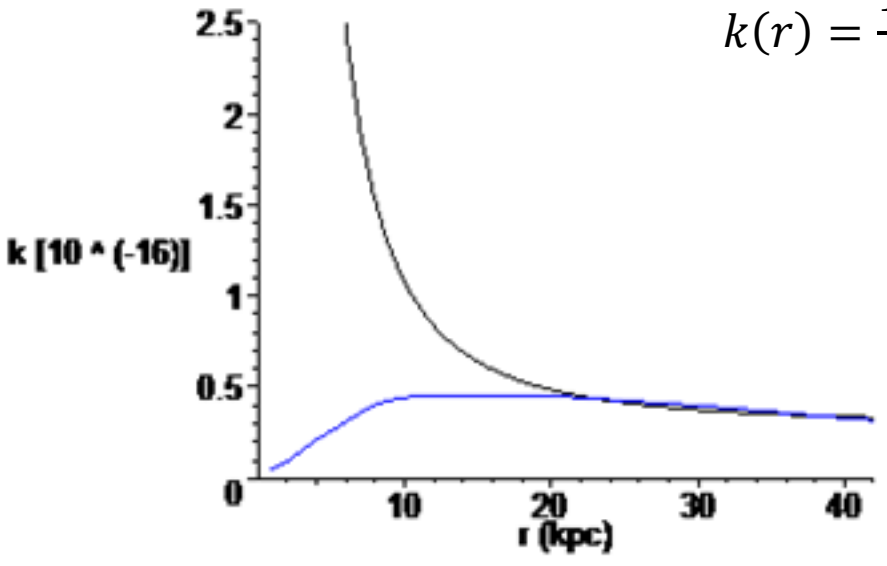

$$k(r) = \frac{10^{24.85}}{r^2} + 10^{-16.54}$$

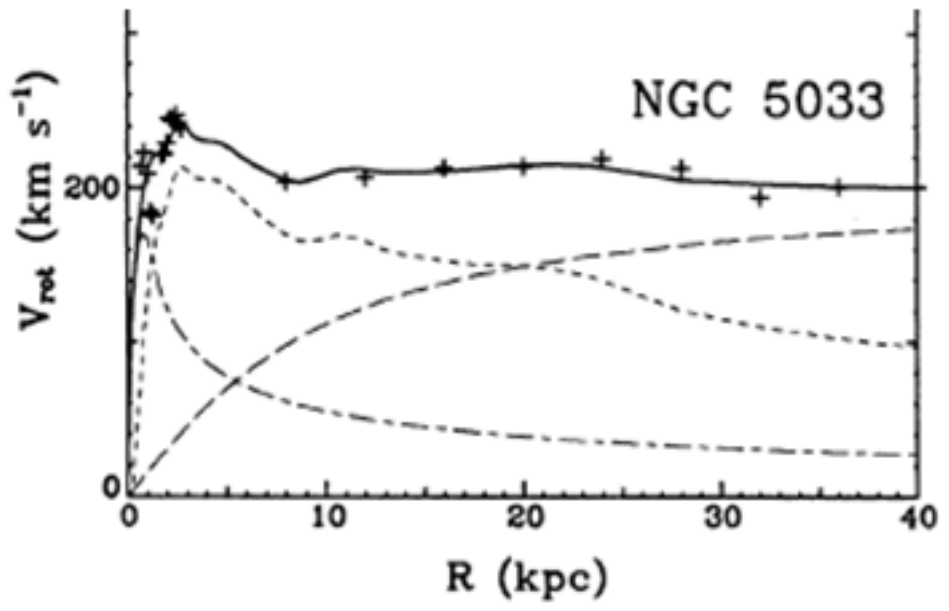

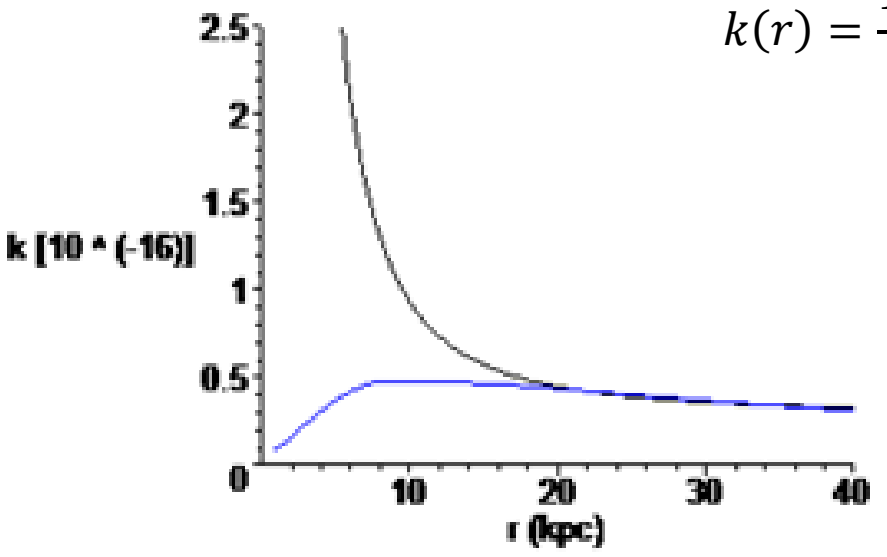

$$k(r) = \frac{10^{24.76}}{r^2} + 10^{-16.54}$$

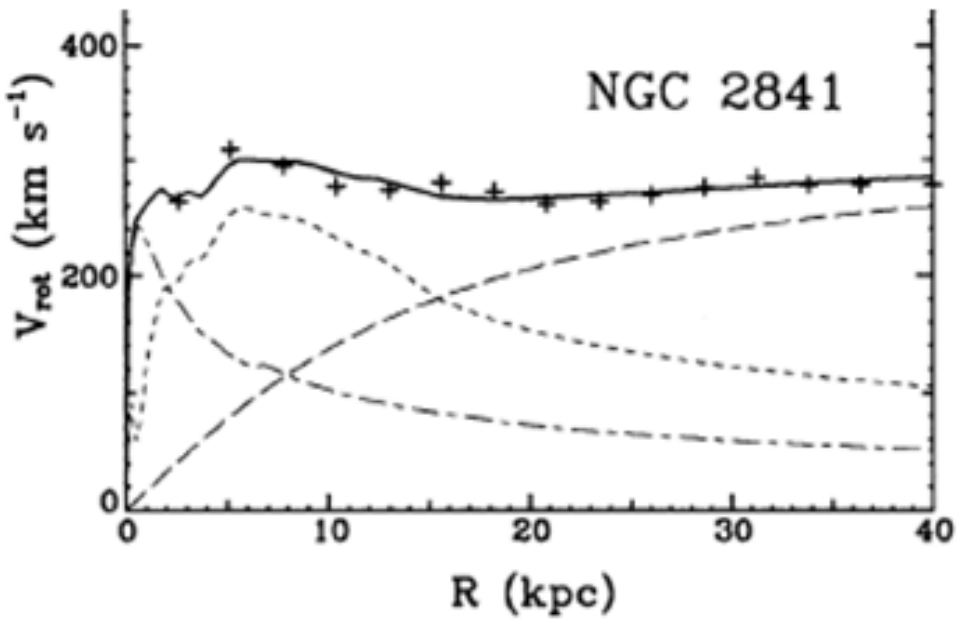

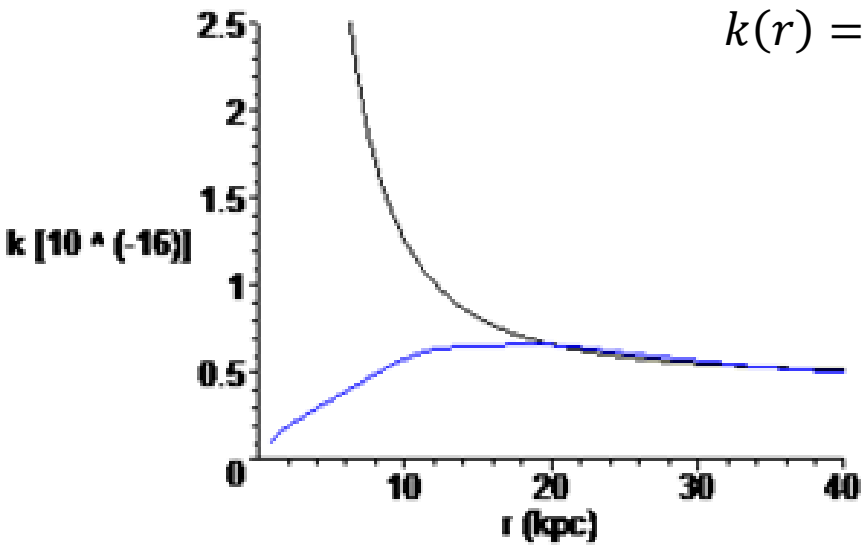

$$k(r) = \frac{10^{24.85}}{r^2} + 10^{-16.33}$$

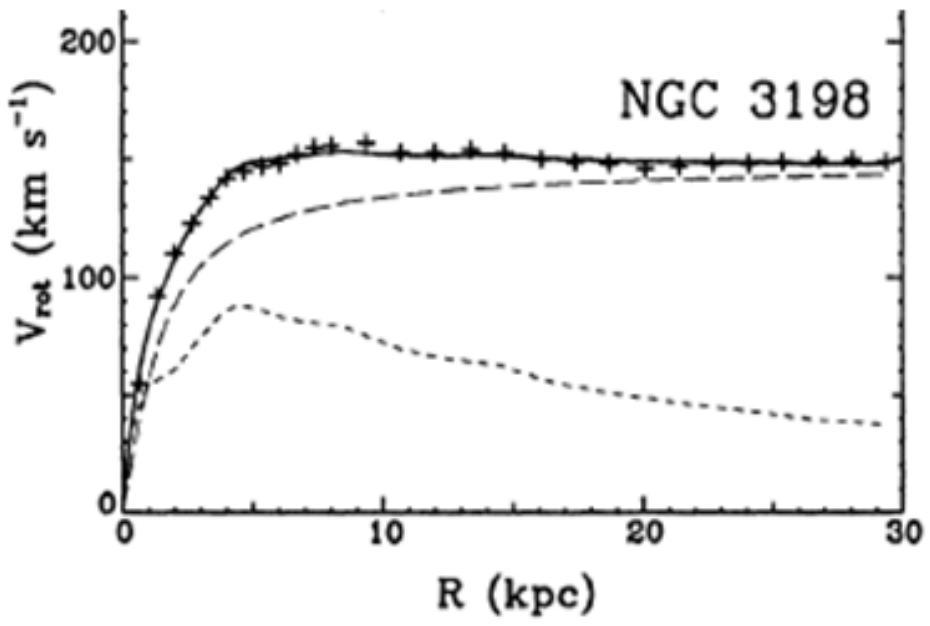

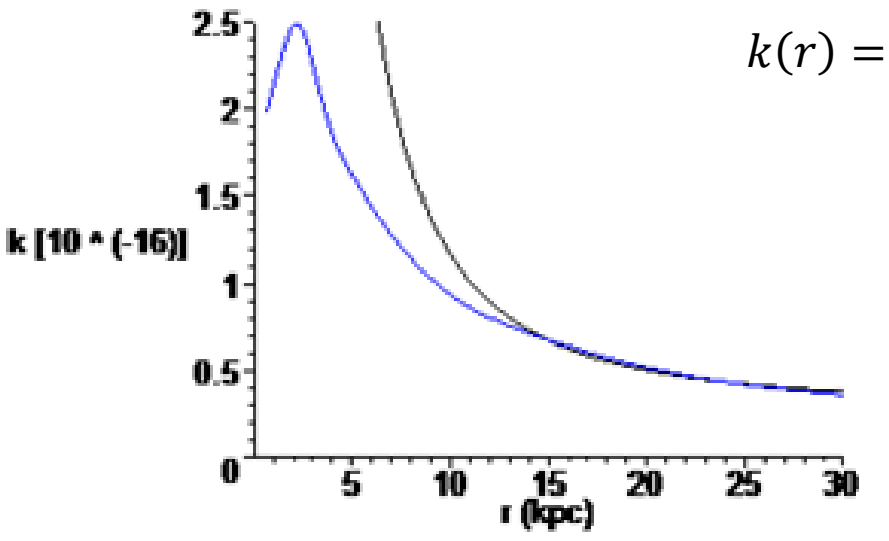

$$k(r) = \frac{10^{24.9}}{r^2} + 10^{-16.55}$$

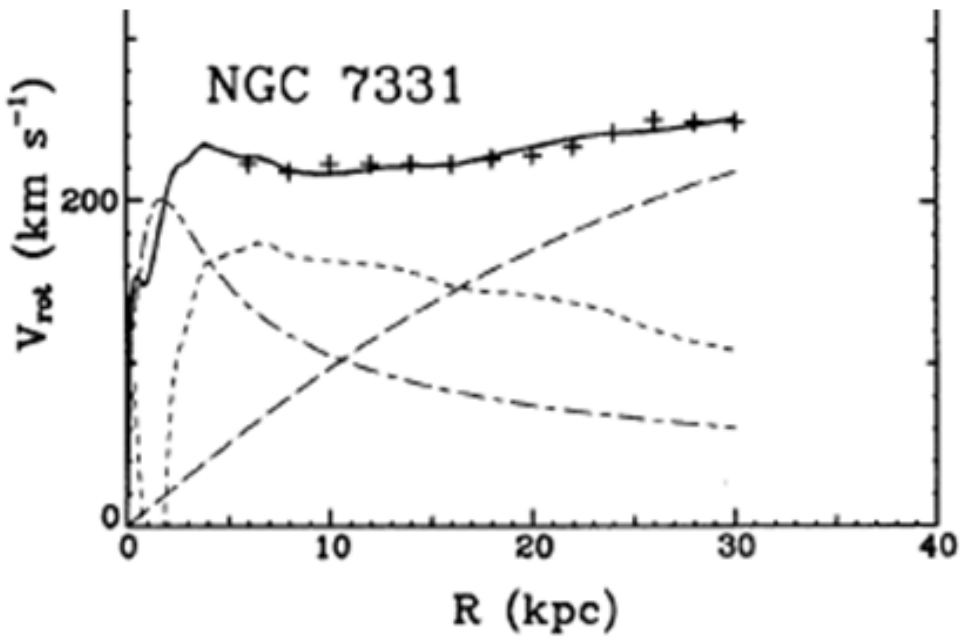

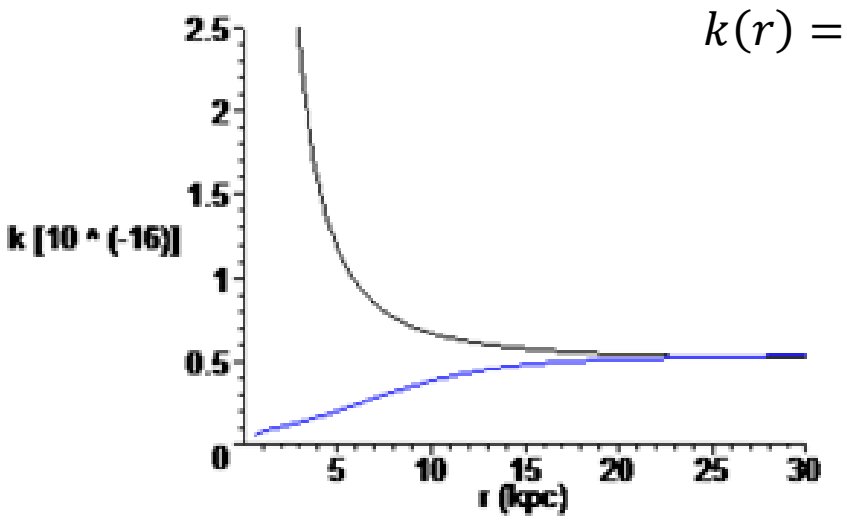

$$k(r) = \frac{10^{24.18}}{r^2} + 10^{-16.3}$$

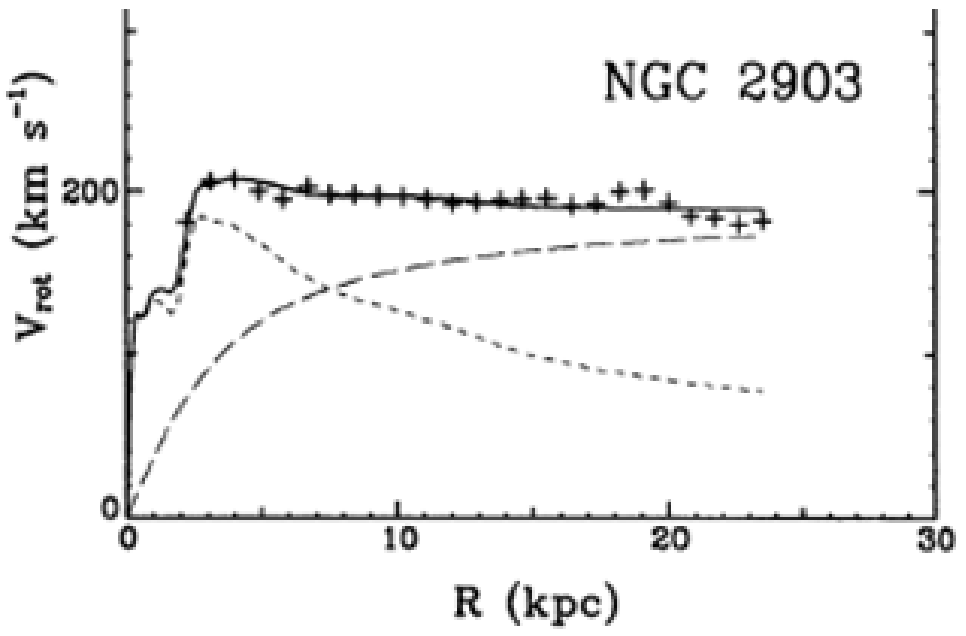

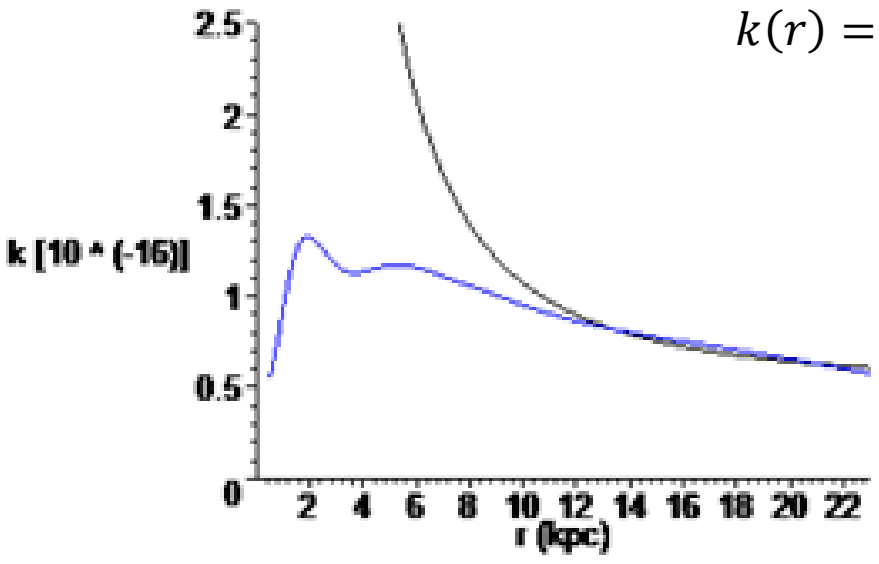

$$k(r) = \frac{10^{24.71}}{r^2} + 10^{-16.3}$$

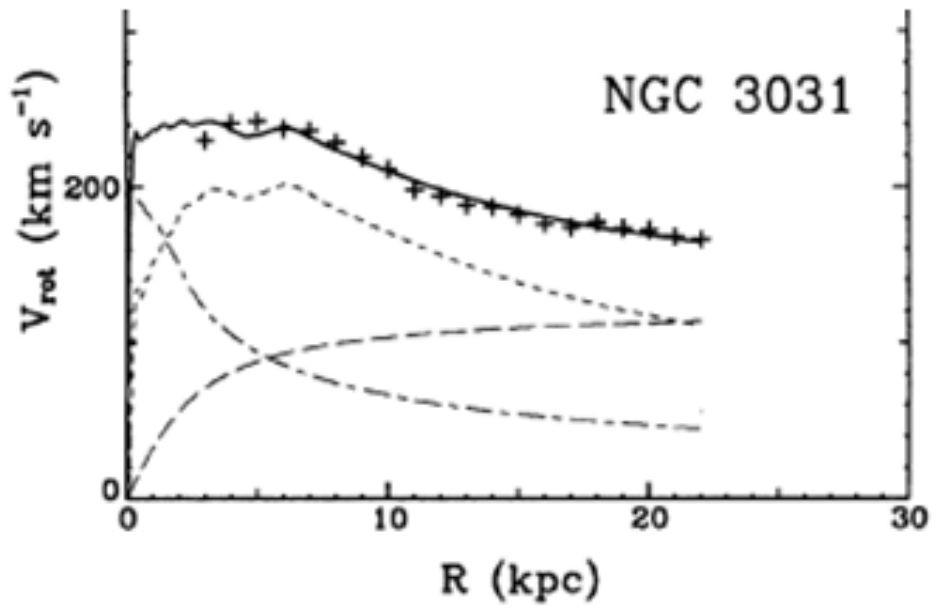

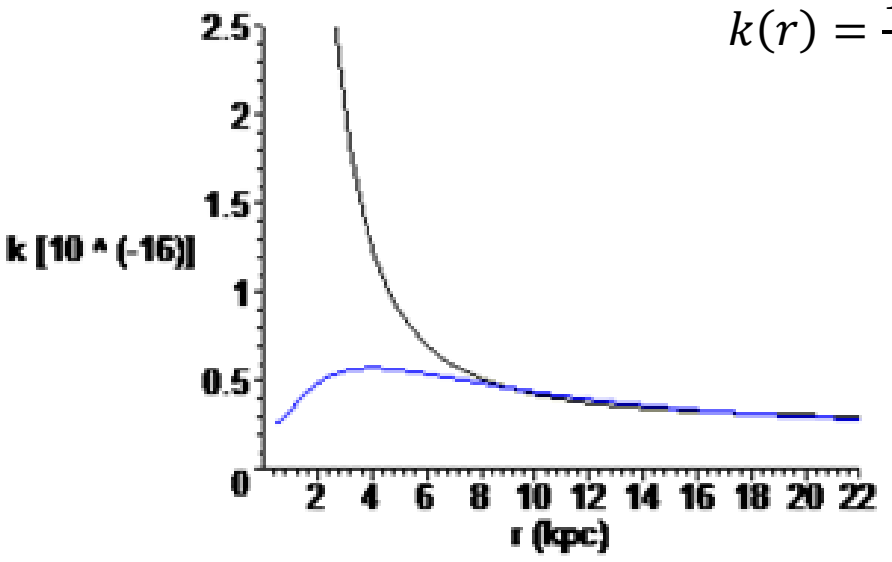

$$k(r) = \frac{10^{24.15}}{r^2} + 10^{-16.57}$$

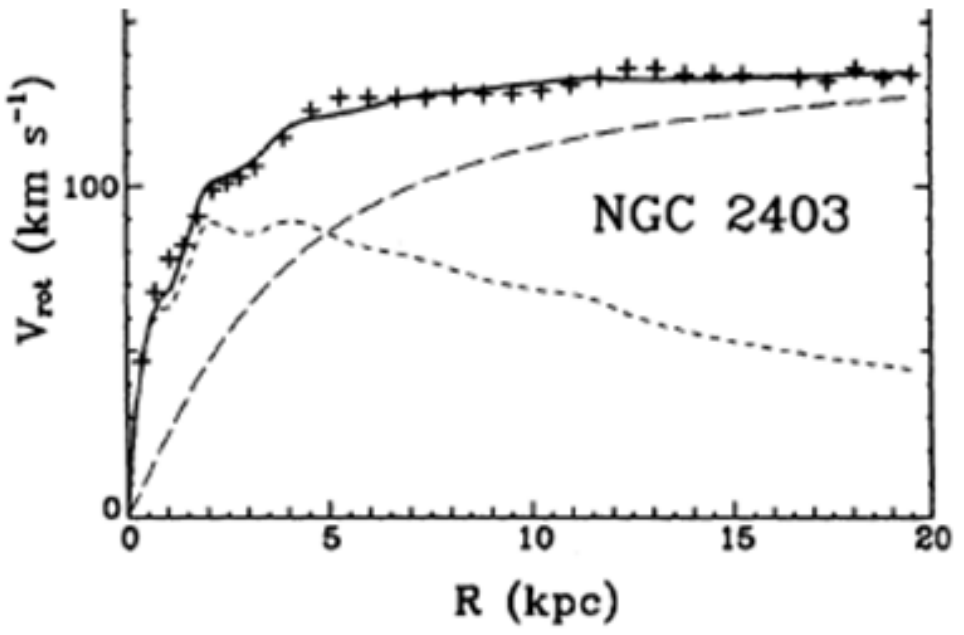

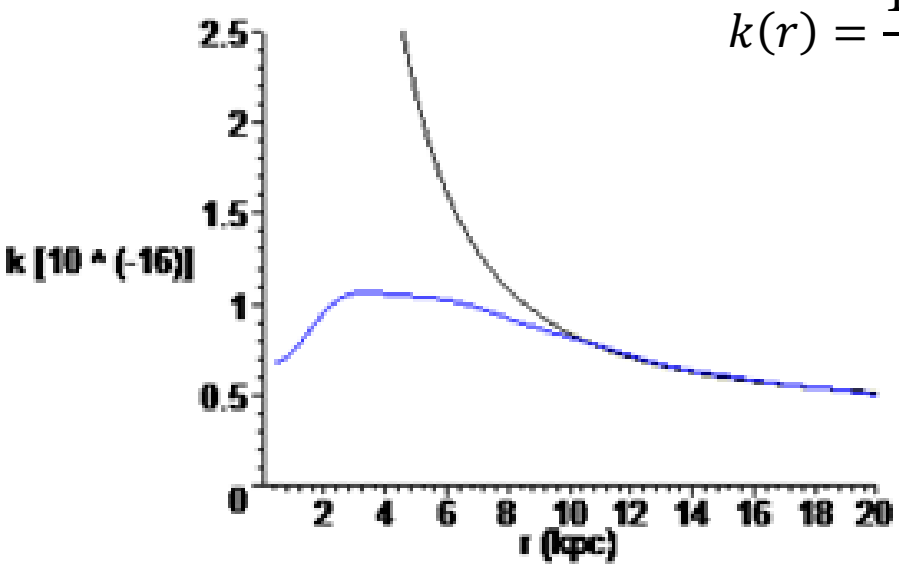

$$k(r) = \frac{10^{24.59}}{r^2} + 10^{-16.39}$$

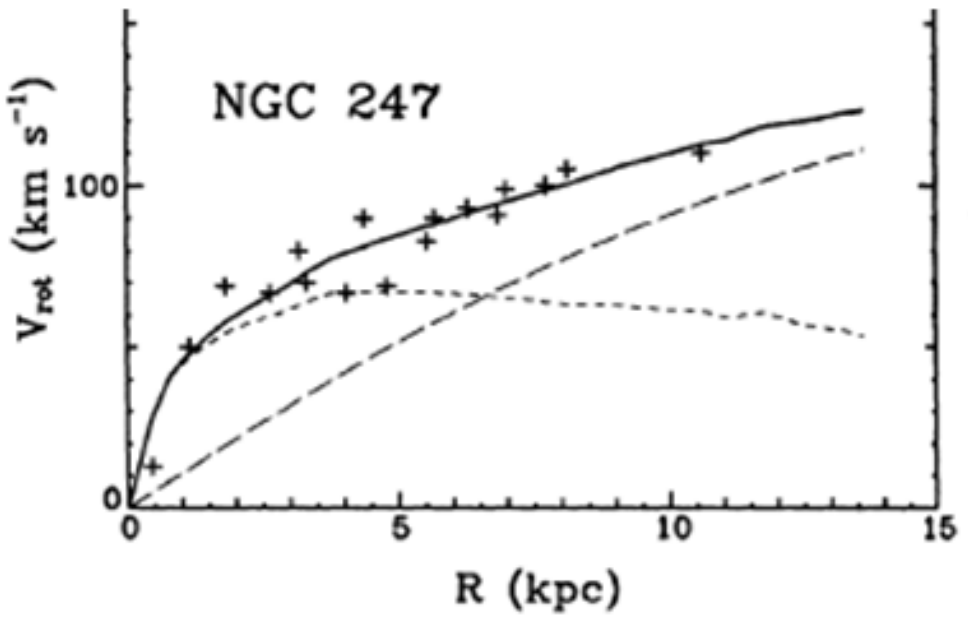

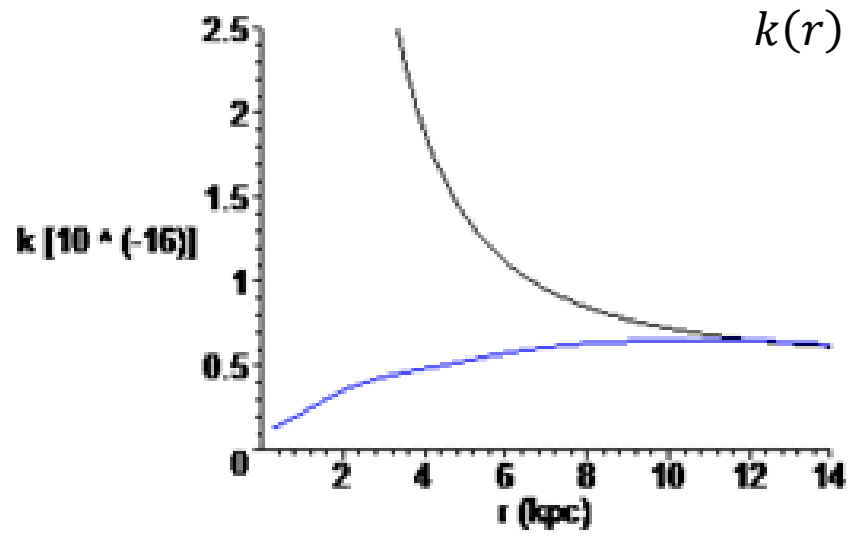

$$k(r) = \frac{10^{24.3}}{r^2} + 10^{-16.3}$$

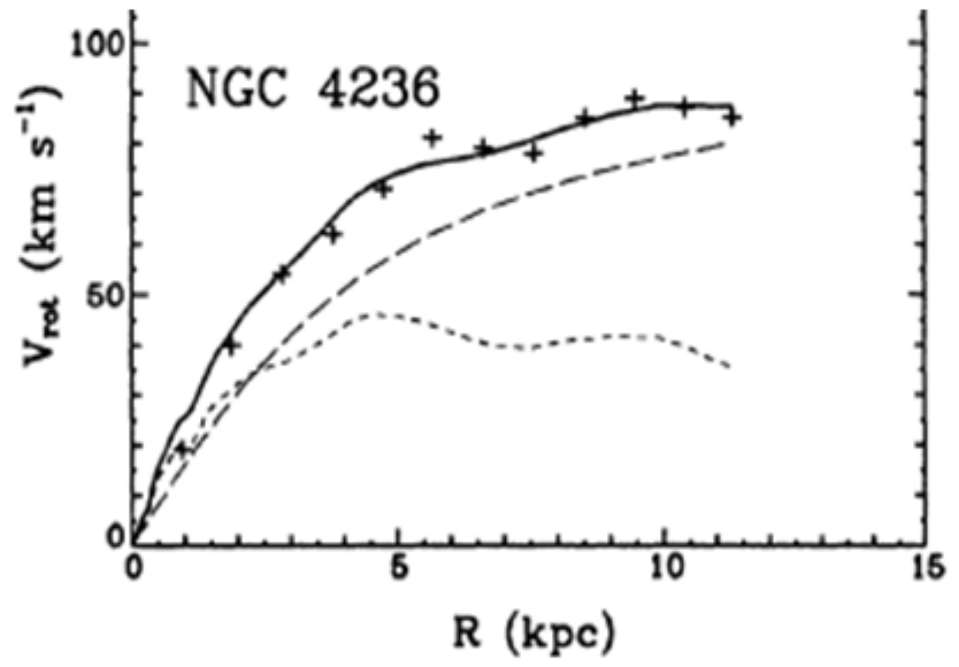

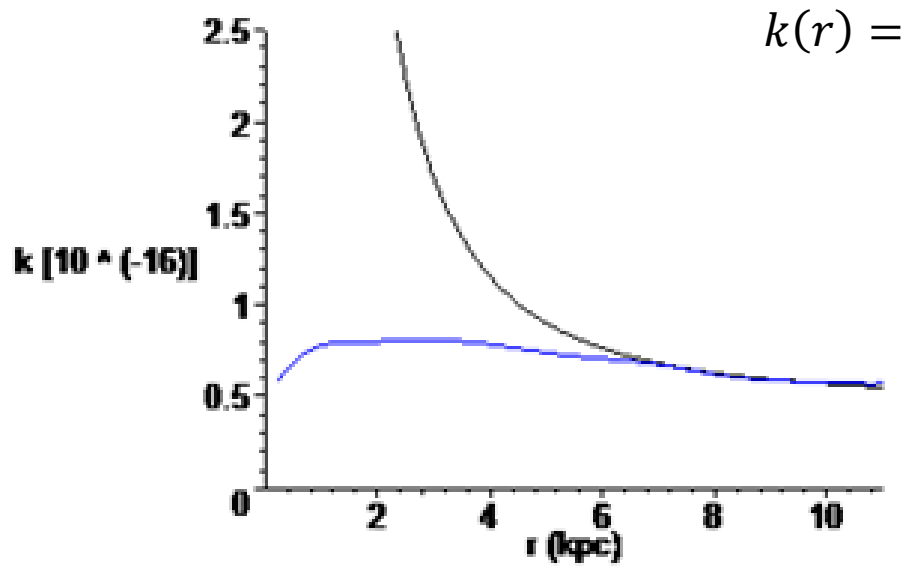

$$k(r) = \frac{10^{24.}}{r^2} + 10^{-16.34}$$

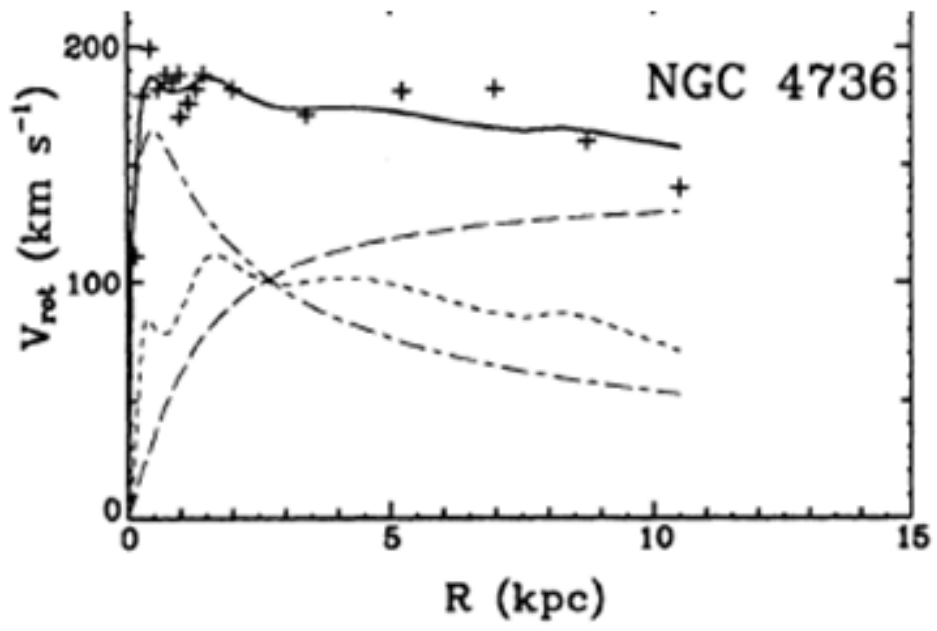

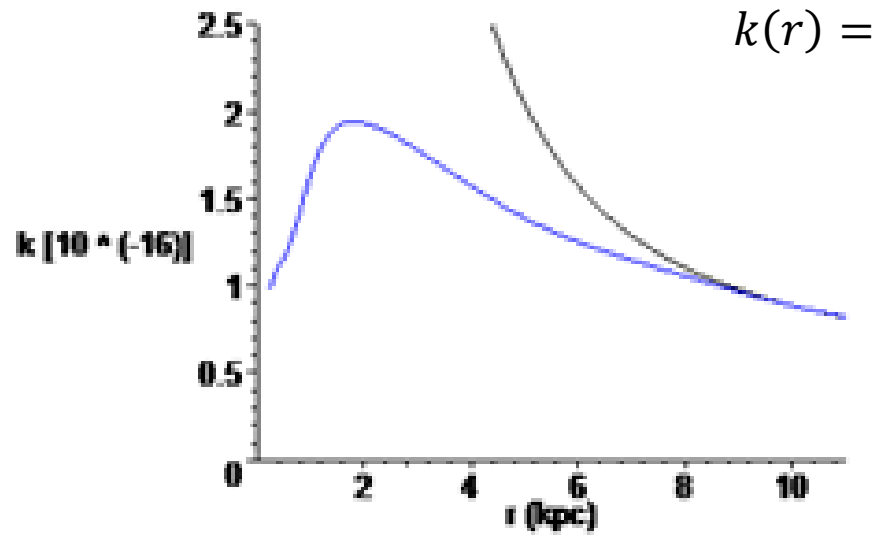

$$k(r) = \frac{10^{24.54}}{r^2} + 10^{-16.3}$$

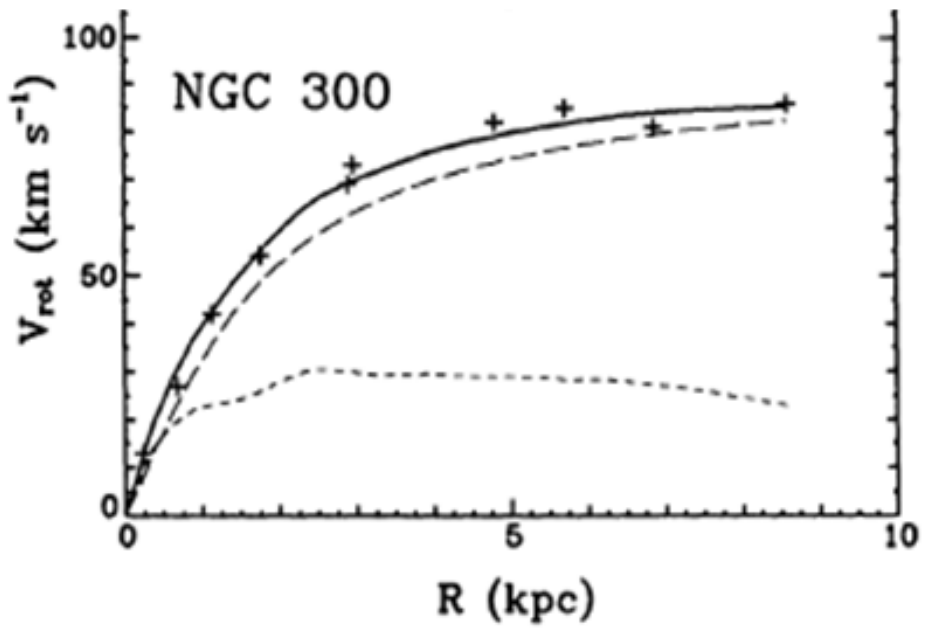

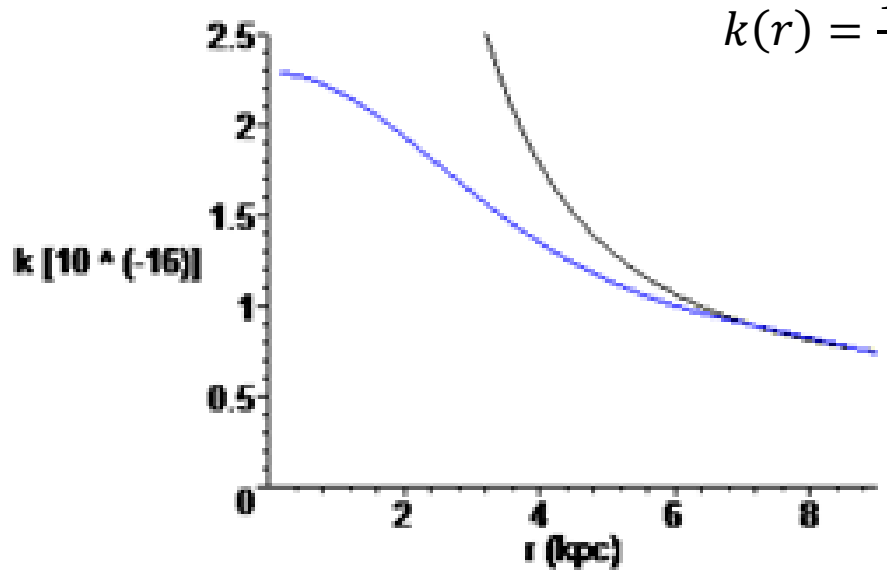

$$k(r) = \frac{10^{24.27}}{r^2} + 10^{-16.31}$$

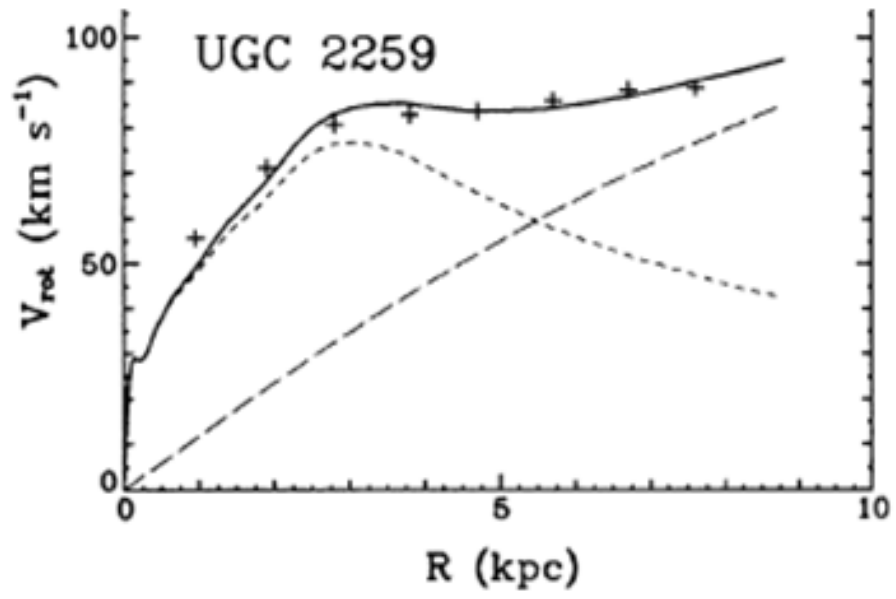

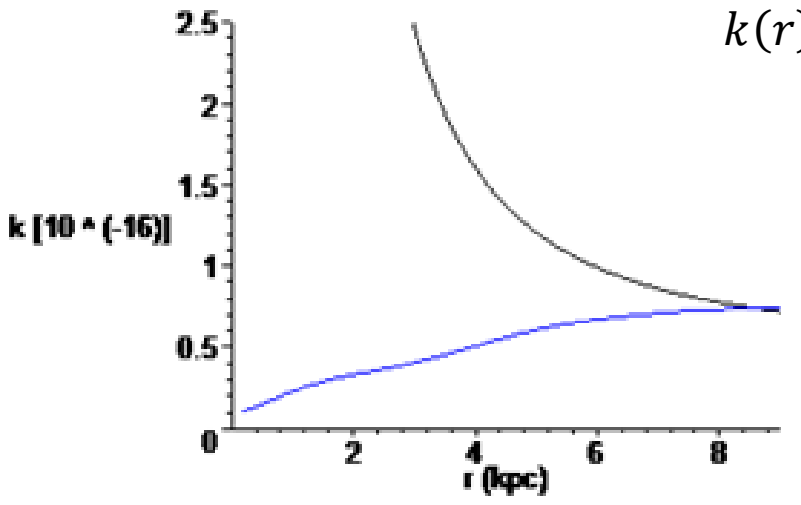

$$k(r) = \frac{10^{24.2}}{r^2} + 10^{-16.3}$$

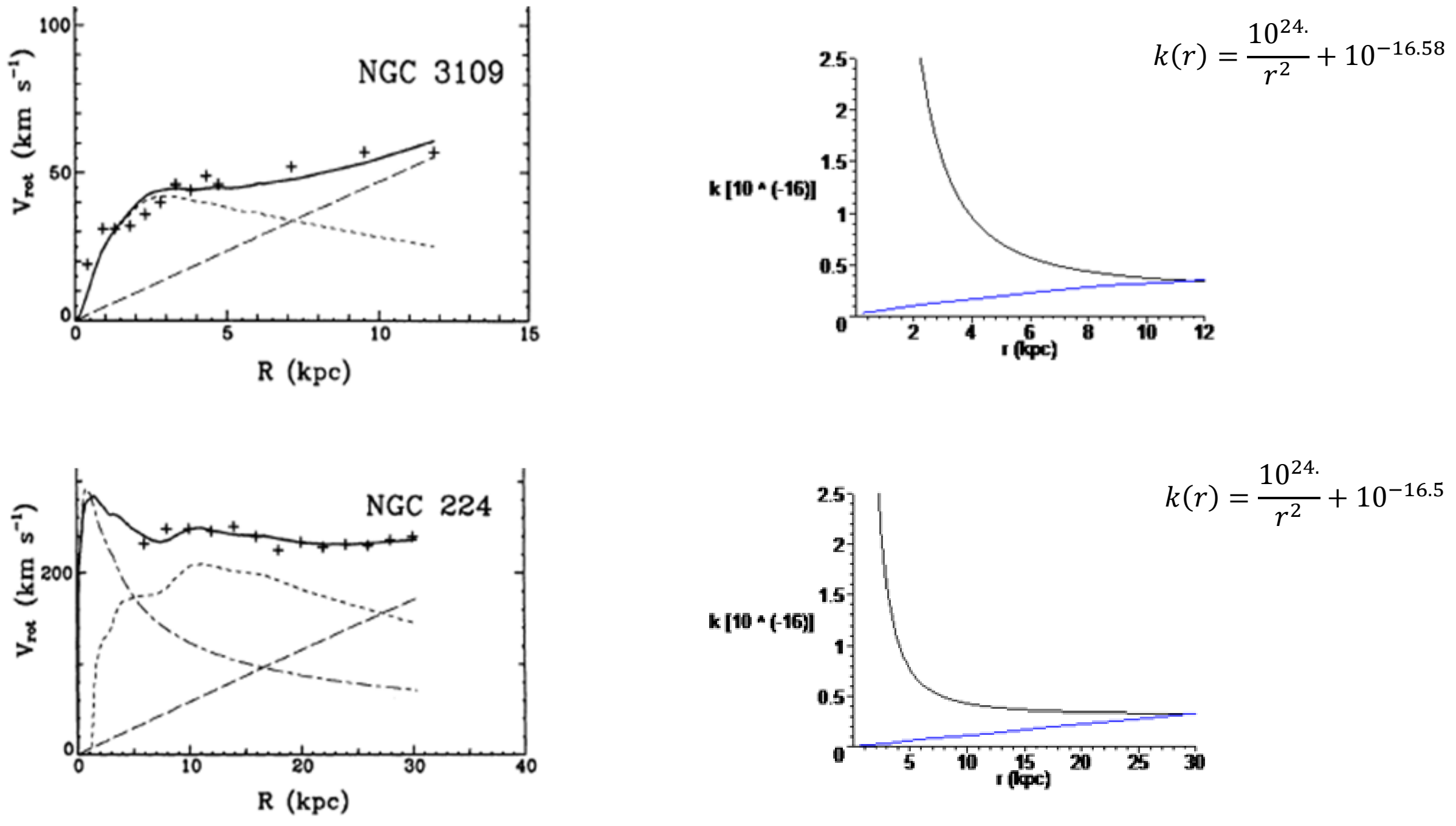

Fig. 7: Left graphs are the basis data from (KENT, 1987); right graphs represent $k_{exp}(r)$ (in blue) and $k(r) = \left(\frac{K_1}{r^2} + k_0\right)$ (in black) explaining ends of rotation speed curves (analytic expression of $k(r)$ is given).

Discussion on these results:

1) Our solution makes the assumption of a constant asymptotic behavior of gravitic field (tending to the external gravitic field $k_0$ at the ends of galaxies). These results confirm the tendency to get a curve $k(r)$ flattens for large $r$. However, there are two exceptions. The last two curves do not flatten in the studied interval of distance. These results also show that the external gravitic field values are in the interval $10^{-16.62} < k_0 < 10^{-16.3}$.

2) Our solution makes the more precise assumption of an evolution of the field as $\frac{K_1}{r^2} + k_0$ far from the center of the galaxy. This hypothesis is verified because theoretical $k(r)$ and experimental $k_{exp}(r)$ curves are equivalent on more than 30 % of their definition (up to 50 % for some), always with the exception of the last two curves. Furthermore, our idealization comes from linearized general relativity approximation that is valid only in zone (III) far from the center of the galaxy and then nearly 100% of the zone (III) is explained. This is the first main result of our study. These results also show that, with our approximation, the internal gravitic field values are in the interval $10^{24.} < K_1 < 10^{24.9}$. We are going to compare these values with published studies.

3) One can note this remarkable agreement with observations. The order of magnitude of $k_0$ is small enough to be undetectable in our solar system (cf. the calculation in paragraph 2) and at the same time it is high enough to explain rotation speed of galaxies. Dark matter has the disadvantage of being almost all the matter of our universe and to be undetectable until now, meaning that all our theories would explain (and would be founded on) only a negligible part of our reality.

4) Validity of our approximation: The value $K_1$ is associated to the internal gravitic field of galaxies. As we said before, it has ever been computed in several studies. One can then compare our approximation to these more accurate studies. For example, NGC 3198 gives the value $K_1 \sim 10^{24.9}$ for a visible mass of about $m_p \sim 6 \times 10^{40} kg$ (VAN ALBADA et al., 1985). One then has:

$$\frac{\|\vec{k}\|}{\|\vec{g}\|} \sim \frac{K_1}{r^2}\left(G\frac{m_p}{r^2}\right)^{-1} \sim \frac{10^{24.9}}{6.67 \times 10^{-11} \times 6 \times 10^{40}} \sim 2 \times 10^{-6}$$

As seen at the beginning of our study, in term of gravitomagnetism, one has $\overrightarrow{B_g} = 4\vec{k}$, it leads to a ratio $\frac{\|\overrightarrow{B_g}\|}{\|\vec{g}\|} = 4\frac{\|\vec{k}\|}{\|\vec{g}\|} \sim 0.8 \times 10^{-5}$. In (BRUNI et al., 2013), the simulation gives a ratio between the vector and scalar potential of about $10^{-5}$. As we said previously, our deduced internal gravitic field must be underestimated compare to the real value because our idealization doesn't take into account the friction's effects ("computed value = real value - dissipation due to the friction's effects"). And secondarily, our idealization doesn't take into account the non linear terms. It then explains that our ratio is slightly inferior to a more accurate simulation. But even with that, our approximation gives a good order of magnitude of the relative values of the gravity and gravitic fields. If one looks at the absolute value of our gravitic field, at $r \sim 1.3 Mpc \sim 10^{22} m$, one has $\|\overrightarrow{B_g}\| = 4\frac{K_1}{r^2} \sim \frac{10^{24.9}}{(1.3 \times 10^{22})^2} \sim 10^{-18.7}$. In (BRUNI et al., 2013), the simulation gives for the power spectra the value $P_B(r \sim 1.3 Mpc) \sim 10^{-18}$. This comparison confirms that our approximation gives a gravitic field that is underestimated. With our example, the factor of correction (due mainly to friction's effects) is about $5 \sim 10^{0.7}$. Its effective gravitic field is around 5 times greater than our approximation. But the order of magnitude is still correct (even without the non linear terms). Let's make a remark. Our explanation of dark matter is focused on the external term $k_0$. The next results will be obtained far from the galaxies' center. We are going to see that far from the center of galaxies only $k_0$ acts. And far from the center of galaxies, the non linear terms (of internal gravitic field) and friction's effects are negligible. For these reasons, with our approximation, the value of external gravitic field $k_0$ should be a more accurate value than our $K_1$.

5) A first reaction could be that the external gravitic field should orient the galaxies in a same direction, which is at odds with observations. With our analysis, one can see that the internal gravitic field of galaxies ($\frac{K_1}{r^2} \sim \frac{10^{24.9}}{r^2}$) is bigger than the embedding gravitic field ($k_0 \sim 10^{-16.55}$) required to explain dark matter for $r < 17kpc$ ($\frac{10^{24.9}}{r^2} > 10^{-16.55} => r^2 < 10^{41.45}$). And if one takes in account the factor of correction seen previously on $K_1$, the internal gravitic field imposes its orientation until $r \sim 40kpc$. With these orders of magnitude, we understand that galaxies' orientation is not influenced by this gravitic field $k_0$ of the environment. The randomly distribution of the galaxies' orientation is then completely in agreement with our solution. We will see further that our solution implies in fact a constraint on the distribution of the galaxies' orientation. Inside a cluster, galaxies' orientation will have to be randomly distributed in a half-space. It will be an important test for our solution.

6) Influence of internal gravitic field far from the galaxy's center: One can note that, for only the internal gravitation, one has in the galaxies $\frac{\|\vec{F_k}\|}{\|\vec{F_g}\|} \sim 4v \frac{\|\vec{k}\|}{\|\vec{g}\|} \sim 10^{-5}v$. And for $r > 10kpc$, one has in the galaxies $v \sim 10^5 m.s^{-1}$. It implies that $\frac{\|\vec{F_k}\|}{\|\vec{F_g}\|} \sim 1$ in this area. So, far from galaxies center, the gravitic force is of the same magnitude than the gravity force. And then, just like the internal gravity force, the internal gravitic force decreases and vanishes far from galaxies center explaining why it cannot explain dark matter.

At this step, to compare in the main lines and in a simplified way our study with already published papers, one can say that in terms of linearized general relativity the already published papers idealize the gravitic field like $k = \frac{K_1}{r^2}$ (but more accurately, taking into account non linear terms) and in our approach we idealize the gravitic field like $k = \frac{K_1}{r^2} + k_0$. The already published papers deduce their expression from mass distribution of galaxies, giving them only the gravitic field of galaxies. Conversely, we deduce our expression from the rotation speeds of galaxies, giving us the gravitic field of galaxies and a supplemented external gravitic field. The rest of our study is going to justify this expression by showing that the computed values (for $k_0$) can be deduced from larger structures than galaxies; that the computed values (for $k_0$) can explain unexplained observations and retrieve some known results; by showing that our expression is consistent with general relativity. It means to show that the uniform $k_0$ verifies the field equations of linearized general relativity (that is our domain of validity) and that $k_0$ has to be associated with a negligible gravity field $g_0$.

## 4. Cluster as possible origin of the gravitic field $k_0$

We are now going to use these values of $k_0$ ($10^{-16.62} < k_0 < 10^{-16.3}$). It will allow obtaining the expected rotation speed of satellite dwarf galaxies and retrieving the expected quantities of "dark matter" in the galaxies. We will also demonstrate that the theoretical expression of gravitic field can explain the dynamic of satellite dwarf galaxies (movement in a plane) and allow obtaining an order of magnitude of the expected quantitiy of "dark matter" in the CMB. But just before we are going to show that the clusters of galaxies could generate these values of $k_0$.

### 4.1. Theoretical relevancies

Before beginning this paragraph, I would like to precise that the goal of this paragraph is not to obtain a value of the gravitic field for large structures but to determine which large structure(s) could be a good candidate(s) to generate our embedding gravitic field. Inside our theoretical frame, we determined the values of the internal gravitic fields of the galaxies. From these values and from the definition of the gravitic field (which depends on the mass and the speed of source), we are going to apply a change of scale to estimate the value of the internal gravitic field of objects larger than galaxies. Of course, this way of approximating can be criticized. But because the same criticism can be assigned to each large structure, one can then hope, by this way, obtaining the relative contribution of these large structures (but not necessarily their actual values). This process will show that the galaxies' clusters are certainly the main contributors to our embedding gravitic field (and astonishingly this way of calculation will also give the good order of magnitude of the expected gravitic field, i.e. not only in relative terms).

*About the mean value $k_0 \sim 10^{-16.5}$:*

What can be the origin of this embedding gravitic field? In our theoretical frame, the gravitic field (just like magnetic field of a charge in electromagnetism) can come from any moving mass. Because of the orders of magnitude seen previously (for particle, Earth, Solar system and galaxies) one can expect that the embedding gravitic field ($k_0$) should be due to a very large astrophysical structure (with a sufficient internal gravitic field "$\frac{K_1}{r^2}$" to "irradiate" large spatial zone). Furthermore, just like magnetic field of magnets (obtained as the sum of spins), this external gravitic field could also come from the sum of several adjacent large structures. Our theoretical solution then implies that our embedding gravitic field can come from the following possibilities:

Case A:
Internal gravitic field of the galaxy
Case A bis:
Sum of internal gravitic fields of several close galaxies
Case B:
Internal gravitic field of the cluster of galaxies
Case B bis:
Sum of internal gravitic fields of several close clusters
Case C:
Internal gravitic field of larger structure (cluster of cluster,…)
Case C bis:
Sum of internal gravitic fields of several close larger structures
…
CaseD:
Internal gravitic field of our Universe

This expected embedding gravitic field should be consistently derived from the rotation of matter within general relativity. It means that if it should come from an internal gravitic field of a large structure, it should evolve like $\frac{K_1}{r^2}$ far from the source (in agreement with linearized general relativity because of Poisson equations $(III)$). So, let's see how large can this term "$\frac{K_1}{r^2}$"be for these large structures.

Case A: We have seen that our solution (in agreement with some other published papers) rejects the possibility of an internal gravitic field of galaxy sufficient to explain dark matter. Let's retrieve this result. We are going to see that, in our approximation, around $r \sim 15\ kpc$, the internal and external gravitic fields have similar magnitudes. But for $r \gg 15\ kpc$, the internal gravitic field of the galaxy can't explain $k_0$ because it becomes too small. And we are going to see that the value of $k_0 \sim 10^{-16.5}$ can explain the rotation speed of satellite dwarf galaxies (for $r > 100\ kpc$). But with an internal gravitic field of galaxy about $K_1 \sim 10^{24.5}$, at $r \sim 100kpc \sim 3 \times 10^{21}m$, one then has the value $\frac{K_1}{r^2} \sim 10^{-18.5}$. At $100\ kpc$ internal gravitic field of galaxy represents only about 1% of $k_0$ (for an average value of $k_0 \sim 10^{-16.5}$). Effectively, the gravitic field of a galaxy can't explain $k_0$.

Case A bis: Even the sum of several close galaxies can't generate our required $k_0$. Roughly in a cluster of galaxies, there are about $N \sim 10^3$ galaxies for a diameter of $D \sim 10^{23}m \sim 5Mpc$. It leads to an average distance between galaxies of about $d \sim 10^{21.5}m$ (for an idealization as a disk: $N * (\frac{d}{2})^2 \sim (\frac{D}{2})^2$). The contribution to the gravitic field of an adjacent galaxy at the distance $\frac{d}{2} \sim 0.5 \times 10^{21.5}m$ with $K_1 \sim 10^{24.5}$ is then $\frac{4K_1}{d^2} \sim \frac{10^{25}}{10^{43}} = 10^{-17.9}$. If we suppose that there are six neighboring galaxies (two galaxies by spatial direction) it leads to a gravitic field of about $10^{-17.1}$. It only represents 25% of the expected external gravitic field.

Case B: Without an explicit observation of the internal gravitic field of a cluster, it is difficult to give its contribution to the embedding gravitic field $k_0$. One can try a roughly approximation. The mass of a cluster is about $10^{44}kg$ (without dark matter). It is about 1000 times the mass of a typical galaxy ($10^{41}kg$) and the typical speed of the matter inside the cluster compare to its center are around 5 times greater than matter in galaxies (in our calculation, it is not the peculiar speed). By definition, the gravitic field is proportional to the mass and to the speed of the source ($\vec{k} \approx \left(-\frac{1}{K}\right) m_p \left(\frac{1}{r^2}\right) \vec{v_s} \wedge \vec{u}$). If we assume that the factor of proportion for matter speed is also applicable for the speed of the source, the internal gravitic field of a cluster could then be about $10^3 \times 5 \times K_1 = 10^{3.7}K_1 \sim 10^{28.2}$. With a typical cluster's size of around $D = 2R \sim 10^{23}m \sim 5Mpc$, it gives at its borders the following value of gravitic field $\frac{10^{3.7}K_1}{R^2} \sim \frac{4\times10^{28.2}}{10^{46}} = 10^{-17.2}$. It then represents around 20% of the expected external gravitic field $k_0$. As in the precedent case A bis, it is not sufficient to obtain $k_0$, but contrary to the case A bis, more we are close to the cluster's center, more the external gravitic field could be entirely explained by internal gravitic field of the cluster (for example at 700kpc it represents about 100%).

Case B bis: Furthermore, the sum of the internal gravitic field of adjacent clusters could maintain this embedding gravitic field on very large distance. For example in the case of the Local Super Cluster, there are around $N \sim 10^2$ clusters for a diameter of $D \sim 10^{24}m \sim 50Mpc$. It leads to an average distance between clusters of about $d \sim 10^{23}m$ (for an idealization as a disk: $N * (\frac{d}{2})^2 = (\frac{D}{2})^2$). We have seen that roughly internal gravitic field of a cluster would be about $10^{3.7}K_1 \sim 10^{28.2}$. The contribution to the gravitic field of an adjacent cluster at the distance $\frac{d}{2} \sim 0.5 \times 10^{23}m$ is then $\frac{4\times10^{3.7}K_1}{d^2} \sim \frac{4\times10^{28.2}}{10^{46}} = 10^{-17.2}$. If we suppose that a cluster has around six close neighbors. It leads to a gravitic field of about $10^{-16.5}$. It represents 100% of the required external gravitic field.

Case C: The mass of a supercluster as the Local Super Cluster is about $10^{45}kg$ without dark matter ($10^{46}kg$ with dark matter). It is around $10^4$ times the mass of a typical galaxy ($10^{41}kg$). Let's take a typical speed of the matter inside the supercluster compare to its center around 5 times greater than matter in galaxies (for Laniakea, the velocities of the matter compare to its center are only around $600km.s^{-1}$). By definition, the gravitic field is proportional to the mass and to the speed. It gives then roughly that the internal gravitic field of a supercluster could be about $10^4 \times 5 \times K_1 \sim 10^{29.2}$. With $D = 2R \sim 10^{24}m \sim 50Mpc$ (size of a supercluster) it would give $\frac{5\times10^4 K_1}{R^2} \sim \frac{4\times10^{29.2}}{10^{48}} = 10^{-18.2}$. It then represents around 2% of the expected external gravitic field.

Case C bis: Even if one considers ten neighbors for this supercluster, one cannot obtain the expected value of $k_0$.

A first conclusion is that with our very elementary analysis (by keeping the proportionalities between different scales), the cluster with its first neighbors could be a relevant structure to generate our embedding gravitic field $k_0$. Inside the cluster, the order of magnitude is correct and more we are close to its border more the contribution of its first neighbors compensate it. But one cannot exclude the possibility that a supercluster could generate a more important gravitic field than what we have computed. In fact there are others reasons that could make the superclusters less appropriate to explain our $k_0$. Let's see these reasons. We just have seen that between several astrophysical structures (larger than galaxies), the clusters seem to be the structures that generate the more important "mean" value of $k_0$ (and surprisingly, it also gives the good order of magnitude). We are now going to see that the interval of expected $k_0$ (not only the mean value as before) also leads to the same conclusion.

*About the interval $10^{-16.62} < k_0 < 10^{-16.3}$:*

As we are going to see below, very far from the galaxy's center, our explanation of the dark matter is given only by the relation $v(r) \sim 4k_0 r$ (with $r$ the distance to the galaxy's center). In this area, one then has a relation between the speed and the embedding gravitic field $k_0$ (for a fixed position $r$, distance to the galaxy's center). Hence, this relation can allow analyzing our solution in distance's terms. Indeed, one can look for the extreme positions $R$ (distance inside a structure, relatively to its center) that allow obtaining the extreme values $10^{-16.62} < k_0 < 10^{-16.3}$. Here $R$ is the distance from the structure's center and not the distance $r$ that appears in the equilibrium's equation used in our study. If we take, for cluster's example, $K_1 \sim 10^{28.5}$ for its internal gravitic field, one obtains at $R \sim 900kpc \sim 2.7 \times 10^{22}$, $k_0 = \frac{K_1}{R^2} = 10^{-16.35}$ and at $R \sim 1.2Mpc \sim 3.6 \times 10^{22}$, $k_0 = \frac{K_1}{R^2} = 10^{-16.61}$. So, the studied galaxies should be situated in an interval of size around $300kpc$ for a typical cluster's size of around 2.5Mpc to obtain our values of $k_0$ ($10^{-16.62} < k_0 < 10^{-16.3}$). Because nearly all the known galaxies have dark matter and that nearly all the known galaxies are in clusters, a representative sample of galaxies should lead to an interval of, at least, the size of the cluster. From this approach, one can then refine our solution. Because our studied galaxies are in several clusters, and because of the diversity of these clusters, the size of this previous interval (around $300kpc$) is likely too small than expected (size of the cluster). If we look at the rotation's speed's curves of our studied galaxies, the main interval is around $150km.s^{-1} < v < 250km.s^{-1}$. If we look at the rotation's speed's curves of all known galaxies, the main interval is at least $100km.s^{-1} < v < 300km.s^{-1}$. One can then expect that our studied galaxies underestimate the interval of around a factor 2. But even with this correction, the interval increases only of a factor $2^{-1/2}$ (because of an evolution in $r^{-2}$) given then an interval of around $450kpc$. It is always too small to be representative (20% of the expected interval).

If the source was the gravitic field of the supercluster, the interval needs to have a typical size of around the size of the supercluster (ten times greater than for the cluster, 25Mpc). Our previous calculated interval would then represent only 2% of the expected interval. The situation is more critical for the supercluster.

How can we enlarge our interval? A way could be to increase the value of $K_1$. Unfortunately, this interval doesn't notably increase with the value of $K_1$ whatever the considered structure. At this step, one can once again note that the cluster is still clearly the candidate that gives the greater interval, but it gives a too small interval (20%). Another way could be to consider an evolution in $r^{-n}$ with $n < 2$ for the gravitic field. But first, in relative terms, it applies the same correction to all the structures (it doesn't then modify the best candidate). And secondly, even with $n = 1$, the interval, for the cluster, still only represents 25% of the expected interval.

Consequently, in our solution, the more likely way to maintain $k_0$ on larger distance is only a case of kind "Bis" (our previous case B bis, C bis…). In other words, our solution implies that the gravitic field of a large structure must be compensated by its neighbors at its border. With our previous calculations, the gravitic field of a cluster could be compensated by the gravitic field of its nearest neighbors, but for the supercluster, it seems unlikely. It would need around 50 superclusters' neighbors (which is not possible) to compensate the littleness of the value of $k_0$ at its border. For the supercluster, the only way to decrease the number of neighbors would be to have a greater $K_1$ (not to enlarge the interval but to have a greater $k_0$ at its border). But then the expected interval of $k_0$ would be moved to its border without being notably enlarged. It would then imply the existence of lots of galaxies with very much more dark matter for a large area close to the supercluster's center (consequence of an increase of $K_1$). The observations don't show such a situation. It is then unlikely that a supercluster has a so great internal gravitic field.

By this way, once again only the clusters (with its neighbors) seem to be able to generate the expected $k_0$ on large areas. Indeed, from 900kpc to the cluster's center, we start to be in the expected interval of $k_0$. And when we are close to the cluster's border, the neighbors compensate the internal $k_0$ (until 2.5Mpc as seen before). It then gives a size of 1.6Mpc. But furthermore, symmetrically, the expected value of $k_0$ still goes on in the neighbors. In other words, between two cluster's centers (around 5Mpc), one has an interval of expected $k_0$ along around 3.2Mpc. It then represents 65% of the expected typical size of interval. And if we apply the correction of $2^{-1/2}$ (for a more generic sample), it leads to 90% of the expected interval. The 10% left should correspond to an area near the cluster's center where there should be more dark matter (in agreement with the observations as we are going to see it).

A second conclusion is that in our solution, the superclusters are irrelevant to generate a sufficient spatial extension for our $k_0$. And once again the clusters with their nearest neighbors can generate it (with this time an astonishing good order of magnitude for the spatial extension of $k_0$).

<u>About the statistic of the dark matter inside the galaxies:</u>

In a complementary way, one can deduce that if the considered structure was larger than a supercluster of kind of Local Super Cluster (as Laniakea for example) and if a mechanism to maintain $k_0$ would be possible on large area (which seems hard to realize as seen previously), the existence of galaxies without dark matter would be very unlikely. The observations show that there are few galaxies without dark matter. Once again, it seem unlikely that larger structures than superclusters can be consistent with our solution.

Inversely, if $k_0$ was explained by only one cluster (without its neighbors), one should find lot of galaxies without dark matter. In fact, all galaxies at the border of a cluster would be without dark matter. The observation doesn't show such a situation.

A third conclusion is that in our solution, larger structures than superclusters and only cluster (without its neighbors) are irrelevant to generate our $k_0$ and to be in agreement with observation. 0nce again clusters with their nearest neighbors can generate it and be in agreement with the observations.

<u>About the negligible $g_0$:</u>

Furthermore, the cases B and B bis are important and relevant because they allow justifying our second constraint on assumption (I). We have seen that there are two constraints on our assumption to be compliant with general relativity. The first one (to retrieve the already calculated and published term $\frac{K_1}{r^2}$) was verified previously. The second one is on the fact that there should be a weak gravity field $g_0$ associated with the gravitic field $k_0$. We have seen that the influence of the gravitic field evolves like $\frac{\|\vec{F_k}\|}{\|\vec{F_g}\|} \sim 4v \frac{\|\vec{k}\|}{\|\vec{g}\|}$. In the galaxies, for $r > 10kpc$ one has $v \sim 10^5 m.s^{-1}$. It implies that, in this area (ends of

galaxies) $\frac{\|\vec{F_k}\|}{\|\vec{F_g}\|} \sim 10^5 \frac{\|4\vec{k}\|}{\|\vec{g}\|}$. For the internal gravitational fields of galaxies, one has seen that one has around $\frac{\|4\vec{k}\|}{\|\vec{g}\|} \sim 10^{-5}$ (giving $\frac{\|\vec{F_k}\|}{\|\vec{F_g}\|} \sim 1$ as ever said). But in the case B of a single cluster, the typical speed is about 5 times greater than in the galaxies. So, for the internal gravitational fields of clusters, one could have $\frac{\|4\vec{k_0}\|}{\|\vec{g_0}\|} \sim 5 \times \frac{\|4\vec{k}\|}{\|\vec{g}\|} \sim 5 \times 10^{-5}$ (because the gravitic field depends on the source's speed). It then gives in the context of the ends of galaxies $\frac{\|\vec{F_{k_0}}\|}{\|\vec{F_{g_0}}\|} \sim 5$. It means that the component of gravitic field of the cluster represents more than 80% of the gravitational effect due to the cluster. In an approximation of first order, the gravity field $\|\vec{g_0}\|$ of the cluster can be neglected ($\|\vec{F_{g_0}}\| \sim 0.2 \times \|\vec{F_{k_0}}\|$). Furthermore, one can add that this order of magnitude is in agreement with the observations.

With a "Newtonian" approximation, one can show the consistency of this origin. The gravitation forces (due to the cluster) are $\|\vec{F_{CL}}\| = m\|\vec{g_0}\| + m4\|\vec{v} \wedge \vec{k_0}\| = \|\vec{F_{g_0}}\| + \|\vec{F_{k_0}}\| = \|\vec{F_{g_0}}\| \left(1 + \frac{\|\vec{F_{k_0}}\|}{\|\vec{F_{g_0}}\|}\right)$. This can be interpreted as a modification of the observed mass of the cluster $\|\vec{F_{CL}}\| = m\left(1 + \frac{\|\vec{F_{k_0}}\|}{\|\vec{F_{g_0}}\|}\right)\|\vec{g_0}\|$. One has seen that for the clusters $\frac{\|\vec{F_{k_0}}\|}{\|\vec{F_{g_0}}\|} \sim 5$ at the ends of the galaxies (because of the dependency with the speed). One then retrieves roughly the expected quantity of dark matter, a little less than ten times greater than what it is observed.

With the case B bis, there is a second reason to neglect the gravity field $\|\vec{g_0}\|$. Associated with the case B bis, our mechanism that could explain the spatial extension of $k_0$ could also allow neutralizing the gravity field. If we take into account the clusters' neighbors, the gravity field must greatly decrease and the gravitic field can increase in the same time, as one can see on this simplified representation (Fig. 8).

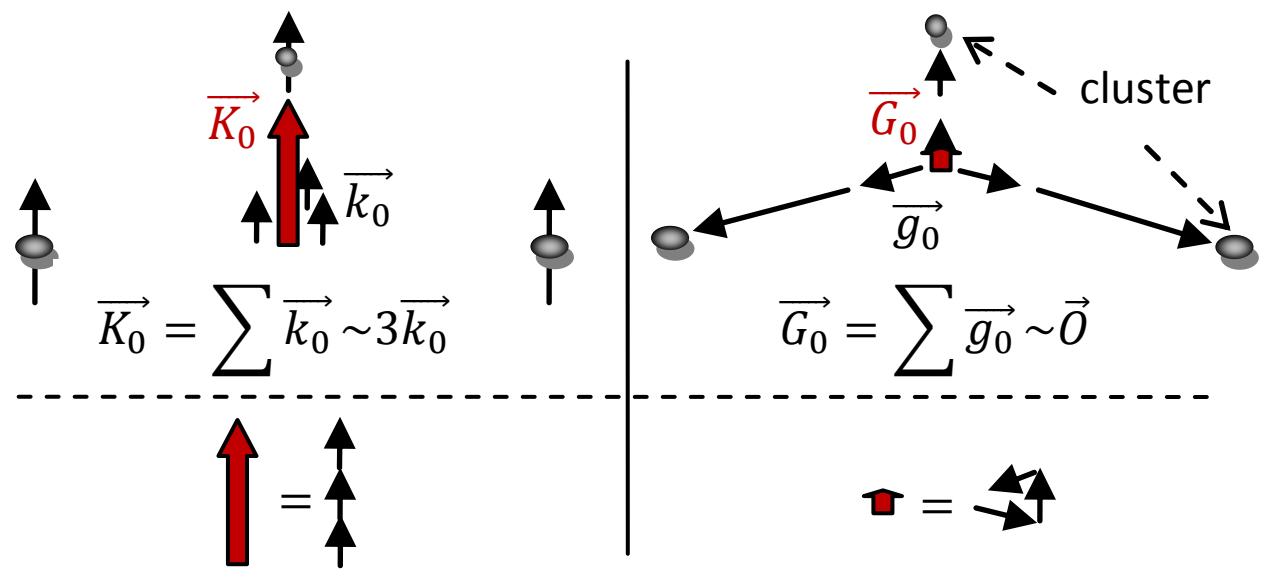

Fig. 8: Influence of the neighbors on the gravitic field (increase of the norm of $\|\vec{k_0}\|$) and on the gravity field (neutralization of $\|\vec{g_0}\|$).

The sum of these vectors allows justifying (once again) that the associated gravity field $\vec{g_0}$ could be neglected compared to $\vec{k_0}$ and also explaining the spatial extension of the value of $k_0$. But, for this latter point, such a situation implies that the internal gravitic fields for clusters close together mustn't have a randomly distributed orientation. In this case, it would mean that for very large astrophysical structures, there should be coherent orientations. This constraint could be in agreement with some published papers (HUTSEMEKERS, 1998; HUTSEMEKERS et al., 2005) that reveal coherent orientations for large astrophysical structures (constraint that will have very important consequences on our solution). This situation is similar, in electromagnetism, to the situation of magnetic materials, from which the atomic spins generate a magnetic field at the upper scale of the material (without an electric field). And I recall that linearized general relativity leads to the same field equations than Maxwell idealization. In other words, this situation is consistent with the general relativity.

So a fourth conclusion is that this case B bis could allow completely justifying our assumption (I) that explains dark matter in the frame of general relativity: negligible $g_0$ and maintain of $k_0$ on large distance.

*Conclusion:*

We have seen that for the four following constraints, to obtain the expected order of magnitude of $k_0$, to obtain the expected interval of $k_0$, to obtain a neglected $g_0$ and to be compliant with some observations, the cluster with its neighbors is the best candidate, and very likely the only candidate.

We will see below that this origin leads to several very important predictions (on the orientations of the galaxies' spin vector and the clusters' spin vector) that allow testing our solution and distinguishing our solution from dark matter assumption and from MOND theories. They are very important because if these predictions are not verified, our solution will certainly not be the solution adopted by the nature to explain dark matter. Inversely, these observations will be difficult to explain in the frame of the others solutions. In fact, they are not predicted by them.

One can make a general remark. With these calculations, we see that general relativity requires that some large astrophysical structures generate significant gravitic fields ($k_0 = 10^{-16.5}$ is significant because sufficient to explain dark matter). And with our roughly calculation, the own gravitic field of a cluster cannot be neglected and even it seems to give the right order of magnitude to explain dark matter. Furthermore, by definition, because $k_0$ depends on the matter speed of the source $\vec{v_s}$ ($\vec{k} \sim \left(-\frac{1}{K}\right) m_p \left(\frac{1}{r^2}\right) \vec{v_s} \wedge \vec{u}$) and because the gravitic force also depends on the speed of matter that undergoes the gravitic field ($\vec{F_k} \sim m_p 4\vec{v} \wedge \vec{k}$), the influence of gravitic field (compare to gravity field) must become more and more important with matter speed ($\frac{\|\vec{F_k}\|}{\|\vec{F_g}\|} \propto v^2$ for $v_s \sim v$). In general, higher is the scale, higher is the typical speed. In other words, higher is the scale, greater is the influence of the gravitic field compared to the gravity field. It is therefore likely that our assumption (I) is less an assumption than a necessary condition that must be taken into account at the scale of galaxies and beyond. We have seen before that at lower scales than galaxies, the gravitic force was negligible but at the scale of galaxies, it begins to be significant, $\frac{\|\vec{F_k}\|}{\|\vec{F_g}\|} \sim 1$. For a cluster, with our previous approximation ($v_{CL} = 5 \times v_{GAL}$), it would give $\frac{\|\vec{F_k}\|}{\|\vec{F_g}\|} \sim 25$.

A last remark on the case D, which is beyond the scope of our paper, the gravitic field at the scale of the universe could also lead to an explanation of dark energy but it implies a new fundamental physical assumption (LE CORRE, 2015). Some experiments at CERN are testing the possibility of such a fundamental assumption.

## 4.2. Experimental relevancies

If we consider the own gravitic field of the cluster, the maximal value of this field must be around the center of the mass's distribution of the cluster. So, with our previous deduction on the origin of the external gravitic field, one can deduce that $k_0$ should be maximal at the center of the cluster. In other words, our solution makes a first prediction:

> The external gravitic field (our "dark matter") should decrease with the distance to the center of the cluster.

Recent experimental observations (JAUZAC et al., 2014) in MACSJ0416.1-2403 cluster reveals that the quantity of dark matter decreases with the distance to the center of the cluster, in agreement with our origins of external gravitic field. This is the second main result of our study.

And for the same reason, a second prediction is:

> More, we are close to the center of the cluster, more the galaxies should have dark matter in very unusual and great proportions.

Very recent publications have just revealed the existence of ultra diffuse galaxies (UDG) in the coma cluster (KODA et al., 2015; VAN DOKKUM et al., 2015). These unsuspected galaxies show a distribution concentrated around the cluster's center and seem to have a more important quantity of dark matter than usually. These UDG would be in agreement with the expected constraint of our solution.

Our solution leads to another consequence due to the origin of our $k_0$. Because our explanation doesn't modify general relativity, the solution of a declining rotation curve (at the ends of a galaxy) is always possible. For such a situation, the galaxy must not be under the influence of the external gravitic field. This means that the galaxy must be isolated. More precisely, our third prediction is:

> The galaxies without dark matter (with a declining rotation curve) must be at the ends of the cluster (far from its center) and far from others clusters (i.e. à priori at the borders of a superclusters).

Some observations corroborate this prediction. For example NGC 7793 (CARIGNAN & PUCHE, 1990) has a truly declining rotation curve. As written in (CARIGNAN & PUCHE, 1990), NGC 7793 is effectively "the most distance member of the Sculptor group" (if the visible member is distributed on the whole group, it also means that it is at the end of the group) in agreement with our solution.

I recall that, as said before, if a larger structure than supercluster (as Laniakea) was the origin of the external gravitic field, the spatial extension of such a structure would make nearly impossible to detect galaxies not under the influence of the expected $k_0$. So, finding few cases of galaxies with a truly declining rotation curve is also in agreement with the fact that the good scale that contributes to the external gravitic field is between the clusters and the supercluster.

# 5. Gravitic field of cluster on galaxies

## 5.1. Application on satellite dwarf galaxies

We are now going to look at satellite dwarf galaxies and retrieve two unexplained observed behaviors (rotation speed values and movement in a plane). First, one deduces an asymptotic expression for the component of gravitic field (our "dark matter"). We know that, to explain rotation speed curve, we need two essential ingredients (internal gravitic field $\frac{K_1}{r^2}$ and embedding external gravitic field $k_0$). Let's calculate how far one has $\frac{K_1}{r^2} \sim k_0$ for our previous studied galaxies.

Tab. 1: Distance $r_0$ where internal gravitic field $\frac{K_1}{r^2}$ becomes equivalent to external gravitic field $k_0$.

| | $K_1$ | $k_0$ | $r_0\left[\frac{K_1}{r^2} \sim k_0\right]$ | $r_0[kpc]$ |
|---|---|---|---|---|
| NGC 5055 | $10^{24.6}$ | $10^{-16.62}$ | $10^{20.61}$ | 13 |
| NGC 4258 | $10^{24.85}$ | $10^{-16.54}$ | $10^{20.695}$ | 16 |
| NGC 5033 | $10^{24.76}$ | $10^{-16.54}$ | $10^{20.65}$ | 15 |
| NGC 2841 | $10^{24.85}$ | $10^{-16.33}$ | $10^{20.59}$ | 13 |
| NGC 3198 | $10^{24.9}$ | $10^{-16.55}$ | $10^{20.725}$ | 18 |
| NGC 7331 | $10^{24.18}$ | $10^{-16.3}$ | $10^{20.24}$ | 6 |
| NGC 2903 | $10^{24.71}$ | $10^{-16.3}$ | $10^{20.505}$ | 11 |
| NGC 3031 | $10^{24.15}$ | $10^{-16.57}$ | $10^{20.36}$ | 8 |
| NGC 2403 | $10^{24.59}$ | $10^{-16.39}$ | $10^{20.49}$ | 10 |
| NGC 247 | $10^{24.3}$ | $10^{-16.3}$ | $10^{20.3}$ | 7 |
| NGC 4236 | $10^{24.}$ | $10^{-16.34}$ | $10^{20.17}$ | 5 |
| NGC 4736 | $10^{24.54}$ | $10^{-16.3}$ | $10^{20.42}$ | 9 |
| NGC 300 | $10^{24.27}$ | $10^{-16.31}$ | $10^{20.29}$ | 6 |
| NGC 2259 | $10^{24.2}$ | $10^{-16.3}$ | $10^{20.25}$ | 6 |
| NGC 3109 | $10^{24.}$ | $10^{-16.58}$ | $10^{20.29}$ | 6 |
| NGC 224 | $10^{24.}$ | $10^{-16.5}$ | $10^{20.25}$ | 6 |

The previous values mean that, in our approximation, for $r \gg r_0 \sim 15\, kpc$ the galactic dynamic is dominated by the external embedding gravitic field $k_0$.

And, very far from the center of galaxies, formula $(VIII)$ can be written $k(r) \sim k_0$. So far, one can only consider external gravitic field $k_0$, formula $(VI)$ gives:

$$v_{halo}{}^2(r) \sim 4k_0 v(r) r$$

Very far from center of galaxies, $v_{disk}(r)$ the speed component due to internal gravity tends to zero. If we neglect $v_{bulge}(r)$, one has $v_{halo}^2(r) = v^2(r) - v_{disk}^2(r) - v_{bulge}^2(r) \sim v(r)$

In our solution, very far from galaxies center ($r > 100\, kpc$), one has (if we neglect $v_{bulge}$):

$$v(r) \sim 4k_0 r \qquad (IX)$$

One can then deduce the rotational speed of satellite galaxies for which the distances are greater than $100\, kpc$. In this area, the own gravitation of galaxies is too weak to explain the measured rotational speed, but the external gravitic field embedding the galaxy can explain them. Our previous study gives:

$$10^{-16.62} < k_0 < 10^{-16.3}$$

With the relation $v(r) \sim 4k_0 r$, one has (in kilo parsec):

$$2.88 \times 10^3 r[kpc] < v(r) < 6.03 \times 10^3 r[kpc]$$

It then gives for $r \sim 100\, kpc$ :

$$288\, km.s^{-1} < v(r \sim 100\, kpc) < 603\, km.s^{-1}$$

This range of rotation speed is in agreement with experimental measures (ZARITSKY et al., 1997) (on satellite dwarf galaxies) showing that the high values of the rotation speed continues far away from the center. This is the third main result of our study.

## 5.2. About the direction of $\overrightarrow{k_0}$ for a galaxy

Our previous study reveals that, inside the galaxy, external gravitic field $k_0$ is very small compare to the internal gravitic field $\frac{K_1}{r^2}$. But this internal component decreases sufficiently to be neglected far from galaxy center, as one can see in the following simplified representation (Fig. 9).

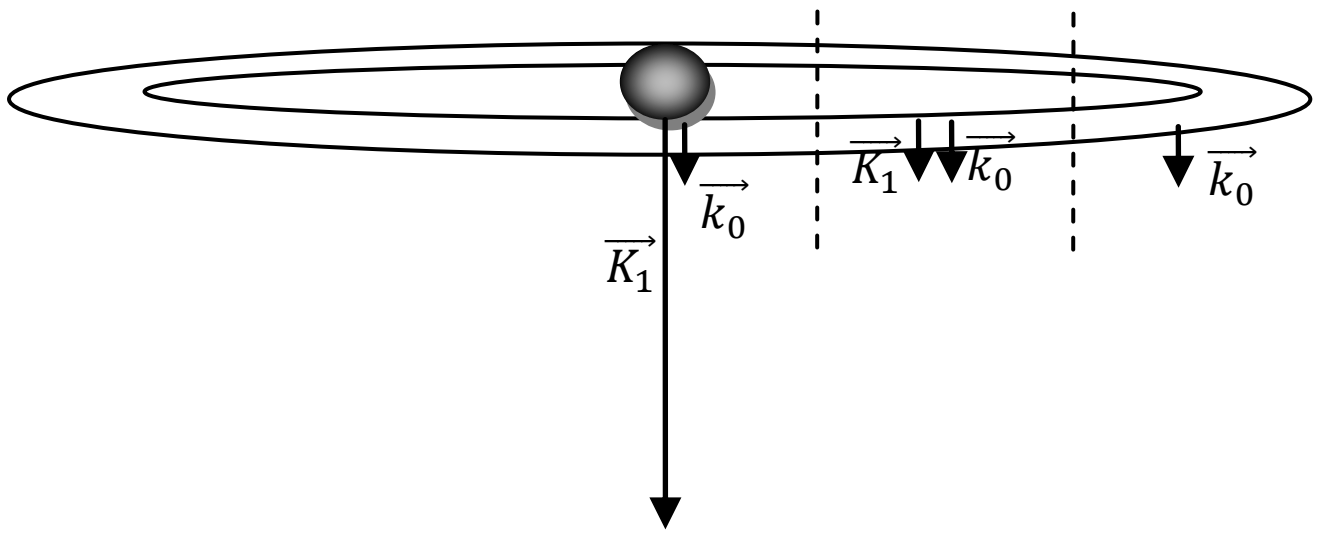

Fig. 9: Evolution of internal gravitic field $\frac{K_1}{r^2}$ compare with external gravitic field $k_0$ defining three different areas.

One can then define three areas with specific behavior depending on the relative magnitude between the two gravitic fields; very far from center of galaxy for which only direction of $\overrightarrow{k_0}$ acts, close around center of galaxy for which only direction of $\overrightarrow{K_1}$ acts and a transition zone for which the two directions act. Just like in electromagnetism with magnetic field, at the equilibrium, a gravitic field implies a movement of rotation in a plane perpendicular to its direction. With these facts, one can make several predictions.

Very far from center of galaxy, there are the satellite dwarf galaxies. One can then make a fourth prediction:

| |
|---|
| The movement of satellite dwarf galaxies, leaded by only external $\overrightarrow{k_0}$, should be in a plane (perpendicular to $\overrightarrow{k_0}$). |

This prediction has been recently verified (IBATA et al., 2014) and is unexplained at this day. In our solution, it is a necessary consequence. This is the fourth main result of our study.

One can also make statistical predictions. The external gravitic field for close galaxies must be relatively similar (in magnitude and in direction). So, close galaxies should have a relatively similar spatial orientation of rotation's planes of their satellite dwarf galaxies. The fifth prediction is:

| |
|---|
| Statistically, smaller is the distance between two galaxies; smaller is the difference of orientation of their satellite dwarf galaxies' planes. |

We saw previously that the cluster of galaxies is likely the good candidate to generate our external gravitic field $\overrightarrow{k_0}$. If this assumption is true, one can expect that the orientations of the satellite dwarf galaxies' planes should be correlated with the direction of the internal gravitic field of the cluster (which is our $\overrightarrow{k_0}$). It is quite natural (just like for the galaxies) to assume that, at the equilibrium, the direction of $\overrightarrow{k_0}$ is perpendicular to the supergalactic plane. It leads to our sixth prediction:

| |
|---|
| Statistically, inside a cluster, the satellite dwarf galaxies' planes should be close to the supergalactic plane. |

These predictions are important because it can differentiate our solution from the dark matter assumption. Because the dark matter assumption is an isotropic solution, such an anisotropic behavior of dwarf galaxies (correlation between dwarf galaxies' directions of two close galaxies and correlation with supergalactic planes) is not à priori expected. As said before, these coherent orientations could be another clue of a more general property on very large structures at upper scales. One can recall that there are some evidences of coherent orientations for some large astrophysical structures (HUTSEMEKERS, 1998; HUTSEMEKERS et al., 2005). Furthermore, in our solution, there is no alternative. Such correlations are imposed by our "dark matter" explanation.

Very recent observations verify these three predictions (TULLY et al., 2015). The conclusion is:"The present discussion is limited to providing evidence that almost all the galaxies in the Cen A Group lie in two almost parallel thin planes embedded and close to coincident in orientation with planes on larger scales. The two-tiered alignment is unlikely to have arisen by chance". It means that the satellites in the Centaurus A group are distributed in planes (and then also their movement). It is our fourth prediction. It means that the two planes have very close orientations. It is our fifth prediction. And it means that these planes have very close orientations to supergalactic plane. It is our sixth prediction. And this observation validates that the cluster is likely the good candidate to generate our external $\overrightarrow{k_0}$. These observations are certainly one of the main evidences of the relevance of our solution (with also the ultra diffuse galaxies). And one can go further to compare our solution with the non baryonic dark matter assumption. As we have seen previously, $\overrightarrow{k_0}$ is too small to influence the orientation of the rotation plane of a galaxy. It means that our solution implies a decorrelation between the galaxy's rotation planes and the satellite dwarf galaxies' planes. While the galaxy's rotation planes are distributed randomly, the satellite dwarf galaxies' planes have a specific orientation imposed by the cluster's orientation. For example in the Centaurus A group, contrary to their satellite dwarf galaxies' planes, the rotation planes of the galaxies are à priori distributed randomly (we will see that our solution implies that they should be randomly distributed in a half-space). This situation imposed by our solution and which is verified by the observation is more difficult to justify in the traditional dark matter assumption. Because, à priori, if non baryonic matter is distributed as ordinary matter, the gravitation laws should lead to the same solution, i.e. the same orientation of the both planes (plane of the galaxy and plane of the satellite dwarf galaxies). It is not the case, meaning that with this assumption dark matter and ordinary matter should have different behaviors. Inversely, if we assume a possible specific distribution of dark matter to explain the plane of the satellite dwarf galaxies, it leads to a problem. One must define a specific distribution for each galaxy (several orientations between the two planes are possible). For a galaxy, then what is the constraint that would favor one configuration over another? Once again a way to solve this problem would be that baryonic and non baryonic matters undergo differently the gravitation laws (to converge to two different mathematical solutions, two different physical distributions). I recall that one of the main interests of the dark matter assumption is to not modify the gravitation laws. With these observations, specific physical laws should make this assumption a little more complex. Of course, non trivial complex solutions can certainly be imagined to explain these situations, but our solution is very natural to explain them. To sum up, these observations are unexpected in the dark matter's assumption but required by our solution.

One can imagine two others configurations. When the internal gravitic field of the galaxy and the external $\overrightarrow{k_0}$ is in the opposite direction:

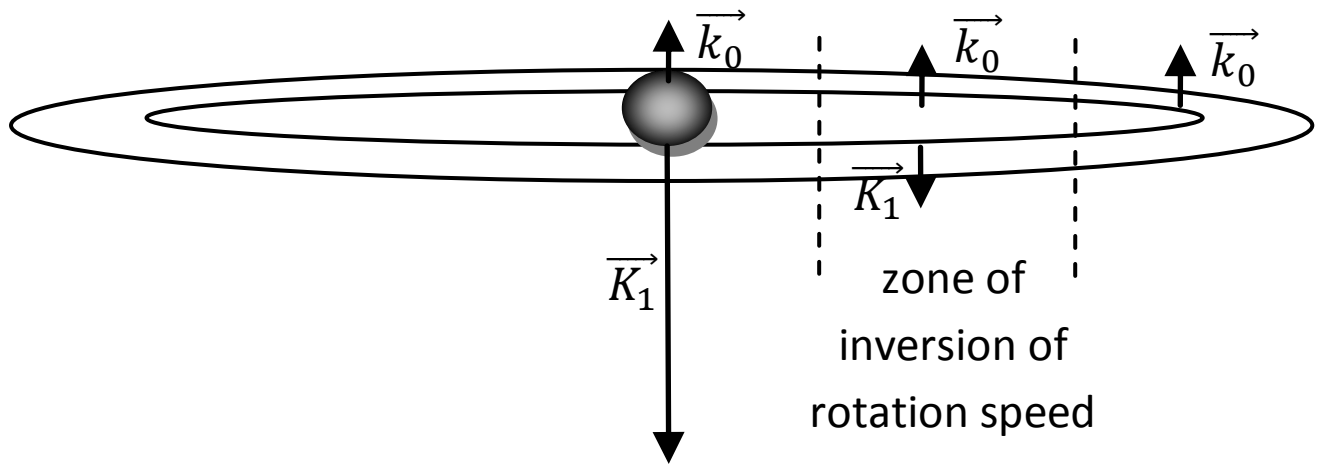

Fig. 10: Internal gravitic field $\frac{K_1}{r^2}$ in opposite direction with external gravitic field $k_0$ implying two zones of opposite rotation speed.

With this situation (Fig. 10) there are two zones inside the galaxy with two opposite directions of gravitic field. It leads to the seventh prediction:

| |
|---|
| A galaxy can have two portions of its disk that rotate in opposite directions to each other. |

This prediction is in agreement with observations (SOFUE & RUBIN, 2001).

The other configuration leads to a new possible explanation of the shape of some galaxies. The warped galaxies could be due to the difference of direction between internal and external gravitic field, as one can see in this other simplified 2D representation (Fig. 11).

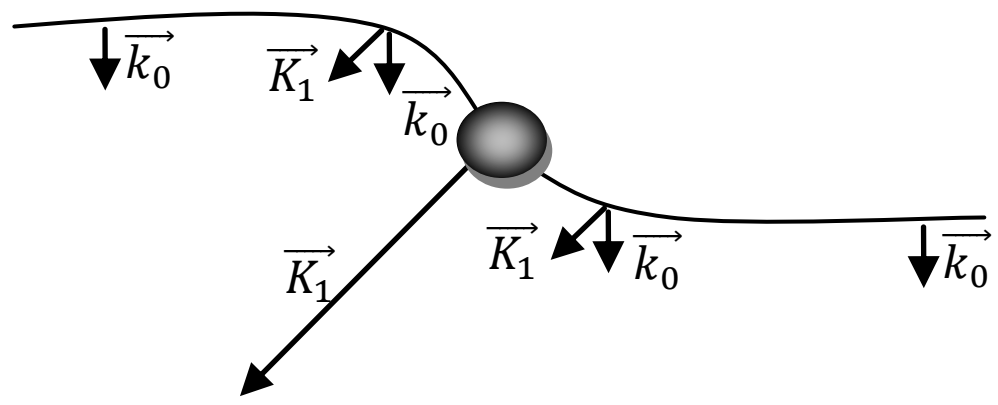

Fig. 11: Warped galaxy due to the difference of direction between internal gravitic field $\frac{K_1}{r^2}$ and external gravitic field $k_0$.

It would lead to our eighth prediction:

| |
|---|
| If the gravitic fields would be effectively the main cause of the wrapping, there should be a correlation between the value of the wrapping and the angular differences of galaxies' and clusters' spin's vector. |

## 5.3. About the galaxies' spins' direction in a cluster

We have seen that in our solution (to maintain $\overrightarrow{k_0}$ on a large area) the neighbors of a cluster must intervene by adding their own gravitic field. This vectorial sum can maintain $\overrightarrow{k_0}$ only if the gravitic fields of the neighbors of a cluster are nearly parallel. This constraint on our solution leads to a new very important prediction. If it is not verified, it will be very hard for our solution to explain dark matter. Our ninth prediction:

| |
|---|
| The gravitic field of two close clusters must be close to parallel. In other words, the spin vector of these two close clusters must be close to parallel. |

Once again, this prediction is not trivial (not expected by any other theories). But some observations seem to make this prediction possible. For example the results of (HUTSEMEKERS, 1998; HUTSEMEKERS et al., 2005) reveal coherent orientations for large astrophysical structures. But furthermore, in the paper of (HU et al., 2005), their result is that the galaxies' spins of the local super cluster (LSC) point to the center of the LSC. With an hundred of clusters inside the LSC, one can estimate at least about twenty clusters along the border of the LSC. It leads to an angle between two close clusters of around $18° = 360°/20$. If the spin of the cluster follows the same physical influence than the galaxies' spins (point to the center of the LSC), it would mean that the gravitic fields of two close clusters are nearly parallel, justifying our previous calculation. Unfortunately, this paper doesn't give results on the cluster's spins.

If the cluster is effectively the origin of $\overrightarrow{k_0}$ and as we have seen it, if $\overrightarrow{k_0}$ and $\frac{\overrightarrow{K_1}}{r^2}$ are not in the same half-space (inside a cluster), the galaxy must have two portions of its disk that rotate in opposite directions to each other. Statistically, these kinds of galaxies are rare. It leads to the tenth prediction (if it is not checked, our solution can be considered as inconsistent):

| |
|---|
| In our solution $\overrightarrow{k_0}$ (from cluster) and $\frac{\overrightarrow{K_1}}{r^2}$ (from galaxy) must be in the same half-space, for nearly all the galaxies inside a cluster. Say differently, galaxies' spin vector and spin vector of the cluster (that contains these galaxies) must be in a same half-space. |

There are some observations that seem to validate this sine qua none constraint (HU et al., 2005). Their result, as said before, is that the galaxies' spins of the local super cluster (LSC) point to the center of the LSC. With an hundred of cluster in the local super cluster, it means that galaxies (our $\frac{\overrightarrow{K_1}}{r^2}$) inside only one cluster are included inside a half-space (except if the cluster has a banana's shape around the center of the LSC which is not the case). It then validates a part of our prediction. Unfortunately, the paper (HU et al., 2005) doesn't indicate if the clusters' spins (our $\overrightarrow{k_0}$) point also to the center of the local super cluster. If it is the case, it will entirely verify our constraint (and also the ninth prediction).

One can note that if the source of $\overrightarrow{k_0}$ was the supercluster, it will lead to a more constraining situation, in which the galaxies' spin vector and the spin vector of the supercluster should be in a same half-space. This constraint means that all the galaxies would be in a same half-space at the scale of the supercluster. With all the galaxies pointing to the LSC's center, the previous observation makes the realization of this constraint impossible. The supercluster (and all larger structures) cannot be the source of our $\overrightarrow{k_0}$ (in agreement with our previous deductions).

*Initial conditions to obtain these orientations of galaxies*: In our solution, one can imagine a process that could explain these orientations. At the birth of a galaxy, the embedded $\overrightarrow{k_0}$ of the cluster curves the trajectories and makes the matter goes towards or away from this galaxy center, depending on the initial orientation of the matter speed and on the position relatively to the center. It seems reasonable to think that the matter of the cluster globally rotates in a same sense. If it is the case, the gravitic field $\overrightarrow{k_0}$ will globally impose not only the rotation's plane (as ever seen) but also a same orientation for the galaxies' spins. Indeed, at a same radius the matter which is curved towards the galaxy's center will more likely fall down and impose its movement in the future galaxy (because of the global rotation of the clusters that makes the local space not isotropic):

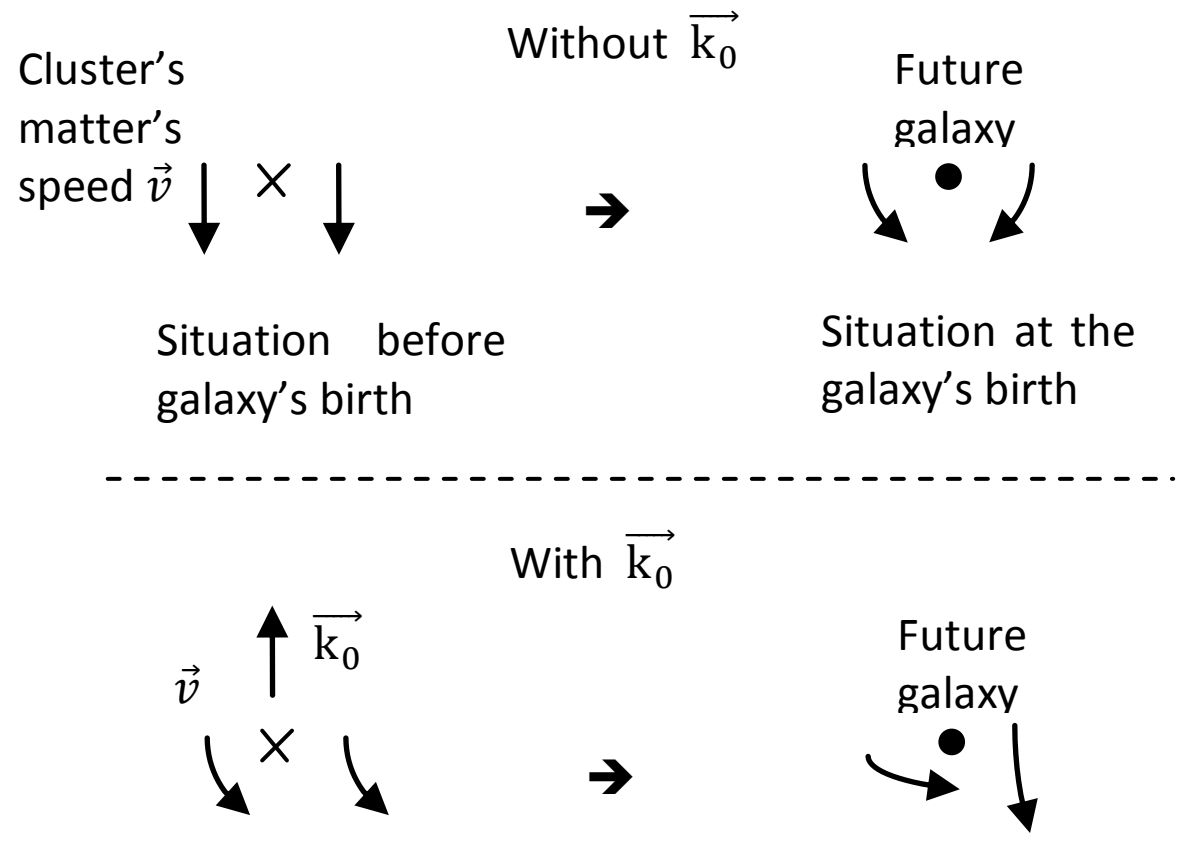

Fig. 12: Influence on the galactic gravity field (and on its spin's vector) at the galaxy's birth with or without the external gravitic field $k_0$.

It doesn't mean necessarily that $\overrightarrow{k_0}$ and $\frac{\overrightarrow{K_1}}{r^2}$ are parallel but only that $\overrightarrow{k_0}$ and $\frac{\overrightarrow{K_1}}{r^2}$ are in a same half-space, because local effects can modify initial speeds and very "quickly" the galaxy's spin become very larger than $\overrightarrow{k_0}$ to finally impose its own direction (one can even imagine a precession phenomenon between these two vectors). So, the more unlikely is that the two vectors are opposite. But, as seen before, it is not impossible (there are some galaxies with two portions of its disk that rotate in opposite directions to each other). With this explanation, this case should be rare. The observation confirms that such galaxies are rare.

*A centripetal gravitic force inside the galaxies*: In the §3.2, we have make the assumption that the force $\overrightarrow{F_k}$ was centripetal. This previous explanation also allows justifying that, over the time the galaxies preferentially orientate themselves with an external gravitic field $\overrightarrow{k_0}$ that generates a centripetal force. As said before, it makes galaxies with two portions of its disk that rotate in opposite directions to each other very rare.

## 5.4. Galaxy composed of two equilibrium solutions

Our solution, for the galaxies, represents an equilibrium state of the physical equations. If we look into details this equilibrium, one can note that the galaxies are then composed of two kind of equilibrium. Near the center of galaxies, the matter is directly maintained by the gravitational forces of the galaxies. But at the ends of the galaxies, the matter is maintained differently. It is essentially due to the external gravitic field. The consequence is then that, in this area, matter can have à priori any speed at any position because, just like in a magnetic field, the radius of curvature of the trajectory is imposed by the speed of the particle, but any speed is possible. So, one must now explain how one has the good speed at the good position. This can be explained by a filtering over the time. If the speed of the matter is smaller than the equilibrium, then the curvature's radius is also smaller than the equilibrium. The matter will be more and more close to the galaxy and will be caught by the internal gravitation field of the galaxy over the time. If the speed of the matter is greater than the equilibrium, then the radius is also greater than the equilibrium. The matter will be more and more far from the galaxy and will escape from the galaxy over the time. Ultimately, there is only matter in equilibrium (with the adequate speed and radius).

From this explanation, one can deduce our eleventh and twelfth predictions in the external area:

| |
|---|
| More a galaxy is young; more the dispersion of the speeds of the satellite dwarf galaxies should increase. |

And because the galaxy has the same years old whatever the position:

| |
|---|
| The speeds' dispersion along the flat curve should be relatively constant. |

*Recall*: It is interesting to recall that our calculations was made in the configuration ($\vec{k} \perp \vec{r}$, $\vec{v} \perp \vec{k}$ and $\vec{v} \perp \vec{r}$ ). Let's also recall that this configuration is a very general expected configuration because of this property of the equilibrium state. In our expression $\vec{k}(r) = \left(\frac{\overrightarrow{K_1}}{r^2} + \overrightarrow{k_0}\right)$, we know from our study that inside the galaxy, the term $k_0$ is negligible, meaning that the term $\vec{k}(r) = \frac{\overrightarrow{K_1}}{r^2}$ imposes its orientation to the galaxy ($\vec{v} \perp \overrightarrow{K_1} \Longrightarrow \vec{v} \perp \vec{k}$). And at the ends of the galaxy, the term $\frac{K_1}{r^2}$ is negligible, meaning that the term $\vec{k}(r) = \overrightarrow{k_0}$ imposes its orientation ($\vec{v} \perp \overrightarrow{k_0} \Longrightarrow \vec{v} \perp \vec{k}$) to the end of the galaxy (and to the dwarf satellites galaxies). Finally, the configuration $\vec{v} \perp \vec{k}$ guarantees the expression in modulus $k(r) = \left(\frac{K_1}{r^2} + k_0\right)$ at the equilibrium, except in the transition zone (where $\frac{K_1}{r^2} \sim k_0$) for which angles should be taken into account, but it is out of the focus of our study. In the frame of our study, the configuration ($\vec{k} \perp \vec{r}$, $\vec{v} \perp \vec{k}$ and $\vec{v} \perp \vec{r}$ ) is then a very general expected situation meaning that each gravitic field ($\overrightarrow{k_0}$ or $\frac{\overrightarrow{K_1}}{r^2}$) imposes its direction to the matter speed in separated areas.

## 6. Gravitic field and quantity of dark matter

We are now going to look at several situations that will give us some values of the quantities of dark matter in agreement with experimental observations.

### 6.1. 1st approximation

As one has seen it before, gravitic field is very weak compared to gravity field at our scale. In first approximation, one can keep only gravity field and neglected gravitic field for many situations (in fact, all situations in which dark matter assumption is useless). This approximation makes the linearized general relativity equivalent to the Newtonian laws (except for the concept of propagation of gravitational wave). Effectively, from the relation $(V)$ of the linearized general relativity:

$$ds^2 = \left(1 + \frac{2\varphi}{c^2}\right)c^2dt^2 - \frac{8\vec{H}.\overrightarrow{dx}}{c}cdt - \left(1 - \frac{2\varphi}{c^2}\right)\sum (dx^i)^2$$

One retrieves the Newtonian approximation that is in low speed ($v \ll c$) and in neglecting gravitic field ($\|\vec{H}\|\sim 0$):

$$ds^2 \sim c^2dt^2\left(1 + \frac{2\varphi}{c^2}\right)$$

In these approximations, the linearized general relativity gives then the same Newtonian expression of the component $g_{00}$ of the metric ($g_{00}\sim 1 + 2\frac{\varphi}{c^2}$).

So, if one can neglect gravitic field, all current computation is always valid and even identical. We are going to see that effectively for many situations gravitic field can be neglected, in particular in solar system (Mercury precession, deviation of light near the sun…).

### 6.2. 2nd approximation

The precedent "Newtonian" approximation completely neglected gravitic field. Let's make an approximation which considers that gravitic field is weak compared to gravity field but not sufficiently to be neglected. We are going to search for an expression of $g_{00}$ containing the new term $g_{0i}$. This approximation will be a simplified way to take into account the term $g_{0i}$ in the traditional relation containing only $g_{00}$. It will allow obtaining orders of magnitude to continue to test the relevancy of our solution.

Our previous study shows that an external gravitic field $\overrightarrow{k_0}$, uniform at the scale of a galaxy, explains the flat rotation's speed. And we have seen that for $r \gg 15kpc$, one can only consider this uniform $\overrightarrow{k_0}$ (the internal gravitic field becomes too small). It is the domain of validity of the linearized general relativity. In electromagnetism, when an atom is embedded in a constant and uniform magnetic field $\vec{B}$, one can take for the potential vector $\vec{A} = \frac{1}{2}\vec{B}\wedge\vec{r}$ (BASDEVANT, 1986). So let's take for the potential vector $\vec{H} = \frac{1}{2}\vec{k}\wedge\vec{r}$. One can note that this definition implies that $\overrightarrow{rot}\,\vec{H} = \vec{k}$ in agreement with Maxwell equations of linearized general relativity, as one can see it in the following calculation of our general configuration ($\vec{k}\perp\vec{r}, \vec{v}\perp\vec{k}$ and $\vec{v}\perp\vec{r}$):

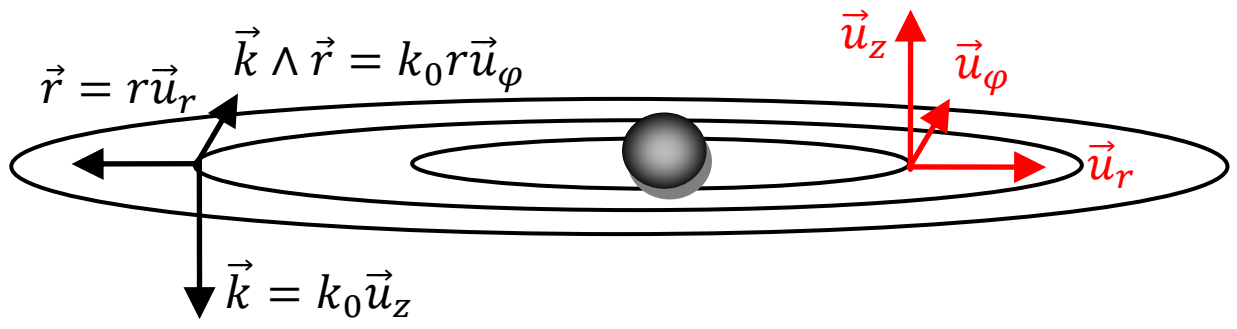

Fig. 13: The potential vector $\vec{H} = \frac{1}{2}\vec{k}\wedge\vec{r}$

Explicitly, in the cylindrical coordinate system $(\vec{u}_r;\vec{u}_\varphi;\vec{u}_z)$ one has the external gravitic field and its potential vector:

$$\vec{k}\sim\overrightarrow{k_0} = \begin{pmatrix}0\\0\\k_0\end{pmatrix}$$

$$\vec{H} = \begin{pmatrix}H_r\\H_\varphi\\H_z\end{pmatrix} = \frac{1}{2}\vec{k}\wedge\vec{r} = \frac{1}{2}\begin{pmatrix}0\\0\\k_0\end{pmatrix}\wedge\begin{pmatrix}r\\0\\0\end{pmatrix} = \frac{1}{2}\begin{pmatrix}0\\k_0 r\\0\end{pmatrix}$$

$$\overrightarrow{rot}\vec{H} = \begin{pmatrix}\frac{1}{r}\frac{\partial H_z}{\partial\varphi} - \frac{\partial H_\varphi}{\partial z}\\ \frac{\partial H_r}{\partial z} - \frac{\partial H_z}{\partial r}\\ \frac{1}{r}\left(\frac{\partial(rH_\varphi)}{\partial r} - \frac{\partial H_r}{\partial\varphi}\right)\end{pmatrix} = \frac{1}{2}\begin{pmatrix}-\frac{\partial(k_0 r)}{\partial z}\\0\\ \frac{1}{r}\left(\frac{\partial(rk_0 r)}{\partial r}\right)\end{pmatrix} = \begin{pmatrix}0\\0\\k_0\end{pmatrix}$$

That shows that one effectively has $\overrightarrow{rot}\vec{H} = \vec{k}$. As announced in our assumption (I), the approximation of a uniform gravitic field $\overrightarrow{k_0}$ is compliant with linearized general relativity.

If we assume that, in the previous cylindrical coordinate system, we have a particle speed $\vec{v} = v\vec{u}_\varphi$ with $v$ constant (one can note that it is approximately the case for the matter in the galaxy for $r \gg 15kpc$), one has $\overrightarrow{grad}(\vec{H}.\vec{v}) = \frac{1}{2}k_0 v\vec{u}_r$:

$$\vec{H}.\vec{v} = \frac{1}{2}\begin{pmatrix}0\\k_0 r\\0\end{pmatrix}.\begin{pmatrix}0\\v\\0\end{pmatrix} = \frac{1}{2}k_0 rv$$

and

$$\overrightarrow{grad}(\vec{H}.\vec{v}) = \begin{pmatrix}\frac{\partial}{\partial r}\\ \frac{1}{r}\frac{\partial}{\partial\varphi}\\ \frac{\partial}{\partial z}\end{pmatrix}\frac{1}{2}k_0 rv = \begin{pmatrix}\frac{1}{2}k_0 v\\0\\0\end{pmatrix}$$

Another explicit calculation gives $\vec{v}\wedge(\overrightarrow{rot}\vec{H}) = k_0 v\vec{u}_r$:

$$\vec{v}\wedge(\overrightarrow{rot}\vec{H}) = \vec{v}\wedge\vec{k} = \begin{pmatrix}0\\v\\0\end{pmatrix}\wedge\begin{pmatrix}0\\0\\k_0\end{pmatrix} = \begin{pmatrix}vk_0\\0\\0\end{pmatrix}$$

Finally, in this configuration, in the galaxy (for $r \gg 15kpc$) with $\vec{H} = \frac{1}{2}\vec{k}\wedge\vec{r}$ and $\vec{v} = v\vec{u}_\varphi$ ($v$ and $k$ constant), one has:

$$\vec{v}\wedge(\overrightarrow{rot}\vec{H}) = 2\overrightarrow{grad}(\vec{H}.\vec{v})$$

By this way, the movement equations become:

$$\frac{d^2\vec{x}}{dt^2} \approx -\overrightarrow{grad}\varphi + 4\vec{v}\wedge(\overrightarrow{rot}\vec{H}) \approx -\overrightarrow{grad}\varphi + 8\overrightarrow{grad}(\vec{H}.\vec{v})$$

$$\frac{d^2\vec{x}}{dt^2} \approx -\overrightarrow{grad}(\varphi - 8\vec{H}.\vec{v})$$

In this configuration, the linearized general relativity modifies the Newtonian potential as:

$$\varphi \to \varphi - 8\vec{H}.\vec{v}$$

And then it leads to an approximation of $g_{00}$ containing $g_{0i}$:

$$g_{00} \sim 1 + 2\frac{\varphi}{c^2} - 16\frac{\vec{H}.\vec{v}}{c^2}$$

Remarks: This relation, valid in the configuration ($\vec{k} \perp \vec{r}$, $\vec{v} \perp \vec{k}$ and $\vec{v} \perp \vec{r}$ ) can be still a good approximation when the discrepancies to these angles are small.

## 6.3. Gravitic field and light deviation

We can use this relation to obtain an order of magnitude of the deviation of light due to gravitic field (our quantities of dark matter). From this new relation $g_{00} \sim 1 + \frac{2\varphi}{c^2} - 16\frac{\vec{v}.\vec{H}}{c^2} = 1 + \frac{2}{c^2}(\varphi - 8\vec{v}.\vec{H})$ the traditional photon deviation expression due to gravitation $\delta_c = \frac{4\varphi}{c^2} = \frac{4GM}{c^2 r}$ is then modified by adding the term $\delta_t = 32\frac{\vec{v}.\vec{H}}{c^2}$ in our specific configuration.

For a photon, one can compute, with $v = c$:

$$\delta_t = 32\frac{\vec{v}.\vec{H}}{c^2} \sim 32\frac{\|\vec{H}\|}{c}$$

As said before, one can take for the potential vector $\vec{H} = \frac{1}{2}\vec{k} \wedge \vec{r}$. The photon deviation expression is then:

$$\delta_t \sim 32\frac{\|\vec{H}\|}{c} \sim 16\frac{kr}{c}$$

Let's apply this relation to our previous studied galaxies. Roughly, at about $r \sim 20kpc \sim 60 \times 10^{19}m$, one has, for nearly all the galaxies, the following value of gravitic field $k \sim 10^{-16.5}$. It gives the order of magnitude of the correction due to gravitic field:

$$\delta_t \propto 16\frac{10^{-16.5} * 60 \times 10^{19}}{3 \times 10^8} \propto 6.3 \times 10^{-5}$$

And the curvature (due to gravity field) is about (for a typical mass of $10^{41}kg$):

$$\delta_c \propto 4\frac{7 \times 10^{-11} * 10^{41}}{9 \times 10^{16} * 60 \times 10^{19}} \propto 0.05 \times 10^{-5}$$

It means that $\delta_t$ represents about 99.2% of the deviation ($\delta_t/(\delta_t + \delta_c)$).

In our solution, gravitic field generates nearly all the curvature of galaxies. In the frame of the dark matter assumption, it means that galaxies would be essentially composed of dark matter, in agreement with experimental data (NEYMAN et al., 1961) which gives at least 90%. This is the fifth main result of our study. One can note that this calculation explains that the phenomenon of gravitational lensing is extremely sensitive to "dark matter", much more than the ordinary matter (the gravity component) for large astrophysical structures.

*About light deviation in solar system:* With solar mass $M = 2 \times 10^{30}kg$ and solar radius $r = 7 \times 10^9$ and with the same previous value of galaxy gravitic field for the sun (which is certainly overestimated because Sun mass ~ $10^{-11}$ Galaxy mass) it gives:

$$\delta_t \propto 16\frac{10^{-16.5} * 7 \times 10^9}{3 \times 10^8} \propto 10^{-14}$$

and

$$\delta_c \propto 4\frac{7 \times 10^{-11} * 2 \times 10^{30}}{9 \times 10^{16} * 7 \times 10^9} \propto 10^{-6}$$

It means that $\delta_t$ represents about $10^{-6}$% of the deviation ($\delta_t/(\delta_t + \delta_c)$). In solar system, the gravitic field of Sun does not modify the light deviation. There is no detectable "dark matter".

## 6.4. Gravitic field and $\Omega_{dm}$

So far, we have treated the problem of dark matter in the context of galaxies. But dark matter is also needed in the description of the CMB. We will always address this problem with linearized general relativity. Despite this approximation, we will see that once again the resulting magnitudes are surprisingly good.

Einstein's equations, with the impulse-energy tensor $T_{kp}$ and the sign convention of (HOBSON et al., 2009), are:

$$G_{kp} = R_{kp} - \frac{1}{2}g_{kp}R = -\frac{8\pi G}{c^4}T_{kp}$$

Let's write these equations in the equivalent form:

$$R_{kp} = -\frac{8\pi G}{c^4}(T_{kp} - \frac{1}{2}g_{kp}T)$$

In weak field and low speed ($T_{00} = \rho c^2 = T$), one can write

$$-\frac{1}{2}\Delta g_{00} = -\kappa(T_{00} - \frac{1}{2}g_{00}T)$$

With the traditional Newtonian approximation:

$$g_{00} = 1 + \frac{2}{c^2}\varphi$$

It gives:

$$\frac{1}{c^2}(\Delta\varphi) = \frac{8\pi G}{c^4}\rho c^2\left(1 - \frac{1}{2}\left(1 + \frac{2}{c^2}\varphi\right)\right)$$

$$\Delta\varphi = 8\pi G\rho\left(\frac{1}{2} - \frac{1}{c^2}\varphi\right)$$

$$\Delta\varphi = 4\pi G\rho\left(1 - \frac{2}{c^2}\varphi\right)$$

In this approximation ($\left|\frac{2}{c^2}\varphi\right| \ll 1$), it gives the Newtonian approximation (HOBSON et al., 2009):

$$\Delta\varphi = 4\pi G\rho$$

Now let's use the Einstein equations with our linearized general relativity approximation:

$$\frac{1}{2}\Delta g_{00} = \kappa\left(T_{00} - \frac{1}{2}g_{00}T\right) \text{ and } g_{00} = 1 + \frac{2}{c^2}(\varphi - 8\vec{v}.\vec{H})$$

It gives:

$$\frac{1}{c^2}\left(\Delta\varphi - 8\Delta(\vec{v}.\vec{H})\right) = \frac{8\pi G}{c^4}\rho c^2\left(1 - \frac{1}{2}\left(1 + \frac{2}{c^2}(\varphi - 8\vec{v}.\vec{H})\right)\right)$$

With the assumption of a uniform $\vec{v}$ (ie $\partial_i\vec{v} \sim 0$) and with Poisson equation ($III$) ($\Delta\vec{H} = \frac{4\pi G}{c^2}\rho\vec{u}$ with $\vec{u}$ the speed of the source), this equation becomes:

$$\left(\Delta\varphi - 32\pi G\rho\frac{\vec{v}.\vec{u}}{c^2}\right) = 8\pi G\rho\left(\frac{1}{2} - \frac{1}{c^2}(\varphi - 8\vec{v}.\vec{H})\right)$$

$$\Delta\varphi = 4\pi G\rho\left(1 - \frac{2}{c^2}(\varphi - 8\vec{v}.\vec{H})\right) + 32\pi G\rho\frac{\vec{v}.\vec{u}}{c^2}$$

In our approximation ($\left|\frac{2}{c^2}(\varphi - 8\vec{v}.\vec{H})\right| \ll 1$), it gives the linearized general relativity approximation:

$$\Delta\varphi = 4\pi G\rho + 32\pi G\rho\frac{\vec{v}.\vec{u}}{c^2}$$

We have then an equation in linearized general relativity approximation that can be interpreted as an idealization of the influence of visible matter $\rho_b$ ("baryonic matter") for the first term "$4\pi G\rho$" and of the gravitic field $\rho_{dm}$ (our dark matter explanation) for the second term "$32\pi G\rho\frac{\vec{v}.\vec{u}}{c^2}$".

$$\Delta\varphi = 4\pi G\left(\rho + 8\rho\frac{\vec{v}.\vec{u}}{c^2}\right) = 4\pi G(\rho_b + \rho_{dm})$$

It is a very interesting result. Even if it is an approximation, our idealization implies naturally to add a component similar to the traditional "ad hoc" dark matter term. One can then try to obtain an approximation of the $\Omega_{dm}$ term. Because of the disparities of the distribution of matter and its speed, one cannot use easily this relation to compute the equivalent dark matter quantities of gravitic field in our universe at current time. But at the time of CMB, distribution of matter can be considered homogeneous and speed of particles can be close to celerity of light $\|\vec{v}\|\sim\|\vec{u}\|\sim\alpha c$ with $\alpha$ a factor that must be close to 1. In this approximation, previous equation gives:

$$\Delta\varphi = 4\pi G(\rho + 8\rho\alpha^2) = 4\pi G(\rho_{b,CMB} + \rho_{dm,CMB})$$

It gives the very interesting ratio of gravitic field (equivalent to dark matter) compared to baryonic matter at the time of CMB:

$$\rho_{dm,CMB} \approx 8\alpha^2\rho_{b,CMB}$$

In term of traditional $\Omega_b$, it means that in the approximation of linearized general relativity, one has:

$$\frac{\Omega_{dm}}{\Omega_b} \approx 8\alpha^2$$

The observations give (PLANCK Collaboration, 2014):

$$\frac{\Omega_b}{\Omega_{dm}} = \frac{\Omega_b h^2}{\Omega_{dm}h^2} \sim \frac{0.022}{0.12} \sim 5.45 \quad => \quad \alpha\sim 0.8$$

With our approximation, this ratio can be obtained with the particles speed $\|\vec{v}\|\sim\|\vec{u}\|\sim 0.8c$, about 80% of the light celerity. The important result is not the accuracy of the value (because of our approximation, but one can note that it is not so bad) but it is the order of magnitude. This order of magnitude is impressive. This is the sixth main result of our study (and I recall that our explanation comes from native components of general relativity, without modification).

## 7. Discusion

### 7.1. Consolidation of our idealization

Let's see a generalization of our study. We have supposed that the internal gravitic field far from its source (idealized as a punctual mass) can be idealized as $k(r) = \frac{K_1}{r^2}$. This idealization can be criticized because if, for example, we look at the magnetic field of the Earth, its source is idealized in part as a magnetic dipole. The field generated by a dipole is then $k(r) = \frac{K_2}{r^3}$. Furthermore, if we always look at the magnetic field of the Earth, the evolution of the field far from its source is in fact not trivial (certainly a mix of $r^{-2}$ and $r^{-3}$ and depending on the direction). To consolidate our solution, we are now going to show that our solution is finally steady (it gives the same results) with an evolution in $r^{-3}$. For that, let's write $k(r) = \left(\frac{K_2}{r^3} + k_0\right)$. Our first result was that, from the rotation's curves of galaxies, one obtained a uniform gravitic field embedding the galaxies. This result is independent of the expression of $k(r)$, it depends only on the expression of the force $\|\vec{F_k}(r)\| = 4k(r)v(r)$. The first important consequence is then that we always can make the assumption of an external gravitic field relatively uniform along the galaxy. After, one has studied the gravitic field with the expression $k(r) = \left(\frac{K_1}{r^2} + k_0\right)$. It becomes with a dipolar source $k(r) = \left(\frac{K_2}{r^3} + k_0\right)$. This modification doesn't modify $k_0$ (value explaining the dark matter in our solution). Concretely, in the case of NGC300 for example, one obtains $k_0 \sim 10^{-16.2}$ instead of $k_0 \sim 10^{-16.31}$. The situation is different for $K_2$. The computation gives $K_2 \sim 10^{20}K_1$. But it doesn't modify our solution. Indeed, far from the center of galaxies (at around $r \sim 3 \times 10^{20}m \sim 10kpc$), the terms $\frac{K_2}{r^3}$ and $\frac{K_1}{r^2}$ becomes equivalent (and equivalent to $k_0$). The increase of $K_2$ is compensated by the evolution in $r^{-3}$. By consequence, $\frac{K_2}{r^3}$ is always too small to explain the dark matter of the ends of the galaxy and it becomes inferior to $k_0$ at around the same distance from the center of the galaxy than $\frac{K_1}{r^2}$ (meaning that the plane of the galaxy is always not influenced by $k_0$). The main result of our study is then still valid: a gravitic field of around $k_0 \sim 10^{-16.5}$ due to a structure of a higher scale than galaxies could explain the dark matter of the ends of the galaxies. And if we apply this dipolar idealization to determine which structure would be a good candidate (as in paragraph 4), the cluster always gives the larger value of internal gravitic field (roughly one obtains for all the cases the same values than the "punctual" idealization with a typical distance ten times lower, i.e. the relative contributions are preserved). For example, with $\frac{K_1}{r^2}$ one obtained $\frac{10^{3.7}K_1}{R^2} \sim \frac{4\times 10^{28.2}}{10^{46}} = 10^{-17.2}$ for a typical distance $R \sim 0.5 \times 10^{23}m \sim 2Mpc$. With $\frac{K_2}{r^3}$ one obtains $\frac{10^{20}\times 10^{3.7}K_1}{R^3} \sim \frac{10^{48.2}}{R^3} = 10^{-17.2}$ for a typical distance $R \sim 10^{21.8}m \sim 0.2Mpc$. One can also note that, this time, to obtain the good order of magnitude for $k_0$, for the cluster, one should have a larger value of $K_2$. Then, the profile of the gravitic field at the center of the cluster would imply very much dark matter than the case $\frac{K_1}{r^2}$. Furthermore, with an evolution as $\frac{K_2}{r^3}$, the interval $10^{-16.6} < k_0 < 10^{-16.2}$ represents, in term of distance, a typical size slightly inferior to the case $\frac{K_1}{r^2}$. In conclusion, an evolution as $\frac{K_2}{r^3}$ is still consistent with our solution but it seems to be a limit case, in particular because of the interval of $k_0$ (too thin) that makes this idealization more unlikely. If in $r^{-n}$, $n$ is too large, the ability to obtain an interval of the size of the cluster is more difficult.

One can show the same results with an evolution in $r^{-1}$ but with a larger interval for $k_0$ (that makes this idealization easier to conciliate with the observations than $\frac{K_2}{r^3}$).

To conclude, our explanation is steady with an idealization of the gravitic field in $r^{-s}$ for $s \in [1;3[$. This stability is interesting because if we look at the idealization of the magnetic field of the Earth, its evolution with the distance is complicated (nor $r^{-2}$ neither $r^{-3}$ but a combination of power associated with the punctual and dipolar idealizations). So, whatever the actual idealizations of the gravitic fields, there are great chances that our solution (around $k_0 \sim 16.5$) is consistent with them.

## 7.2. A way to measure the gravitic fields

Some recent papers (CLOWE et al., 2006; HARVEY et al., 2015) seem to prove the existence of an exotic matter. The main assumption of these studies is that dark matter is either invisible ordinary matter (not yet detected gases) or either exotic matter (non-baryonic). With this assumption, assuming there is no exotic matter one deduces that the gravitational lens effect is mainly due to the invisible ordinary gas. They then show that the spatial offset between the distributions of the gas (which would explain the dark matter) and the observed mass (visible) is so great that they cannot coincide. One then conclude that dark matter cannot be invisible ordinary matter. Finally, what is shown is that:
If dark matter is either invisible ordinary matter or either non-baryonic matter then there is no doubt that it can be only non-baryonic.
The second result of these studies is that if dark matter is exotic matter (previous deduction) then the possibilities of an exotic particle will greatly reduce. Somehow, while supporting the assumption of dark matter, they make it more unlikely (but not actually impossible).
What about our solution? These studies do not address this solution. Indeed, the fundamental assumption of this study is that the gravitational lens effect is mainly due to the gravity field of the matter (ordinary or not). But, as one shows in §6.3, at the scale of the galaxies, the light deviation in our solution is mainly due to the gravitic field. So, what has been attributed previously to an invisible gas must be, in our solution, attributed to the gravitic field (without added mass). It leads to two interesting consequences. First, because the gravitic field depends on the mass but also on the speed, there is no reason that the centroid of the mass coincides with the "centroid" of the gravitic field for which the mass is weighted by the speed and the angle between the speed and the gravitic field. It would then explain the observed asymmetry of the centroids. Secondly (certainly the main consequence), is that the gravitational lens effect become the main way to measure the gravitic field. Just like in electromagnetism, the spectral shift can give several information (the value of the magnetic field for example), the gravitational lens can give several information (the value of the gravitic field for example). So the previous papers, instead of invalidate our solution, open a way to test our solution. To address our solution, the gravitational lens effect has just need to be interpreted not in term of new quantity of mass but in term of gravitic fields (the second component of gravitoelectromagnetism required by general relativity). So, from the gravitational lens effect, one can expect to measure the gravitic field of the galaxies and the gravitic field of the clusters.

## 7.3. Some possible experimental tests of gravitic field

Only direct measures could be a proof of the effective existence of the external gravitic field. On Earth the weakness of this term makes it difficult to detect it before a long time. Certainly, experiments on Lense-Thirring effect with very high precision could lead to direct measures of our gravitic field explaining dark matter (by adding, in a classical point of view, a force "$k_0.v$" with $10^{-16.62} < k_0 < 10^{-16.3}$).

But right now, its existence could be indirectly tested by using the expected computed values of the gravitic fields «$\frac{K_1}{r^2}$» and $\overrightarrow{k_0}$ (explaining dark matter) to explain others phenomena.

For example, gravitic field could play a role in the collisions between galaxies but also in the following situations:
_The dynamics of internal organization of galaxies_: By studying galaxies at different steps of evolution, one could idealize the evolution of this field $\vec{k}$ according with time for a "typical" galaxy and then check that it is coherent with the evolution observed (for example precocity of organization according to the mass, because gravitic field should accelerate galaxies organization).
_The jets of galaxies_: One could also try to correlate, statistically, the evolution of the field $\vec{k}$ within the galaxy and thermal agitation towards its center with the size of the jets of galaxies. A priori these matter jets should appear where the gravitic and gravity forces become sufficiently weak energetically compared to thermal agitation. Very close to the galaxy rotational axis, the gravitic force ($\overrightarrow{F_k} \propto \vec{v}\wedge\vec{k}$) becomes small (even if $k$ grows, $v$ is null on the axis), letting matter escape along this rotational axis (explaining the narrowness of the jets, the position of the jets at the galaxy center of rotation and the two opposite direction along the rotational axis). In fact, even the existence of jets can also be more easily explained with gravitic force. More the matter is close to the center, more it accumulates thermal agitation to compensate gravity and gravitic forces (to avoid collapsing). But very close to the center, gravitic force decreases (contrary to gravity force) because of the low speed in the center. Then this very energetic matter cannot be kept by only gravity force. The only exit for this matter very close to the center is to escape explaining the existence of galactic jets.
_The increase of the gravitational effect_: We have seen that one can write in a Newtonian approximation $\|\overrightarrow{F_{GR}}\| = \|\overrightarrow{F_g}\|\left(1 + \frac{\|\overrightarrow{F_k}\|}{\|\overrightarrow{F_g}\|}\right)$, interpretable as a modification of the mass $\|\overrightarrow{F_{GR}}\| = m\left(1 + \frac{\|\overrightarrow{F_k}\|}{\|\overrightarrow{F_g}\|}\right)\|\vec{g}\|$. For the large astrophysical structures, the correction $\frac{\|\overrightarrow{F_k}\|}{\|\overrightarrow{F_g}\|}$ cannot be neglected and then the effect must be an increase of the effect of the gravitation. In fact, the gravitic field implies a general precocity of organization for all large structures (for example why not for early super massive black holes). Such an effect could explain the following unexplained situations: the disappearance of the pulsars in the center of our Galaxy (because the accumulation of our "dark matter", the gravitic field, could lead to their explosion), the oscillation of the far red giants (because there is a gravitic field very far from the center of the galaxy that tends to maintain a symmetrical movement around the perpendicular plane to the gravitic field), the disappearance of bulbs for the majority of galaxies contrary to the theoretical expectations (because the gravitic field would accelerate the contraction and without new matter there should be less frictions in galaxies' collisions), the existence of large pairs of clusters such as Bullet cluster or El Gordo (because the

precocity could be explained by the great gravitic fields of clusters), the recently discovered fast radio bursts (because a pulsar with its great speed of rotation could own an important gravitic field).

*The influence on the "redshift"*: With the trio {mass, rotation speed and redshift}, one could compare our solution with dark matter assumption. Briefly, with $\varphi_M$ the potential scalar associated with the measured mass of an object, $v$ its rotation speed, $\varphi_{DM}$ its potential scalar associated with the dark matter, $\vec{k}$ and $\|\vec{H}\|$ its gravitic field and its potential vector, one has for the theory with dark matter assumption (forces equilibrium and redshift definition):

$$\nabla\varphi_M + \nabla\varphi_{DM} = \frac{v^2}{r}$$

$$1 + z = \frac{\nu_E}{\nu_R} = \left(\frac{g_{00}(r_R)}{g_{00}(r_E)}\right)^{1/2} = \left(\frac{1 + \frac{2\varphi_M(r_R)}{c^2} + \frac{2\varphi_{DM}(r_R)}{c^2}}{1 + \frac{2\varphi_M(r_E)}{c^2} + \frac{2\varphi_{DM}(r_E)}{c^2}}\right)^{1/2}$$

And for our gravitic field (forces equilibrium and redshift definition) in our approximation:

$$\nabla\varphi_M + 4kv = \frac{v^2}{r}$$

$$1 + z = \frac{\nu_E}{\nu_R} = \left(\frac{g_{00}(r_R)}{g_{00}(r_E)}\right)^{1/2}$$

$$= \left(\frac{1 + \frac{2\varphi_M(r_R)}{c^2} - \frac{16\vec{v}(r_R).\vec{H}(r_R)}{c^2}}{1 + \frac{2\varphi_M(r_E)}{c^2} - \frac{16\vec{v}(r_E).\vec{H}(r_E)}{c^2}}\right)^{1/2}$$

In our solution, there is a coupling between rotation speed and gravitation. In the theory with dark matter assumption, there is not this coupling. These differences of idealization could make appear distinguishable situations with speed rotation. One can add that for the clusters, the spectral shift should be mainly due to the gravitic field (just like the gravitational lens effect). It is not impossible to imagine that it could then explain the unexplained intensity's at the energy 3.5keV observed in the clusters and why not also the anomaly of the cosmic ultraviolet background, five times larger than expected (with a phenomenon of MASER).

## 7.4. External gravitic field versus dark matter

It is now interesting to make the comparison between the more widely accepted assumption (dark matter) and our solution (significant gravitic field).

In traditional approximation (that neglects gravitic field), general relativity can explain very well the inhomogeneities of CMB, but by introducing two new ad hoc unexplainable terms. The most immediate way to interpret them is to postulate, for one of them, the existence of a matter that is not visible (dark matter) and, for the other, an energy embedding our Universe. But these interpretations pose some problems. Let's focus on the dark matter, subject of our study. One of them is that dark matter should be sensitive only to gravitation (and strangely not to electromagnetism). A second one is about its distribution that is significantly different from the ordinary matter (subject of the same gravitation). A third one is about its quantity (five to six times more abundant than visible matter) in contradiction with the experimental results which show no dark matter. Sometimes, one compares the assumption of dark matter with the one that Le Verrier had made for the existence of a new planet (Neptune). This comparison is very interesting because it effectively can emphasize two fundamental differences between these two situations. The first is that in the case of Le Verrier gravitation worked very well for the other planets. There was therefore no reason to modify the idealization of gravitation for one particular planet. In the assumption of dark matter, it is different. Gravitation does not work for nearly all galaxies. The second is that, for the path of Uranus, the discrepancies were sufficiently small to be explained by the presence of a single planet (slight perturbation of the planet trajectory). This explanation was consistent with the observations and the fact that this planet could not yet be detected (small enough effect). For galaxies, the discrepancy with idealization is so great that the visible matter is almost nothing compared to the dark matter, which is surprisingly at odds with current observations. It is then not a slight perturbation of our observations. We are clearly not in the same situation than Le Verrier. So, one can wonder if adopting the same solution is really relevant. The solution I propose leads to a new interpretation of these terms, entirely explained by current general relativity. And furthermore it will avoid the previous problems (in agreement with current experimental results):

*Ad hoc assumption solved*: The assumption of dark matter is an "ad hoc" assumption and the origin of dark matter is, until now, unexplainable. In our study, we have seen that the gravitic field (that is a native general relativity component) of clusters is large enough not to be neglected and to explain "dark matter". Furthermore, we have seen that several observations corroborate (and can be explained by) this origin of the external gravitic field.

*Strange behavior solved*: On one hand, dark matter makes the assumption of a matter with a very strange behavior because, contrary to all known matter, it doesn't interact with electromagnetism. On the other hand, general relativity implies the existence of a significant gravitic field for clusters (and not a "dark" gravitic field with strange behavior). Because gravitic field is a component of gravitation, it explains why it doesn't concern electromagnetism. The problem of a strange behavior of dark matter is then solved.

*Contradiction with experimental observations solved*: The dark matter assumption implies that we are embedded in nearly exclusively dark matter (ordinary matter should represent only a little part of the matter). Until now no experimental observation has revealed the main part of our universe. Inversely, external gravitic field (explaining the dark matter with $10^{-16.62} < k_0 < 10^{-16.3}$) is sufficiently small to be undetectable at our scale and sufficiently large to explain dark matter at large scale, in agreement with observations.

*Strange theoretical behavior*: In a theoretical point of view, one can say that all our known theories have been built with only the ordinary matter. The existence of the dark matter would mean that even if we are embedded in nearly exclusively dark matter, it doesn't influence the theories at our scale. Astonishingly our theories only need dark matter for very large structure. It is a strange theoretical consistency. The gravitic field allows retrieving this theoretical consistency.

*Dark matter distribution solved*: Despite the fact that dark matter, just like visible matter, undergoes gravitation interaction, dark matter distribution is astonishingly different from visible matter. With gravitic field there is no more this problem. Gravitic field is applied on current distribution of visible matter. This component increases the effect of gravity field. It then doesn't modify the distribution of matter, but it accelerates its organization. It even emphasizes the effect of a distribution in a plane perpendicular to gravitic field. As we have seen it, some observations seem to confirm these tendencies.

## 8. Conclusion

The linearized general relativity leads to an approximation of the gravitation, equivalent (in term of field equations) to electromagnetism. And just like the atomic spins can explain the magnetic field of magnets without adding a new term to the Maxwell idealization (or a new "dark charge"), the own gravitic field of large astrophysical structures can explain the dark matter (and certainly the dark energy (LE CORRE, 2015)) without adding new ad hoc terms to the Einstein idealization. Rather, the dark matter (and certainly the dark energy) would reveal this traditionally neglected component of general relativity. With our solution, dark matter and dark energy become two wonderful new proofs of general relativity idealization. Indeed, the classical tests of general relativity are usually associated with space curvature. But with our solution, general relativity reveals another specificity of its idealization, the gravitic field that becomes essential at large scale (about 25% of dark matter and 70% of dark energy).

To summarize, one can say that our solution, instead of adding matter, consists in taking into account gravitic fields. We have shown that to explain rotation speed of the ends of the galaxies, a gravitic field of around $10^{-16.5}$ is necessary. This value would be compliant with the expected gravitic field of clusters with their neighbors. This origin is also in agreement with the recent observation of a decreasing quantity of dark matter with the distance to the center of galaxies' cluster and with the existence of ultra diffuse galaxies at center's cluster. This value applied to satellite dwarf galaxies gives the expected rotation speed and applied to light deviation is equivalent to the expected quantity of "dark matter" inside the galaxies. The theoretical expression also allows explaining the rotation in a plane of the satellite dwarf galaxies and retrieving a quantity of "dark matter" in the CMB with a good order of magnitude. One can add that this value is small enough to explain why it has not been detected until now at our scale. So, our solution reveals no contradiction with the experimental observations and the theoretical expectations.

Our solution also implies several non trivial predictions that differentiate it from dark matter assumption and from MOND theories. If these predictions are not verified, it will mean that our solution fails to explain dark matter. Inversely, it will be an important consolidation of our solution. Here are these twelve predictions:

1) The external gravitic field (our "dark matter") should decrease with the distance to the center of the cluster.
2) More, we are close to the center of the cluster, more the galaxies should have dark matter in very unusual and great proportions.
3) The galaxies without dark matter (with a declining rotation curve) must be at the ends of the cluster (far from its center) and far from others clusters (i.e. à priori at the borders of a superclusters).
4) The movement of satellite dwarf galaxies, leaded by only external $\overrightarrow{k_0}$, should be in a plane (perpendicular to $\overrightarrow{k_0}$).
5) Statistically, smaller is the distance between two galaxies; smaller is the difference of orientation of their satellite dwarf galaxies' planes.
6) Statistically, inside a cluster, the satellite dwarf galaxies' planes should be close to the supergalactic plane.
7) A galaxy can have two portions of its disk that rotate in opposite directions to each other.
8) If the gravitic fields would be effectively the main cause of the wrapping, there should be a correlation between the value of the wrapping and the angular differences of galaxies' and clusters' spin's vector.
9) The gravitic field of two close clusters must be close to parallel. In other words, the spin vector of these two close clusters must be close to parallel.
10) In our solution $\overrightarrow{k_0}$ (from cluster) and $\frac{\overrightarrow{K_1}}{r^2}$ (from galaxy) must be in the same half-space, for nearly all the galaxies inside a cluster. Say differently, galaxies' spin vector and spin vector of the cluster (that contains these galaxies) must be in a same half-space.
11) More a galaxy is young; more the dispersion of the speeds of the satellite dwarf galaxies should increase.
12) The speeds' dispersion along the flat curve should be relatively constant.

One can add that in our solution, as seen in our study, the following situations can be naturally explained: galaxies with decreasing rotation's speeds (i.e. without dark matter), galaxies with two portions of its disk that rotate in opposite directions to each other, galaxies with warped shape. Inversely, such situations are not natural expectations of dark matter assumption and MOND theories. And more generally, all the predictions due to the directions of the gravitic fields are out of the scope of these current theories which lead to an isotropic idealization. One must precise that our solution doesn't imply an anisotropic Universe. It implies anisotropy at the "local" (intermediary) scale of the clusters and superclusters. But globally, at the scale of all the clusters and superclusters (i.e. scale of the Universe), we can find all the possible directions. This gravitic component certainly also intervenes in galaxies organization (precocity according to the mass, narrowness of the basis of the jets, collisions…).

Of course, this solution need to be further tested and developed but these first results are very encouraging. In particular, it has to be studied in the general frame of general relativity (not only in its linearized approximation as in our paper), i.e. in the area where grvaitic field is very large (as the center of clusters) for which space's curvature makes linearized approximation not sufficient. One can wonder why the traditional calculations that take into account this term of gravitic field never revealed this solution (and we saw that this solution is a solution of general relativity, just like the magnetic field of magnets is a solution of Maxwell equations). An explanation can be that the simulations are not enough detailed or complete. One can be skeptical of this explanation but first one cannot exclude this possibility. In our solution, dark matter comes from the gravitic field of the clusters and our knowledge of these clusters is, at this day, not enough accurate (in particular about the direction and intensity of their spin's vectors). Secondly, to clarify my thinking, if we look at the situation of the magnetic field of the Earth, one has a similar problem. In a first approximation, the Earth is globally neutral and the matter's movement doesn't change its direction (rotation around the Sun and on itself). A conclusion could be that the magnetic field of the Earth should be negligible and it should never change its direction. Actually, the magnetic field is not negligible and its direction evolves with time. In fact, it can be explained by the complex movement of the matter inside the Earth (that is then not easy to detect and to idealize). Once again, our knowledge of the center of the clusters is not sufficient, for example we just have discovered the existence of ultra diffuse galaxies in this area (which seem be composed of only 1% of baryonic matter). And third, the gravitic field (as a sum of several gravitic fields) can depend on another parameter, just like in a phase's transition. For example with magnetic materials, depending on the agitation, the spins can be in a state where there is no global field and when the temperature decreases they can make appear a global field. In one case, the simulations don't give any embedding magnetic field, in the other they do. So, one cannot exclude that we haven't enough knowledge of the complex movement of internal matter for the large astrophysical structure to make completely relevant simulations. I think it would be a mistake to dismiss the proposed solution only on the fact that the traditional calculations have never revealed such a solution.

Taking into account significant gravitic fields should deeply change cosmology at greater scales than the galaxies, as the magnetic field at the microscopic level. For example, one may wonder what role the gravitic field could have on the geometry at the origin of the universe (flatness of the universe). One can also imagine that the value of this external gravitic field could become a signature of the space localization of galaxies (as a digital print). Another interesting point is that gravitic field can open a way to explain dark energy, creating by the same time a link between dark matter and dark energy.

## 9. Acknowledgements

I thank Dr Eric Simon for his encouragement and useful correspondence on this topic.

Tab. 2: Numerical approximations for $v_{halo}(r)$ and $v(r)$ used to compute $k(r)$ of Fig.4.

| | $v_{halo}(y)$ | $v(y)$ |
|---|---|---|
| Curves in Fig.3 | $-\frac{5}{54}y^6+\frac{115}{18}y^5-\frac{54065}{378}y^4+\frac{64285}{42}y^3$<br>$-\frac{1828390}{189}y^2+\frac{3041650}{63}y$ | $.07526914441y^8-3.199934769y^7+54.98000686y^6$<br>$-486.9847011y^5+2305.246690y^4$<br>$-4872.331582y^3-3569.566352y^2$<br>$+49571.78060y+50000$ |

Tab. 3: Numerical approximations for $v_{halo}(r)$ and $v(r)$ used to compute $k_{exp}(r)$ of Fig. 7.

| | $v_{halo}(y)$ | $v(y)$ |
|---|---|---|
| NGC 5055 | $\frac{3726000}{313}*y-\frac{11320}{313}*y^3, y<5$<br>$-\frac{2230000}{313}+\frac{5064000}{313}*y-\frac{267600}{313}*y^2+\frac{6520}{313}$<br>$*y^3, y<10$<br>$\frac{2815000}{313}+\frac{3550500}{313}*y-\frac{116250}{313}*y^2+\frac{1475}{313}$<br>$*y^3, y<20$<br>$\frac{8495000}{313}+\frac{2698500}{313}*y-\frac{73650}{313}*y^2+\frac{765}{313}$<br>$*y^3, y<30$<br>$24830000/313+1065000/313*y-19200/313$<br>$*y\text{^}2+160/313$<br>$*y\text{^}3, otherwise$ | $71068.3101*y+275726.759*y^3, y<.5$<br>$274745.7897-1577406.427*y+3296949.476*y^2$<br>$-1922239.558*y^3, y<.6$<br>$-123337.8881+413011.9624*y-20414.5077*y^2$<br>$-79259.5671*y^3, y<1$<br>$-373366.8873+1163098.961*y-770501.5063*y^2$<br>$+170769.4325*y^3, y<1.5$<br>$202116.4129+12132.36060*y-3190.439287*y^2$<br>$+255.8621300*y^3, y<5$<br>$262248.7924-23947.06710*y+4025.446258*y^2$<br>$-225.1969059*y^3, y<8$<br>$-49697.3974+93032.75412*y-10597.03140*y^2$<br>$+384.0729965*y^3, y<10$<br>$411968.2980-45466.95452*y+3252.939462*y^2$<br>$-77.59269895*y^3, y<15$<br>$89680.1135+18990.68234*y-1044.236329*y^2$<br>$+17.90009639*y^3, y<20$<br>$232804.0362-2477.906045*y+29.1930908*y^2$<br>$+.96060556e-2*y^3, y<30$<br>$260115.2720-5209.02962*y+120.2305432*y\text{^}2$<br>$-1.001921193*y\text{^}3, otherwise$ |
| NGC 4258 | $\frac{127220328}{13325}*y+\frac{1507418}{13325}*y^3, y<2$ | $202541.4800*y-254148.0049*y^3, y<.1$<br>$-263.8226943+210456.1609*y-79146.80830*y^2$<br>$+9674.68944*y^3, y<3$ |
| | $\frac{32381064}{13325}+\frac{78648732}{13325}*y+\frac{24285798}{13325}*y^2-\frac{508043}{2665}*y^3, y<4$<br>$-\frac{1301896}{123}+\frac{41742716}{2665}*y-\frac{8230414}{13325}*y^2+\frac{508408}{39975}*y^3, y<10$<br>$-\frac{968}{13}+\frac{33339924}{2665}*y-\frac{4029018}{13325}*y^2+\frac{147114}{66625}*y^3, y<15$<br>$-\frac{2199256}{533}+\frac{35499492}{2665}*y-\frac{365298}{1025}*y^2+\frac{227098}{66625}*y^3, y<20$<br>$-\frac{22448600}{533}+\frac{10137300}{533}*y-\frac{8545626}{13325}*y^2+\frac{543494}{66625}*y^3, y<25$<br>$\frac{79122400}{533}-\frac{2051220}{533}*y+\frac{3642894}{13325}*y^2-\frac{20698}{5125}*y^3, y<30$<br>$-\frac{15028544}{533}+\frac{36819372}{2665}*y-\frac{4203018}{13325}*y^2+\frac{33362}{13325}*y^3, y<40$<br>$53347520/533+11178348/2665*y-199578/2665*y^2+33263/66625*y^3, otherwise$ | $322666.1470-112473.8089*y+28496.51494*y^2-2285.679817*y^3, y<5$<br>$-135294.1316+162302.3576*y-26458.71827*y^2+1378.002402*y^3, y<7$<br>$409186.8580-71046.63794*y+6876.852493*y^2-209.4057277*y^3, y<10$<br>$324159.1553-45538.32711*y+4326.021405*y^2-124.3780247*y^3, y<13$<br>$-3937.00148+30176.17064*y-1498.170729*y^2+24.96023518*y^3, y<20$<br>$195373.7921+279.551521*y-3.3397692*y^2+.4638564710e-1*y^3, y<25$<br>$196243.5202+175.18414*y+.8349260*y^2-.9276955555e-2*y^3, otherwise$ |
| NGC 5033 | $\frac{4823500}{313}*y-\frac{17660}{313}*y^3, y<5$<br>$-\frac{3855000}{313}+\frac{7136500}{313}*y-\frac{462600}{313}*y^2+\frac{13180}{313}*y^3, y<10$<br>$\frac{8470000}{313}+\frac{3439000}{313}*y-\frac{92850}{313}*y^2+\frac{855}{313}*y^3, y<20$<br>$\frac{4590000}{313}+\frac{4021000}{313}*y-\frac{121950}{313}*y^2+\frac{1340}{313}*y^3, y<30$<br>$39555000/313+524500/313*y-5400/313*y^2+45/313*y^3, otherwise$ | $127886.6993*y+208453.2027*y^3, y<.5$<br>$219738.9134-1190546.782*y+2636866.962*y^2-1549458.105*y^3, y<.6$<br>$-95482.75032+385561.5368*y+10019.76474*y^2-90098.55117*y^3, y<1$<br>$-356911.4272+1169847.569*y-774266.2682*y^2+171330.1266*y^3, y<1.5$<br>$220238.0172+15548.67987*y-4733.67545*y^2+322.8837593*y^3, y<5$<br>$229445.0169+10024.47988*y-3628.835446*y^2+249.2277590*y^3, y<8$<br>$615622.9469-134792.2439*y+14473.25503*y^2-505.0260109*y^3, y<10$<br>$61694.0144+31386.43587*y-2144.612949*y^2+48.90292179*y^3, y<15$<br>$272319.0307-10738.56739*y+663.7206015*y^2-13.50449044*y^3, y<20$<br>$95433.49906+15794.26233*y-662.9208847*y^2+8.606201007*y^3, y<30$<br>$428274.4118-17489.82894*y+446.5488241*y^2-3.721240201*y^3, otherwise$ |
| NGC 2841 | $\frac{11433125}{679}*y-\frac{22765}{679}*y^3, y<5$<br>$-\frac{3493750}{679}+\frac{13529375}{679}*y-\frac{419250}{679}*y^2+\frac{5185}{679}*y^3, y<10$<br>$-\frac{333750}{679}+\frac{12581375}{679}*y-\frac{46350}{97}*y^2+\frac{2025}{679}*y^3, y<15$<br>$-\frac{40327500}{679}+\frac{20580125}{679}*y-\frac{857700}{679}*y^2+\frac{13875}{679}*y^3, y<20$<br>$\frac{96312500}{679}+\frac{84125}{679}*y+\frac{167100}{679}*y^2-\frac{3205}{679}*y^3, y<25$<br>$\frac{62718750}{679}+\frac{4115375}{679}*y+\frac{5850}{679}*y^2-\frac{1055}{679}*y^3, y<30$ | $243232.3588*y+169191.0307*y^3, y<.2$<br>$4060.584737+182323.5877*y+304543.8552*y^2-338382.0614*y^3, y<.6$<br>$-122346.7788+814360.4056*y-748850.8411*y^2+246837.2144*y^3, y<1$<br>$124174.0771+74797.83757*y-9288.27327*y^2+316.3585146*y^3, y<4$<br>$118154.4151+79312.58412*y-10416.95991*y^2+410.4157346*y^3, y<10$<br>$711596.8253-98720.13905*y+7386.312414*y^2-183.0266764*y^3, y<15$<br>$-300751.9836+103749.6228*y-6111.671710*y^2+116.9285264*y^3, y<20$<br>$1312175.526-138189.5036*y+5985.284610*y^2-84.68741227*y^3, y<25$<br>$-352020.0201+61513.96177*y-2002.854001*y^2+21.82110251*y^3, y<30$<br>$202090.4505+6102.91471*y-155.8190990*y^2+1.298492492*y^3, otherwise$ |
| | $-\frac{23748750}{97}+\frac{27011375}{679}*y-\frac{757350}{679}*y^2+\frac{7425}{679}*y^3, y < 35$<br>$30825000/97-5732875/679*y+178200/679*y^2-1485/679*y^3, otherwise$ | |
| NGC 3198 | $\frac{293412740}{6367}*y-\frac{9683185}{6367}*y^3, y < 2$<br>$-\frac{146442880}{6367}+\frac{513077060}{6367}*y-\frac{109832160}{6367}*y^2+\frac{8622175}{6367}*y^3, y < 4$<br>$\frac{1151080000}{19101}+\frac{115474900}{6367}*y-\frac{10431620}{6367}*y^2+\frac{1016390}{19101}*y^3, y < 10$<br>$\frac{753526000}{6367}+\frac{4525100}{6367}*y+\frac{663360}{6367}*y^2-\frac{31036}{6367}*y^3, y < 15$<br>$\frac{451261000}{6367}+\frac{64978100}{6367}*y-\frac{3366840}{6367}*y^2+\frac{58524}{6367}*y^3, y < 20$<br>$\frac{913981000}{6367}-\frac{4429900}{6367}*y+\frac{103560}{6367}*y^2+\frac{684}{6367}*y^3, y < 25$<br>$1085981000/6367-25069900/6367*y+929160/6367*y^2-10324/6367*y^3, otherwise$ | $93824.86109*y-16552.19458*y^3, y < .7$<br>$-8178.94368+128877.4768*y-50075.16525*y^2+7293.12221*y^3, y < 2.5$<br>$166259.5811-80448.75314*y+33655.32672*y^2-3870.943384*y^3, y < 3.5$<br>$-68364.34133+120657.4660*y-23803.59303*y^2+1601.334687*y^3, y < 5$<br>$137821.3623-3053.95617*y+938.691490*y^2-48.150949*y^3, y < 10$<br>$47486.56233+24046.48382*y-1771.352509*y^2+42.18385099*y^3, y < 15$<br>$218829.5039-10222.10438*y+513.2200344*y^2-8.58442774*y^3, y < 20$<br>$148923.5942+263.780383*y-11.07416018*y^2+.1538077657*y^3, y < 25$<br>$151807.4887-82.28696*y+2.7685335*y^2-.3076148333e-1*y^3, otherwise$ |
| NGC 7331 | $\frac{4794600}{443}*y-\frac{3952}{443}*y^3, y < 5$<br>$-\frac{306000}{443}+\frac{4978200}{443}*y-\frac{36720}{443}*y^2-\frac{1504}{443}*y^3, y < 10$<br>$-\frac{3078000}{443}+\frac{5809800}{443}*y-\frac{119880}{443}*y^2+\frac{1268}{443}*y^3, y < 20$<br>$-39074000/4873+64690200/4873*y-1357800/4873*y^2+14600/4873*y^3, otherwise$ | $212182.6079*y-128730.4317*y^3, y < .5$<br>$-36547.82379+431469.5507*y-438573.8856*y^2+163652.1587*y^3, y < 1$<br>$172982.5391-197121.5382*y+190017.2033*y^2-45878.20428*y^3, y < 1.5$<br>$13425.08720+121993.3659*y-22726.06625*y^2+1398.077914*y^3, y < 5$<br>$151016.1452+39438.73131*y-6215.139392*y^2+297.3494651*y^3, y < 8$<br>$358311.0499-38296.85847*y+3501.809375*y^2-107.5234025*y^3, y < 10$<br>$253301.4346-6793.973966*y+351.5209292*y^2-2.51378785*y^3, y < 15$<br>$332142.0734-22562.10177*y+1402.729450*y^2-25.87397720*y^3, y < 20$<br>$85247.19501+14472.12994*y-448.9821363*y^2+4.987882573*y^3, y < 30$<br>$219854.5902+1011.39042*y-.2908189*y^2+.2423490834e-2*y^3, otherwise$ |
| NGC 2903 | $21379/143640*y^5-850483/71820*y^4+1486603/4104*y^3-78143515/14364*y^2+52108475/1197*y$ | $415459.9988*y-1545999.876*y^3, y < .1$<br>$-1730.507376+467375.2200*y-519152.2128*y^2+184507.5001*y^3, y < 1$<br>$185740.5740-95038.02464*y+43261.03254*y^2-2963.58188*y^3, y < 2$<br>$305413.2309-274547.0112*y+133015.5265*y^2-17922.66429*y^3, y < 3$<br>$-326611.8774+357478.0971*y-77659.50959*y^2+5485.673053*y^3, y < 5$<br>$485961.3992-130065.8686*y+19849.28352*y^2-1014.913154*y^3, y < 7$<br>$-147538.6900+141434.1694*y-18936.43614*y^2+832.0258747*y^3, y < 8$<br>$376318.2422-55012.18008*y+5619.357539*y^2-191.1321953*y^3, y < 10$<br>$179676.8668+3980.232569*y-279.8837259*y^2+5.509180172*y^3, y < 15$<br>$163516.5331+7212.29966*y-495.3548729*y^2+10.29742797*y^3, y < 20$<br>$354776.6712-21476.72119*y+939.0961703*y^2-13.61008942*y^3, otherwise$ |
| NGC 3031 | $26743.83559 * y - 439.013694 * y^3, y < 2.5$<br>$-11157.53400 + 40132.87638 * y - 5355.61631 * y^2 + 275.0684805 * y^3, y < 5$<br>$13520.54625 + 25326.02821 * y - 2394.246683 * y^2 + 77.6438387 * y^3, y < 10$<br>$83547.94313 + 4317.808584 * y - 293.4246740 * y^2 + 7.616438449 * y^3, y < 15$<br>$123123.2679 - 3597.256355 * y + 234.2463215 * y^2 - 4.109583667 * y^3, y < 20$<br>$83671.29132 + 2320.53991 * y - 61.6434925 * y\text{^}2 + .8219132331 * y\text{^}3, otherwise$ | $189876.7738 * y + 253080.6547 * y^3, y < .2$<br>$4823.935713 + 117517.7382 * y + 361795.1784 * y^2 - 349911.3094 * y^3, y < .6$<br>$-115914.8568 + 721211.7004 * y - 644361.4254 * y^2 + 209064.5817 * y^3, y < 1$<br>$91465.97151 + 99069.21553 * y - 22218.94053 * y^2 + 1683.753411 * y^3, y < 4$<br>$190434.0526 + 24843.14000 * y - 3662.419121 * y^2 + 137.3764595 * y^3, y < 10$<br>$331893.2604 - 17594.62234 * y + 581.357113 * y^2 - 4.08274834 * y^3, y < 15$<br>$463698.3927 - 43955.64878 * y + 2338.758875 * y^2 - 43.13612084 * y^3, y < 20$<br>$-14408.3742 + 27760.36621 * y - 1247.041875 * y\text{^}2 + 16.62722500 * y\text{^}3, otherwise$ |
| NGC 2403 | $-2 * y\text{^}4 + 356/3 * y\text{^}3 - 2650 * y\text{^}2 + 83800/3 * y$ | $-282354289225649567/12952950506496 * y\text{^}5 - 22573477007046319/269853135552 * y\text{^}3 + 133187390454658231/25905901012992 * y\text{^}6 + 101042261599165/3747960216 * y\text{^}2 + 3929289082412429/67463283888 * y\text{^}4 - 5292189132515423/6476475253248 * y\text{^}7 + 53566280982301/138164805402624 * y\text{^}10 - 162623638202939/23027467567104 * y\text{^}9 + 4686121933585051/51811802025984 * y\text{^}8 - 2190475429/414494416207872 * y\text{^}13 + 150173866691/414494416207872 * y\text{^}12 + 44202063850/503217 * y + 549071/15942092931072 * y\text{^}14 - 6069909731069/414494416207872 * y\text{^}11$ |
| NGC 247 | $-12500/49 * y\text{^}2 + 82500/7 * y$ | $-575/1482624 * y\text{^}10 + 169175/5189184 * y\text{^}9 - 2065225/1729728 * y\text{^}8 + 10776925/432432 * y\text{^}7 - 1125775775/3459456 * y\text{^}6 + 431509825/157248 * y\text{^}5 - 38626919725/2594592 * y\text{^}4 + 65166877975/1297296 * y\text{^}3 - 10701265525/108108 * y\text{^}2 + 15901450/143 * y$ |
| NGC 4236 | $-55/756 * y\text{^}6 + 110/63 * y\text{^}5 - 1480/189 * y\text{^}4 - 1685/42 * y\text{^}3 - 477865/756 * y\text{^}2 + 1975475/126 * y$ | $-10/693 * y\text{^}10 + 78425/99792 * y\text{^}9 - 604075/33264 * y\text{^}8 + 3859925/16632 * y\text{^}7 - 1420195/792 * y\text{^}6 + 40909925/4752 * y\text{^}5 - 31346025/1232 * y\text{^}4 + 277117825/6237 * y\text{^}3 - 367945145/8316 * y\text{^}2 + 29263225/693 * y$ |
| NGC 4736 | $\frac{46336750}{691} * y - \frac{4876750}{691} * y^3, y < 1$<br>$-\frac{8530500}{691} + \frac{71928250}{691} * y - \frac{25591500}{691} * y^2 + \frac{3653750}{691} * y^3, y < 2$<br>$\frac{15685500}{691} + \frac{35604250}{691} * y - \frac{7429500}{691} * y^2 + \frac{626750}{691} * y^3, y < 3$<br>$\frac{27136875}{691} + \frac{24152875}{691} * y - \frac{3612375}{691} * y^2 + \frac{202625}{691} * y^3, y < 5$ | $550647.7605 * y - 5064776.046 * y^3, y < .1$<br>$-10388.65628 + 862307.4487 * y - 3116596.883 * y^2 + 5323880.231 * y^3, y < .2$<br>$35233.72684 + 177971.7020 * y + 305081.8504 * y^2 - 378917.659 * y^3, y < .6$<br>$-115393.8023 + 931109.3480 * y - 950147.5600 * y^2 + 318432.0135 * y^3, y < 1$<br>$203629.0993 - 25959.35679 * y + 6921.144988 * y^2 - 590.8875212 * y^3, y < 5$<br>$84986.1054 + 45226.43958 * y - 7316.014285 * y^2 + 358.2564302 * y^3, y < 8$<br>$398938.2925 - 72505.63059 * y + 7400.494489 * y^2 - 254.9314355 * y^3, y < 10$ |
| | $\frac{47360000}{691}+\frac{12019000}{691}*y-\frac{1185600}{691}*y^2+\frac{40840}{691}*y^3, y < 10$<br>$90840000/691-1025000/691*y+118800/691*y^2-2640/691*y^3, otherwise$ | $127510.2856+8922.77147*y-742.345718*y^2+16.49657151*y^3, otherwise$ |
| NGC 300 | $.8201058201*y^6-23.25396825*y^5+212.4074073*y^4-316.9047616*y^3-5996.084656*y^2+38123.01587*y$ | $-115/189*y^6+2965/126*y^5-142595/378*y^4+137455/42*y^3-911945/54*y^2+476725/9*y$ |
| NGC 2259 | $325/12852*y^7-415/378*y^6+15785/918*y^5-388660/3213*y^4+4719565/12852*y^3-4030265/6426*y^2+13242700/1071*y$ | $-.1005701747e-1*y^{10}+.6559331859*y^9-18.29303697*y^8+285.4915926*y^7-2735.333803*y^6+16578.82267*y^5-63092.40395*y^4+145111.6377*y^3-190899.9559*y^2+147769.3889*y$ |
| NGC 3109 | $14000/3*y$ | $.1172414586e-1*y^9-.6726765728*y^8+16.43331861*y^7-223.1214090*y^6+1841.805727*y^5-9488.700957*y^4+30135.90051*y^3-56649.97586*y^2+62368.31962*y$ |
| NGC 224 | $\frac{241000}{39}*y-\frac{280}{39}*y^3, y < 5$<br>$-\frac{31000}{13}+\frac{296800}{39}*y-\frac{3720}{13}*y^2+\frac{464}{39}*y^3, y < 10$<br>$\frac{233000}{13}+\frac{59200}{39}*y+\frac{4200}{13}*y^2-\frac{328}{39}*y^3, y < 15$<br>$-\frac{388000}{13}+\frac{431800}{39}*y-\frac{4080}{13}*y^2+\frac{224}{39}*y^3, y < 20$<br>$\frac{60000}{13}+\frac{230200}{39}*y-\frac{720}{13}*y^2+\frac{56}{39}*y^3, y < 25$<br>$1060000/13-129800/39*y+4080/13*y^2-136/39*y^3, otherwise$ | $\frac{9329878000}{46689}*y-\frac{925858000}{46689}*y^3, y < 1$<br>$-\frac{1262020000}{46689}+\frac{13115938000}{46689}*y-\frac{1262020000}{15563}*y^2+\frac{112054000}{15563}*y^3, y < 4$<br>$\frac{21329980000}{46689}-\frac{3828062000}{46689}*y+\frac{149980000}{15563}*y^2-\frac{16838000}{46689}*y^3, y < 10$<br>$\frac{31020000}{15563}+\frac{2543014000}{46689}*y-\frac{62389200}{15563}*y^2+\frac{4398920}{46689}*y^3, y < 15$<br>$\frac{4883730000}{15563}-\frac{368612000}{46689}*y+\frac{2313600}{15563}*y^2+\frac{85400}{46689}*y^3, y < 20$<br>$\frac{7792530000}{15563}-\frac{1677572000}{46689}*y+\frac{24129600}{15563}*y^2-\frac{1005400}{46689}*y^3, y < 25$<br>$1508780000/15563+584578000/46689*y-6032400/15563*y^2+201080/46689*y^3, otherwise$ |